\documentclass[]{gji}
\usepackage{timet,color}
\usepackage[urlcolor=blue,citecolor=black,linkcolor=black]{hyperref}
\usepackage{url} 
\usepackage{graphicx}
\usepackage{url}
\usepackage{lineno}
\usepackage{mathtools}
\usepackage{soul}
\usepackage{amsmath, amssymb}
\usepackage{blindtext}
\usepackage{indentfirst}
\usepackage{wrapfig}
\usepackage{cleveref}
\usepackage{gensymb} 
\usepackage{multirow}

\newcounter{supfig}
\newcounter{suptab}
\newcounter{supsec}
\DeclareUnicodeCharacter{1EA1}{\d{a}}
\DeclareUnicodeCharacter{030C}{\v}
\DeclareUnicodeCharacter{0301}{\'}
\let\bibhang\relax
\usepackage{natbib}
\makeatletter

\providecommand{\bibhang}{1.5em}
\makeatother
\title[Theory errors in moment tensor inversion with SBI]
  {Improving moment tensor solutions under Earth structure uncertainty with simulation-based inference}
\author[A. A. Saoulis et al.]
  {A. A. Saoulis$^{1,2}$\thanks{a.saoulis@ucl.ac.uk}, T.-S. Phạm$^3$, A. M. G. Ferreira$^2$ \\
   $^1$ Department of Physics \& Astronomy, University College London, Gower Street, London, WC1E 6BT, United Kingdom \\
   $^2$ Department of Earth Sciences, University College London, 5 Gower Place, London, WC1E 6BS, United Kingdom \\
   $^3$ Research School of Earth Sciences, The Australian National University, Canberra, ACT, Australia
}
\pagerange{\pageref{firstpage}--\pageref{lastpage}}
\pubyear{{\the\year{}}}

\begin{document}

\label{firstpage}

\maketitle

\begin{summary}
Bayesian inference represents a principled way to incorporate Earth structure uncertainty in full-waveform moment tensor inversions, but traditional approaches generally require significant approximations that risk biasing the resulting solutions. We introduce a robust method for handling theory errors using simulation-based inference (SBI), a machine learning approach that empirically models their impact on the observations. This framework retains the rigour of Bayesian inference while avoiding restrictive assumptions about the functional form of the uncertainties. We begin by demonstrating that the common Gaussian parametrisation of theory errors breaks down under minor ($1-3 \%$) 1-D Earth model uncertainty. To address this issue, we develop two formalisms for utilising SBI to improve the quality of the moment tensor solutions: one using physics-based insights into the theory errors, and another utilising an end-to-end deep learning algorithm. We then compare the results of moment tensor inversion with the standard Gaussian approach and SBI, and demonstrate that Gaussian assumptions induce bias and significantly under-report moment tensor uncertainties. We also show that these effects are particularly problematic when inverting short period data and for shallow, isotropic events. On the other hand, SBI produces more reliable, better calibrated posteriors of the earthquake source mechanism. Finally, we successfully apply our methodology to two well studied moderate magnitude earthquakes: one from the 1997 Long Valley Caldera volcanic earthquake sequence, and the 2020 Zagreb earthquake. 

\end{summary}


\section{Introduction}

Focal mechanism solutions of full-waveform moment tensor inversion have been shown to rely on high-quality forward modelling, which primarily depends on accurate Earth structure models (\citealp{hardebeck2002new,hjorleifsdottir2010effects, takemura2020centroid, rosler2021uncertainties, song2022moment,simute2023bayesian}, see \citealp{pham2025global} for a recent review). These effects are particularly pronounced in regions with strongly heterogeneous or poorly constrained crustal structure  \citep{abercrombie2001earthquake,kvrivzova2013resolvability,petersen2021regional}. Reliable moment tensor solutions are also critical for downstream applications such as seismic tomography \citep{blom2023mitigating}. 

Beyond characterising individual events, heterogeneities can lead to systematic biases in 1-D Earth model derived source catalogs \citep{hingee2011seismic, hejrani2017centroid, rosler2024global}. These inaccuracies can induce biases in, and specious trade-offs between, moment tensor components \citep{dziewonski1983experiment, ferreira2006long, weston2011global, ferreira2011global}. Such effects have been shown to cause, for instance, systematic spurious non-double-couple components in retrieved seismic source solutions for tectonic events \citep{sawade2022global, rosler2024apparent}. Moreover, unresolved Earth structure uncertainty continues to pose significant challenges for interpreting seismic source solutions \citep[e.g.,][]{akuhara2025non}.  An appealing solution is Bayesian inference, which provides a framework for treating theory or mismodelling errors directly in the uncertainty estimation procedure \citep{tarantola2005inverse}. We assume the uncertain 1-D Earth model to be the primary source of theory errors in the context of this work.

Including the effect of theory errors in source inversions can significantly improve inversion quality \citep[e.g.,][]{yagi2011introduction,minson2013bayesian, minson2014bayesian, duputel2015iquique, phạm2021toward}. This is often achieved by treating theory errors as an additional term in a Gaussian likelihood function, following \citet{tarantola2005inverse}. Several studies approximate the theory errors by sampling over the modelled Earth structure uncertainty empirically and estimating the resulting variability in the observations \citep{duputel2012uncertainty, vasyura2021accounting, phạm2021toward,poppeliers2021effects, phạm2024towards}, while others rely on theoretical approximations for the effect of theory errors on the observation covariance structure \citep{duputel2014accounting, hallo2016fast}. All these prior works show that incorporating theory error contributions yields improved solutions over the naïve approach. However, they perform significant approximations to make the inverse problem tractable by traditional means. The question of whether the resulting solutions are genuinely accurate and trustworthy thus still remains.

There are many other hybrid or empirically motivated approaches. Grid-search strategies have proven effective for tracing likelihood contours for uncertainty quantification \citep{sokos2013evaluating,vackavr2017bayesian,zahradnik2025isola2024, thurin2025mtuq}, though likelihood contours often do not have a clear probabilistic interpretation. Additionally, making this approach tractable for holistically treating theory errors may prove challenging. Another family of approaches relies on building solution ensembles through hybrid techniques, including bootstrapping \citep{weber2006probabilistic, valentine2012assessing} and by injecting empirically motivated theory errors \citep{poppeliers2022efficient}. A relevant example to this study is \citet{stahler2014fully, stahler2016fully}, which used earthquake source catalogues to build empirical likelihood functions that better account for mismodelling errors. While these approaches remain valuable, particularly as a means for dealing with unknown (or challenging to model) theory uncertainties, it is desirable to continue to develop a complementary approach that directly integrates our ability to model theory errors.

Simulation-based inference (SBI) is a promising alternative to fully treat the effects of theory errors without compromising (or abandoning) a rigorous Bayesian approach to moment tensor inversion.  SBI uses machine learning (ML) models to learn the observational uncertainties empirically through direct simulation, in principle allowing arbitrarily complex sources of uncertainty to be modelled without making parametric assumptions about their form \citep{alsing2019fast, Cranmer2020, zammit2025neural}. \citet{saoulis2025full} demonstrated that standard Gaussian likelihood assumptions about full-waveform data errors yielded inaccurate moment tensor solutions. SBI, on the other hand, produced accurate results at a fraction of the computational cost. This work applies SBI to full-waveform moment tensor inversion while incorporating theory errors.

SBI has received significant attention across the physical sciences for enabling fast and accurate Bayesian inference \citep{gonccalves2020training, stockman2024sb, von2025kids,atlas2025implementation, dax2025real}. A key practical challenge in adapting SBI for a given domain is preparing the observational data to make it amenable to ML-based probabilistic modelling \citep[see e.g.][for a practical perspective]{deistler2025sbi}. This is typically straightforward in low dimensions, but high-dimensional observations, such as full-waveform multi-station seismic data, can pose problems. The choice of data representation therefore plays a central role in determining both the accuracy and efficiency of SBI-based inversions \citep{gerardi2024optimal,park2025dimensionality}.

Within this context, two broad strategies have emerged for handling high-dimensional observations in SBI. One approach uses carefully selected lower dimensional statistics or generic linear compression algorithms to reduce the dimensionality \citep[e.g.,][]{alsing2019fast}. An alternative approach relies on data-driven ML methods to learn compressed representations directly from the observations. Increasingly, SBI has turned to deep learning to extract complex, non-linear information from high-dimensional data, improving the constraining power, accuracy, and efficiency of the inverse problem \citep[e.g.,][]{gloeckler2024all, lanzieri2025optimal,jeffrey2025dark, dax2025real}. Our work fuses this strand of SBI with a broad trend in seismology utilising deep learning for processing full-waveform data (\citealt{zhu2019phasenet, zhu2019seismic, mousavi2020earthquake, munchmeyer2021transformer, sun2023phase,mcbrearty2023earthquake, si2024seisclip}; see e.g. \citealt{mousavi2022deep, kubo2024recent} for reviews).

We propose two SBI frameworks for full-waveform moment tensor inversion that differ in how the observations are compressed. We assess a physics-motivated linear compression method, adapted from \citet{saoulis2025full} to address theory errors, and a data-driven deep learning approach. We explore the advantages of each approach, and compare both SBI frameworks with the most recent attempts using explicit-likelihood Bayesian inversion technique for treating theory errors \citep{phạm2021toward,phạm2024towards}.

The manuscript is laid out as follows. \Cref{sec:theory_intro} sets out the theoretical background, elucidating the approximations often made in parametric likelihood-based treatments of the theory errors, and explaining how SBI can bypass these limitations. \Cref{sec:gl_methods} presents the Gaussian likelihood approach for full-waveform moment tensor inversions, which serves as a benchmark for SBI. \Cref{sec:sbi_methods} presents the technical and architectural details for our two SBI frameworks. \Cref{sec:evaluation} introduces methods to evaluate the accuracy of Gaussian likelihood assumptions for a given problem, and how we evaluate the moment tensor solution quality of each approach. 

We then explore a range of applications. \Cref{sec:synthetic_inversions} presents a systematic evaluation of each of the approaches across three case-studies: how each approach fares under moderate Earth structure uncertainty; how Gaussian likelihood assumptions produce inaccurate results across a wide range of experimental configurations; and how SBI more accurately recovers moment tensors, including with short-period data and shallow, isotropic sources. We present the results of two real data inversions in \Cref{sec:real_data}, before discussing the broader limitations and implications of this work in \Cref{sec:discussion} and \Cref{sec:conclusions}.

\section{Theoretical background}
\label{sec:theory_intro}

\subsection{Posterior inference under Earth structure uncertainty}

For a given observation $\mathbf{D}$, the joint posterior distribution over the
source parameters $\mathbf{m}$ and the uncertain Earth model parameters $\Omega$
is given by
\begin{equation}
p(\mathbf{m}, \Omega \mid \mathbf{D}) 
= \frac{p(\mathbf{D} \mid \mathbf{m}, \Omega)\, p(\mathbf{m})\, p(\Omega)}
{p(\mathbf{D})},
\label{eq:joint_posterior}
\end{equation}
where $p(\mathbf{D} \mid \mathbf{m}, \Omega)$ is the likelihood (e.g. obtained
via the deterministic forward model $g(\Omega, \mathbf{m})$),
$p(\mathbf{m})$ and $p(\Omega)$ are the prior distributions over the source and
Earth model parameters respectively, and $p(\mathbf{D})$ is the marginal
likelihood (evidence).

Our ultimate goal, however, is to infer the posterior over the source parameters
$\mathbf{m}$ while properly accounting for unresolved uncertainty in the Earth
model. Analytically, this requires marginalising over $\Omega$:
\begin{align}
p(\mathbf{m} \mid \mathbf{D})
&= \int p(\mathbf{m}, \Omega \mid \mathbf{D})\, d\Omega, 
\label{eq:marginal_posterior_1} \\[6pt]
&= \frac{p(\mathbf{m}) \int p(\mathbf{D} \mid \mathbf{m}, \Omega)\, p(\Omega)\, d\Omega}
{p(\mathbf{D})}.
\label{eq:marginal_posterior_2}
\end{align}

A conceptually straightforward route would be to sample from the joint posterior in \Cref{eq:joint_posterior} directly using Markov Chain Monte Carlo (MCMC) or other joint-sampling algorithms, and then
marginalise the empirical sample set over $\Omega$ as in \Cref{eq:marginal_posterior_1}. This approach is exact in the limit of many posterior samples, and it extends to hierarchical formulations (e.g. by introducing
hyperparameters in the prior for $\Omega$ and sampling them jointly). The
principal drawback for seismological problems is cost: each proposed
$\Omega^{(i)}$ typically requires recomputing Green's functions, and the posterior dimensionality $\dim\{\mathbf{m}, \Omega\}$ may become very large, so joint sampling can be extremely expensive.

The more popular approach is to attempt to approximate the marginalisation of the likelihood over all
prior Earth structures in \Cref{eq:marginal_posterior_2} \citep{tarantola2005inverse}:
\begin{equation}
p(\mathbf{D} \mid \mathbf{m})
= \int p(\mathbf{D} \mid \mathbf{m}, \Omega)\, p(\Omega)\, d\Omega,
\label{eq:marginal_likelihood}
\end{equation}
which is typically analytically intractable. Most prior work approximates \Cref{eq:marginal_likelihood} by assuming that the variability introduced by $\Omega$ can be
represented as an additional stochastic term in the likelihood \citep{tarantola1982inverse}. In particular, assuming a Gaussian likelihood with data covariance matrix $\mathbf{C}_d$, we may write:
\begin{equation}
p(\mathbf{D} \mid \mathbf{m})
= \int\mathcal{N}\!\big(\mathbf{D}\,\big|\,g(\mathbf{m}, \Omega),\; \mathbf{C}_d\big)p(\Omega)d\Omega,
\label{eq:marginal_likelihood_gaussian}
\end{equation}
whereafter $g(\mathbf{m}, \Omega)$ can be expanded perturbatively around some fiducial model $\Omega^*$ to linear order. This yields
\begin{equation}
g(\mathbf{m}, \Omega) \approx g(\mathbf{m}, \Omega^*) + K_\Omega\mid_{\Omega^*} \delta\Omega +  ...
\label{eq:expansion}
\end{equation}
where $K_\Omega=\nabla_\Omega g(\mathbf{m}, \Omega)$  are the gradients (i.e., sensitivity kernels) with respect to the Earth model parameters. This can be substituted into \Cref{eq:marginal_likelihood_gaussian}:
\begin{equation}
p(\mathbf{D} \mid \mathbf{m})
= \int\mathcal{N}\!\big(\mathbf{D}\,\big|\,g(\mathbf{m}, \Omega^*) + K_\Omega \delta\Omega,\; \mathbf{C}_d\big)p(\Omega)d\Omega.
\label{eq:marginal_likelihood_expansion}
\end{equation}
To make progress, one can then assume that the Earth model perturbations are \textit{also} Gaussian distributed following $\delta\Omega \sim \mathcal{N}(0, \Sigma_\Omega)$. Applying simple Gaussian convolution rules produces:
\begin{equation}
p(\mathbf{D} \mid \mathbf{m})
\approx \mathcal{N}\!\big(\mathbf{D}\,\big|\,g(\mathbf{m}, \Omega^*),\; \mathbf{C}_d + \mathbf{C}_t(\mathbf{m})\big),
\label{eq:final_gaussian}
\end{equation}
where $\mathbf{C}_t(\mathbf{m}) = K_\Omega (\mathbf{m}) \Sigma_\Omega K_\Omega^T(\mathbf{m})$ is the theory covariance matrix. Note that kernels $K_\Omega$ may in general be expensive to compute. \Cref{eq:final_gaussian} can also be reached without assuming Gaussianity of $\delta \Omega$ \citep[e.g. as in][]{duputel2014accounting}. In that case, however, the induced distribution $p(\mathbf{D} \mid \mathbf{m})$ is not strictly Gaussian, and a Gaussian approximation is being introduced at the level of the likelihood.

In practice, $\mathbf{C}_{t}(\mathbf{m})$ can be estimated by Monte Carlo sampling. One can draw $N$ samples from the prior $p(\Omega)$, evaluate the forward model, and compute the empirical covariance of the modelled waveforms:
\begin{equation}
\Omega^{(n)} \sim p(\Omega), \qquad 
g^{(n)}(\mathbf{m}) \equiv g(\Omega^{(n)},\mathbf{m}), \quad n=1,\dots,N,
\label{eq:mc_omega_samples}
\end{equation}
\begin{equation}
\widehat{\mathbf{C}}_{t}(\mathbf{m}) \equiv \frac{1}{N-1}\sum_{n=1}^{N}
\big(g^{(n)}(\mathbf{m})-\overline{g}(\mathbf{m})\big)
\big(g^{(n)}(\mathbf{m})-\overline{g}(\mathbf{m})\big)^{\!T},
\label{eq:mc_theory_cov}
\end{equation}
where $\overline{g}(\mathbf{m})$ is a mean estimate of the synthetic samples $g^{(n)}(\mathbf{m})$.  The Gaussian likelihood in \Cref{eq:final_gaussian} can then be used for maximum likelihood estimation and posterior inference. 

At this stage, it is worth taking stock of the significant simplifications required to make previous approaches tractable. We made the following approximations:
\begin{enumerate}
    \item Perturbations to structure lead to small and additive perturbations on the forward model operator.
    \item Both the prior $p(\Omega)$ and the data noise model $p(\mathbf{D}\,\big|\,g(\mathbf{m}, \Omega),\; \mathbf{C}_d)$ must be Gaussian.
    \item The theory error contribution $\mathbf{C}_{t}(\mathbf{m})$ is only valid locally for a given set of source parameters $\mathbf{m}$, and is estimated empirically to avoid direct computation of Earth model kernels $K_\Omega$. 
    \item The data and theory error contributions are independent and additive.
\end{enumerate}
We note that these are not novel observations; (i) and (ii) are discussed in \citet[Sec.~5.8.6]{tarantola2005inverse}, for instance.

Of these approximations, (i)-(iii) all pose significant constraints and may introduce inaccuracies during posterior inference. In particular, (i) may be violated for realistic Earth structure perturbations that induce non-linear and multiplicative effects on the full waveforms, such as phase offsets and amplitude anomalies. (ii) is highly restrictive since in general $p(\Omega)$ may not be Gaussian (for instance, if $p(\Omega)$ is taken from a tomographic ensemble), and the data noise errors may not be Gaussian \citep[see e.g.][]{saoulis2025full}. For (iii), the fact that the covariance varies over parameter space $\mathbf{m}$ poses several practical problems that increase the computational costs of inversions. In addition, estimating and using the dense matrix $\widehat{\mathbf{C}}$ is often computationally challenging or intractable, so typically practitioners make simplifications (e.g. block sparse, per-station covariance matrices).  

\begin{figure*}
 \centering
    \includegraphics[width=\textwidth]{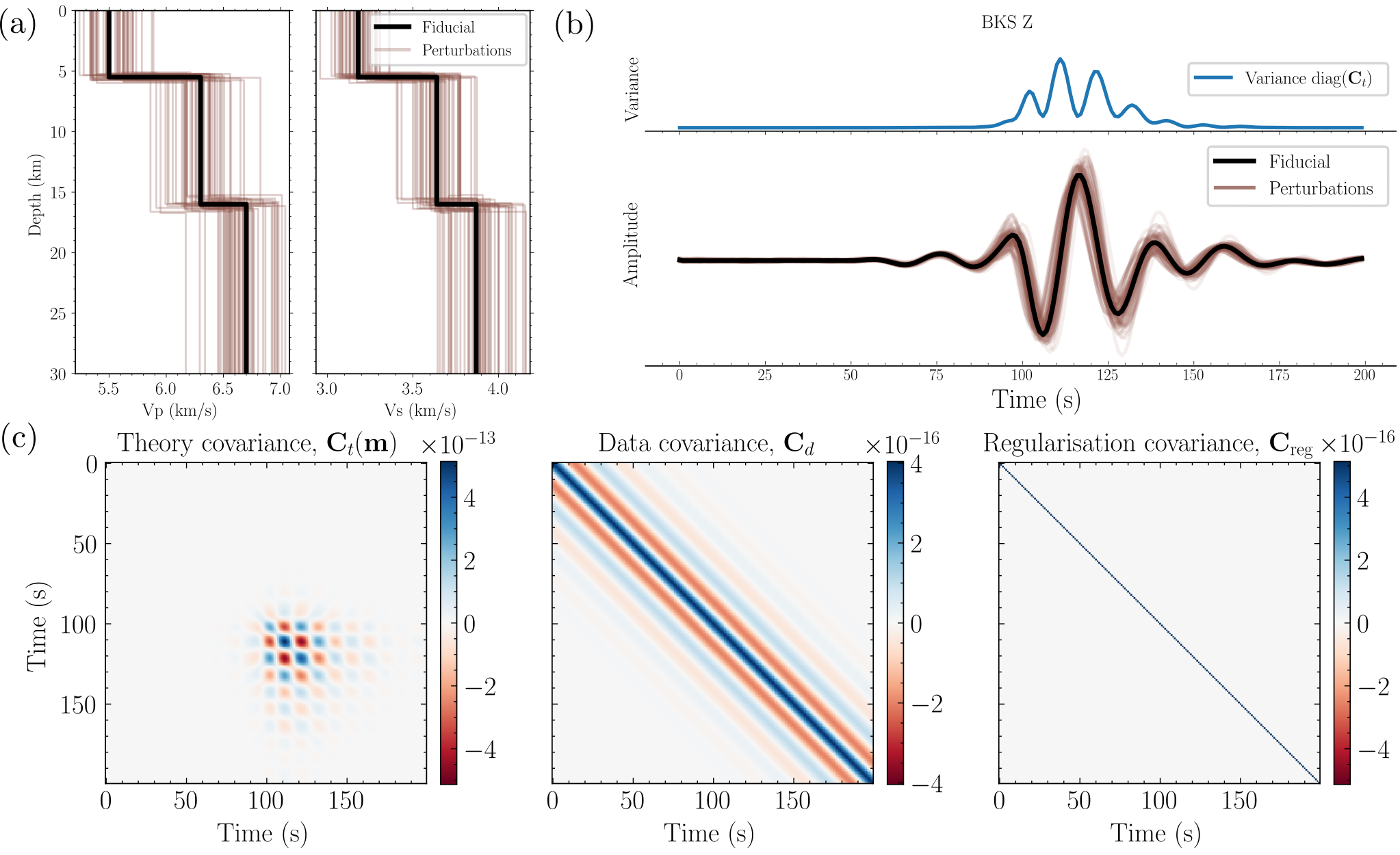}
   \caption{Panel (a) shows the effect of $\kappa=3\%$ fractional perturbations to the layer depths, $V_p$ and $V_s$ of the Southern California 1-D velocity model \citep{dreger1990broadband}. Panel (b) shows the resulting uncertainty in the observations for a single station component, as well as the variance in the seismograms (i.e. the diagonal of $\mathbf{C}_t$ in (c)). Panel (c) shows the three independent contributions to the Gaussian covariance $\mathbf{C}$. $\mathbf{C}_t(\mathbf{m})$ is computed for an event and receiver configuration similar to the LV2 earthquake studied in this work (see Sections \ref{sec:synthetic_inversions} \& \ref{sec:LV2}). Note the differences in the colourbar scales, where the theory errors dominate for this moderate magnitude event. }
   \label{fig:covariances}
\end{figure*}

\subsection{Simulation-based inference for theory errors}

Simulation-based inference (SBI) offers an alternative to such approximate analytic or
sampling-based marginalisation. SBI draws simulations from the joint prior predictive model:
\begin{equation}
\mathbf{m} \sim p(\mathbf{m}), \quad 
\Omega \sim p(\Omega), \quad 
\mathbf{D} \sim p(\mathbf{D} \mid \mathbf{m}, \Omega).
\label{eq:prior_predictive}
\end{equation}
This builds a realistic data set of observations $\mathcal{D}$: 
\begin{equation}
\mathcal{D} = \{\mathbf{m}, \mathbf{D}\}
\label{eq:training_data}
\end{equation}
which implicitly encodes the key quantities required for posterior inference, like the marginalised posterior $p(\mathbf{m}\mid\mathbf{D})$ from \Cref{eq:marginal_posterior_1} or marginalised likelihood $p(\mathbf{D}\mid\mathbf{m})$ from \Cref{eq:marginal_likelihood}.

Within this framework, a variety of probabilistic quantities may be modelled using machine learning, including the joint distribution, the likelihood (or ratios between likelihoods) or the posterior itself \citep{lueckmann2019likelihood, lueckmann2021benchmarking, durkan2020contrastive, hermans20nlre, Cranmer2020}. For example, an empirical likelihood can be combined with classical sampling-based approaches such as MCMC to perform posterior inference \citep{papamakarios2019sequential}. In this work, however, we focus on neural posterior estimation (NPE) in which the posterior density is modelled directly, enabling rapid and amortised posterior sampling once the model has been trained \citep{papamakarios2016fast, greenberg2019automatic,deistler2022truncated}.

SBI proceeds by using $\mathcal{D}$ to train a machine learning model $q_{\phi}$ with learnable parameters $\phi$ to approximate one of the key probabilistic quantities. This is achieved by training a neural network to minimise some loss function quantifying the divergence between the true and modelled probability distributions ($p^*(\mathbf{x})$ and $q_{\phi}(\mathbf{x})$, respectively).  A typical choice of loss function is the Kullback-Liebler (KL) divergence \citep{papamakarios2021normalizing}:
\begin{equation}
\begin{aligned}
\mathcal{L} &= \mathrm{KL}\big(p^*(\mathbf{x}) \, \| \, q_{\phi}(\mathbf{x})\big) = \mathbb{E}_{p^*(\mathbf{x})} \left[ \log p^*(\mathbf{x}) - \log q_{\phi}(\mathbf{x}) \right] \\
&= \text{const.} -\mathbb{E}_{p^*(\mathbf{x})} \left[ \log q_{\phi}(\mathbf{x}) \right] ,
\end{aligned}
\end{equation}
where the constant term can be ignored in an optimisation routine. The resulting objective function to be minimised for, say, the posterior $p(\mathbf{m}\mid\mathbf{D})$ is then: 
\begin{equation}
\mathcal{L}=-\mathbb{E}_{(\mathbf{m}_i, \mathbf{D}_i) \sim \mathcal{D}} \left[ \log q_{\phi}(\mathbf{m}_i \mid \mathbf{D}_i) \right].
\label{eq:sbi_training}
\end{equation}
This has a simple interpretation: given observation $\mathbf{D}_i$, the model $q_{\phi}$ is trained to maximise the posterior density assigned to the each ``true'' source parameter $\mathbf{m}_i$. Since $\Omega$ is sampled from its prior during dataset generation, the learned conditional density $q_{\phi}(\mathbf{m} \mid \mathbf{D})$ implicitly marginalises over Earth model uncertainty, avoiding the need for costly joint MCMC sampling or simplifying assumptions about the likelihood.

With this framework at hand, applying SBI hinges on the practical question of how to perform robust empirical density estimation. Much of the success of SBI over the last decade has been driven by the development of specialised ML models, referred to as neural density estimators (NDEs), which have proven capable of reliable and flexible density modelling \citep[see e.g.,][]{alsing2019fast, Cranmer2020, deistler2025sbi}. This work makes use of several popular classes of NDEs to empirically model the required density functions, with details presented in \Cref{sec:sbi_methods}. 

\section{The Gaussian likelihood}
\label{sec:gl_methods}

\subsection{Approximating the likelihood}

We intend to make comparisons between moment tensor inversions performed with a Gaussian likelihood, parametrised by some covariance $\widehat{\mathbf{C}}$, and inversions using SBI. We follow prior work by splitting the covariance into a theory error $\mathbf{C}_t(\mathbf{m}) $ and data error $\mathbf{C}_d$ contributions \citep{tarantola2005inverse}:
\begin{equation}
\widehat{\mathbf{C}} = \mathbf{C}_d + \mathbf{C}_t(\mathbf{m}).
\label{eq:data_theory_covariance}
\end{equation}
We estimate $\mathbf{C}_t(\mathbf{m})$ using Monte Carlo sampling outlined in Section 2, \Cref{eq:mc_theory_cov}. \citet{phạm2024towards} demonstrated that estimating $\overline{g}(\mathbf{m})$ as a mean of the Monte Carlo samples in \Cref{eq:mc_theory_cov} leads to bias. Instead, one should use the fiducial observation $\overline{g}(\mathbf{m}) =g(\mathbf{m}, \Omega^*)$ for a better estimate.

There is a wide range of approaches for specifying $\mathbf{C}_d$. Prior work has demonstrated that several of these, such as constant diagonal or exponentially tapered covariances, are poor representations of the measurement noise that can bias the inversions \citep{vasyura2021accounting,saoulis2025full}. Here, we use the empirically motivated exponential-tapered cosine covariance from \citet{kolb2014receiver}:
\begin{equation}
\mathbf{C}_d^{ij} =\sigma^2e^{-\lambda \tau_{ij}}
\cos\bigl(\lambda \omega_0 \tau_{ij}\bigr),
\label{eq:Cd_single}
\end{equation}
where $\tau_{ij} = |t_j - t_i|$ is the time lag between samples and $\sigma^2$ is the noise variance. We use the empirical choice from \citet{kolb2014receiver} for $\omega_0=4.4$ and use $\lambda=0.05$ to match maximum frequency of the filtering band for our data. The noise level $\sigma^2$ is estimated from the filtered noise prior to an event for each receiver–component. For the high signal-to-noise events studied here, theory errors dominate and the data covariance parametrisation $\mathbf{C}_d$ does not impact the quality of the inversions.

Note that, following prior work, we only model the per station-component covariances in the Gaussian likelihood covariance  $\widehat{\mathbf{C}}$, leading to a block sparse form that assumes all cross-component and cross-station covariances are zero. This hugely reduces the computational complexity of estimating (and inverting) $\widehat{\mathbf{C}}$, making the problem tractable. However, velocity model perturbations will produce non-zero cross-component, cross-station covariances (as will realistic noise, to a lesser extent). All of these effects are therefore neglected in the Gaussian likelihood approach. In contrast, the SBI approach does not use an explicit covariance matrix for inference and can therefore capture and model these correlations implicitly.

\subsection{Application to real data}

In practice, we found the covariance in \Cref{eq:data_theory_covariance} failed to provide an adequate model of the uncertainties for the real data inversions. Since we did not encounter such issues during the synthetic experiments, we hypothesise that this is the result of further, unmodelled uncertainties. For instance, 3-D heterogenieties will likely cause theory errors in the observations that are not captured by our forward model and Earth structure perturbations. 

In order to stabilise the inversions, we added a small diagonal regularisation term to the covariance matrix:
\begin{equation}
\widehat{\mathbf{C}} = \mathbf{C}_d + \mathbf{C}_t(\mathbf{m}) + \mathbf{C}_\mathrm{reg},
\label{eq:regularised_final_covariance}
\end{equation}
where $\mathbf{C}_\mathrm{reg} =\alpha\mathbf{I}$. Each covariance term in \Cref{eq:regularised_final_covariance} is visualised in \Cref{fig:covariances}. We choose $\alpha$ empirically to be a small fraction of the maximum magnitude of the theory errors: $\alpha=\epsilon\max \mathbf{C}_t(\mathbf{m})$, where $\epsilon$ is chosen empirically. This approach has the advantage of encoding the heuristic that the unmodelled theory errors will be related to the magnitude of the modelled theory errors. This regularisation term is a form of covariance shrinkage \citep{ledoit2004well}, a widely used approach for mitigating covariance mismodelling effects. 

We note that our covariance $\widehat{\mathbf{C}}$ takes a similar form to \citet{phạm2021toward}, where instead a diagonal contribution to the covariance was motivated by and interpreted as data noise. We argue that this interpretation is unphysical, in the sense that uncorrelated noise is not a realistic model of the data noise covariance \citep{duputel2012uncertainty, vasyura2021accounting, saoulis2025full}. Instead, the diagonal term should be understood as a way of regularising against unmodelled theory errors.

\section{Applying SBI to theory errors}
\label{sec:sbi_methods}
\subsection{Neural density estimation}

A central requirement of SBI is the ability to accurately model target probability densities such as the posterior  $p(\mathbf{m}\mid\mathbf{D})$, which may exhibit strong non-Gaussian structure, complex correlations, and multimodality. Neural density estimators (NDEs) address this problem by learning flexible, parameterised distributions that can be trained with standard ML techniques.

Normalizing flows provide a particularly effective class of NDEs for SBI \citep{alsing2019fast}. The key idea is to represent a complex target density by learning a transformation that maps samples from a simple base distribution to the target. Concretely, let $\mathbf{z}\sim p_0(\mathbf{z})$ denote a draw from a simple base distribution (typically a standard multivariate Gaussian). A normalizing flow defines an invertible transformation $f_\phi$, parametrised by neural networks, such that
\begin{equation}
\mathbf{m} = f_\phi(\mathbf{z}; \mathbf{D}),
\end{equation}
where explicit dependence on the observations $\mathbf{D}$ allows the transformation to model the conditional density $q_\phi(\mathbf{m}\mid\mathbf{D})$. The resulting density over $\mathbf{m}$ is obtained via the change-of-variables formula,
\begin{equation}
\log q_\phi(\mathbf{m}\mid\mathbf{D})
= \log p_0(\mathbf{z}) + 
\log \left| \det \frac{\partial f_\phi^{-1}}{\partial \mathbf{m}} \right|.
\label{eq:normalizing_flows_obj}
\end{equation}
with $\mathbf{z}=f_\phi^{-1}(\mathbf{m};\mathbf{D})$. This formulation yields a simple, computationally tractable expression for the loss in \Cref{eq:sbi_training}, allowing the neural network parameters of the transformation $f_\phi$ to be learned directly by optimisation.

\begin{figure*}
 \centering
    \includegraphics[width=\textwidth]{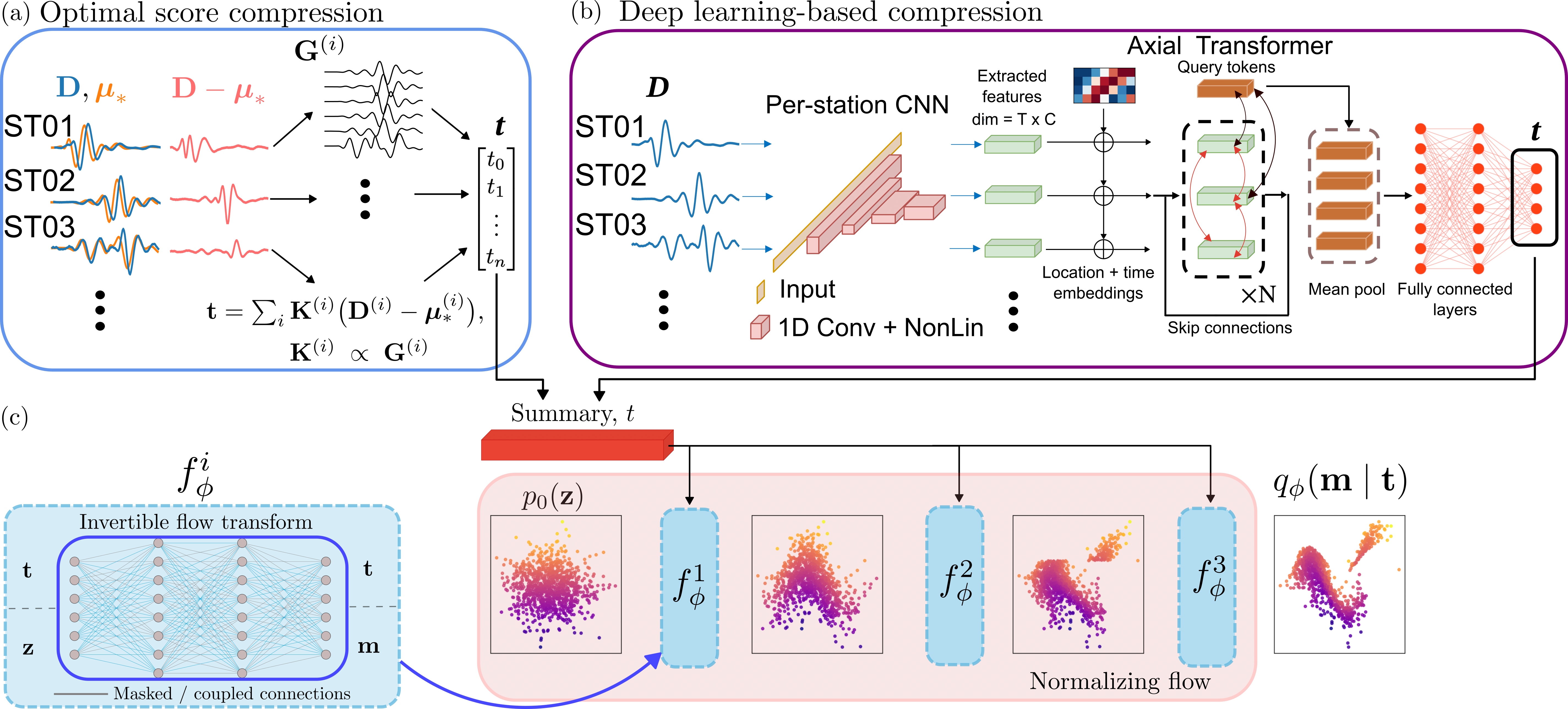}
   \caption{Illustration of the two SBI frameworks introduced in this work. Panel (a) shows optimal score compression, which projects the residuals $\mathbf{D} - \boldsymbol{\mu}_*$ for stations ST01, ST02, ST03, etc. onto error weighted  sensitivity kernels $\mathbf{G}$ to produce compressed observations  $\mathbf{t}$. Panel (b) instead uses a deep learning algorithm to learn a flexible, non-linear compression operation. A shared CNN processes full-waveforms into dense features with T time-steps and C feature channels. These station-level features are passed to an N-block axial transformer to aggregate information across the station array before producing a final compressed representation. Panel (c) visualises a normalizing flow architecture, used by both compression frameworks to perform empirical density modelling. The normalizing flow is trained to model the posterior over source parameters $q_\phi(\mathbf{m} \mid \mathbf{t})$ by transforming samples from a simple latent distribution $p_0(z)$ using learnable transformations $f^i_\phi$. }
   \label{fig:sbi_approaches}
\end{figure*}

In practice, the transformation $f_\phi$ is constructed as a composition of multiple simple transformations $f^i_{\phi}$ (see \Cref{fig:sbi_approaches}c, allowing the overall model to represent highly complex densities while remaining tractable for both density evaluation and sampling. This hinges on designing $f_\phi$ with an easily computable Jacobian (the $\partial f_\phi^{-1}/\partial \mathbf{m}$ term in \Cref{eq:normalizing_flows_obj}). Different normalizing flow architectures correspond to different choices of these transformations and their parameterisations, trading off expressivity, computational efficiency, and numerical stability. 

In this work, we employ masked autoregressive flows \citep[MAFs;][]{papamakarios2017masked} and neural spline flows \citep{durkan2019neural}, which are both well-established flow classes that have demonstrated strong performance in SBI and related conditional density estimation tasks. We defer to prior work for a more detailed exposition of normalizing flows and their variations \citep[e.g.,][]{papamakarios2021normalizing}. 

\subsection{Compression algorithms}
In practice, modelling the distribution $p(\mathbf{m}\mid\mathbf{D})$ directly is challenging. This largely results from the high-dimensionality of full-waveform seismic observations, which contain complex latent relationships that are non-trivial to model. Prior work in SBI addresses this issue by compressing the high dimensional observations $\mathbf{D}$ into some lower dimensional summary statistics $\mathbf{t}$. 

Once the compression algorithm is defined, the training dataset $\mathcal{D}$ (Equation \ref{eq:training_data}) can be compressed to a set pairs of source parameters $\mathbf{m}$ and compressed observations $\mathbf{t}$. SBI then proceeds as usual, modelling the posterior proxy $p(\mathbf{m}\mid\mathbf{t})$. A diagram of this workflow is shown in \Cref{fig:sbi_approaches}. Note that while this is not the ``true'' posterior $p(\mathbf{m}\mid\mathbf{D})$, this approach has the advantage of constructing a posterior solution $p(\mathbf{m}\mid\mathbf{t})$ that is statistically consistent with the true source parameters $\mathbf{m}$. This is not the case when significant approximations are made in standard likelihood-based inference techniques. 

We propose two compression frameworks for applying SBI in the context of theory errors.

\subsubsection{Optimal score compression}
\label{sec:optimal_score_compression}

The first relies on a physically-motivated compression algorithm to reduce the dimensionality of observations $\mathbf{D}$ to a set of informative summary statistics $\mathbf{t}$. The goal of a compression algorithm is to preserve as much discriminative power in the summaries $\mathbf{t}$ with respect to the source parameters $\mathbf{m}$ as possible. One popular and theoretically motivated approach is optimal score compression \citep{Alsing2017, Alsing2018, alsing2019fast}. 

Optimal score compression requires a known, analytic expression for the likelihood $p(\mathbf{D} \mid \mathbf{m})$. For instance, we may re-use our approximate Gaussian likelihood from \Cref{eq:final_gaussian} with some unknown covariance $\mathbf{C}$:
\begin{equation}
p(\mathbf{D} \mid \mathbf{m})
=\mathcal{N}\!\big(\mathbf{D}\,\big|\,g(\mathbf{m}, \Omega^*),\; \mathbf{C}).
\label{eq:compression_gaussian}
\end{equation}
Score compression then proceeds by Taylor expanding the likelihood $p(\mathbf{D} \mid \mathbf{m})$ about a fiducial set of source parameters $\mathbf{m}^*$ to first order. The data sensitivities with respect to the source parameters, $\mathbf{G}_{\mathbf{m_*}}= \nabla_\mathbf{m_*}\mathbf{D}$ can then be used to construct an optimal compression algorithm \citep{Alsing2017}. Specifically, \citet{Alsing2018} demonstrate that one such optimal algorithm is
\begin{linenomath*}
\begin{equation}
\mathbf{t} = \mathbf{m}_* + \mathbf{F}^{-1}_* \mathbf{G}^T_{\mathbf{m_*}} \mathbf{C}^{-1} (\mathbf{D} - \boldsymbol{\mu}_*), 
\label{eq:optimal_score_mle}
\end{equation}
\end{linenomath*}
where $\boldsymbol{\mu}_*$ is the fiducial data vector $\boldsymbol{\mu}_*=g(\mathbf{m}_*, \Omega_*)$, and the Fisher matrix $\mathbf{F}_* = \mathbf{G}^T_{\mathbf{m_*}} \mathbf{C}^{-1}\mathbf{G}_{\mathbf{m_*}}$. \Cref{eq:optimal_score_mle} has a particularly simple interpretation: the summary $\mathbf{t}$ is the (local) maximum likelihood estimate (MLE) of the source parameters $\mathbf{m}_{\textrm{MLE}}$ given an observation $\mathbf{D}$ \citep{tarantola2005inverse}. For a detailed mathematical derivation we defer to prior work \citep{Alsing2017, Alsing2018}, including a previous application of optimal score compression to seismic source inversion \citep{saoulis2025full}. We provide some discussion on our exclusion of higher order terms in Section S\ref{SI:sec_1}. 

There are, however, some limitations of optimal score compression as applied to seismic source inversions. As highlighted in \citet{saoulis2025full}, nonlinearity in the forward model leads to rapid degradation of compression algorithm, and therefore the useful information in summaries $\mathbf{t}$. While the forward model is linear with respect to the moment tensor, perturbations in the Earth structure introduce nonlinearities \citep{phạm2021toward, phạm2024towards}. 

We therefore adopt the two-stage approach utilised in \citet{saoulis2025full}. First, we utilise the approximate Gaussian-likelihood to find a maximum likelihood estimate of the source parameters $\mathbf{m}_{\textrm{MLE}}$. This is achieved through an iterative least squares algorithm: at each iteration, we estimate the local theory covariance $\widehat{\mathbf{C}}_t(\mathbf{m})$, and then apply the maximum likelihood estimate in \Cref{eq:optimal_score_mle}. The second step is to then truncate the prior $p(\mathbf{m})$ over a region where the compression is accurate. We follow the procedure taken in \citet{saoulis2025full} to use a Gaussian estimate of the posterior to propose a prior that should support the target posterior $p(\mathbf{m}\mid\mathbf{t})$. In this approximation, the posterior covariance is given by the inverse Fisher information matrix $\mathbf{F}^{-1}$ \citep{tarantola2005inverse, Alsing2017}. The diagonal of this covariance provides an estimate of the characteristic per-parameter uncertainties $\boldsymbol{\sigma}$, which we use to define a weakly informative truncation of the prior. We define a uniform prior centred on $\mathbf{m}_{\textrm{MLE}}$ with width $N\boldsymbol{\sigma}$ along each parameter direction, where $N$ is a free parameter to be tuned: 
\begin{equation}
    p_{\text{Fisher}}(\mathbf{m};N)= \mathcal{U}(\mathbf{m}_{\textrm{MLE}} - N\cdot\boldsymbol{\sigma}, \mathbf{m}_{\textrm{MLE}} + N\cdot\boldsymbol{\sigma}) * p(\mathbf{m}).
    \label{eq:prior_fisher}
\end{equation}
SBI then proceeds as usual by sampling from the truncated prior region $\mathbf{m}\sim p_{\text{Fisher}}(\mathbf{m};N)$. 

This approach has several downsides. For one, at all stages it relies on a Gaussian likelihood to compute a local maximum likelihood estimate. While \citet{phạm2021toward, phạm2024towards} demonstrated that the approximate Gaussian likelihood is sufficiently accurate to recover an improved MLE, it may also introduce bias. By then truncating the prior around a potentially biased estimate of the MLE, we run the risk of biasing the entire posterior inference process. We probe the degree to which this effect can bias our posteriors in \Cref{sec:synthetic_inversions}. Nonetheless, an improved formulation for utilising SBI is desirable. 

\subsubsection{Machine learning-based compression}

Optimal score compression is appealing as a generic and simple compression algorithm. However, as its core assumptions break down --- local linearity and a well-specified likelihood --- its effectiveness at preserving information in the compressed representation $\mathbf{t}$ becomes restricted \citep{gerardi2024optimal,park2025dimensionality, saoulis2025full}. 

Machine learning-based compression has now received significant attention across the field of SBI \citep{charnock2018automatic,prelogovic2024informative,lanzieri2025optimal, lehman2025learning}. The highly flexible parametrisation of deep neural networks ensures that they can learn globally valid, expressive compressed representations. Unlike in unsupervised compression (e.g. variational autoencoders), the goal in SBI is to preserve as much discriminative power in the compressed statistics $\mathbf{t}$ with respect to the model parameters $\mathbf{m}$. 

\citet{jeffrey2021likelihood} demonstrated that the optimal objective function to achieve this amounts to training the compression model and NDE together to predict the posterior density. Specifically, given a compression model $F_\theta$ and a density estimation model $q_\phi$, the networks can be trained jointly end-to-end on the neural posterior estimation (NPE) objective function from \Cref{eq:sbi_training}:
\begin{equation}
\mathcal{L} =
-\mathbb{E}_{(\mathbf{m}, \mathbf{D}) \sim \mathcal{D}}
\left[
\log q_{\phi}\!\left(\mathbf{m} \mid F_\theta(\mathbf{D})\right)
\right].
\label{eq:compression_training}
\end{equation}
One can then use the trained model $F_\theta$ to produce informative, non-linear compressed summaries of the observation for downstream tasks, or use the chained models $q_{\phi}(\mathbf{m} \mid F_\theta(\mathbf{D}))$ for posterior estimation directly. For the purposes of this work the two networks can be thought of as a single model performing compression and posterior inference jointly, as in \Cref{fig:sbi_approaches}. 

The remaining challenge is to design a compression architecture, $F_\theta$, that is capable of extracting the relevant information from multi-station seismic observations $\mathbf{D}$. 

\subsubsection{Deep learning architecture for multi-station full-waveform data}

Deep learning for processing multi-station seismic data has received significant attention. A broad range of model architectures have been explored, such as graph neural networks \citep[GNNs][]{mcbrearty2022earthquake, mcbrearty2023earthquake,si2024all, hourcade2025pegsgraph}, transformer-based architectures \citep{munchmeyer2021transformer, song2025foconet, huang2026msep}, Fourier neural operators \citep{sun2023phase}, and aggregation across foundation model embeddings \citep{si2024seisclip}. In this work, we design an architecture inspired by \citet{munchmeyer2021transformer}, which combines a per-station CNN-based feature extractor with a transformer-based approach to perform multi-station feature aggregation. 

A simplified illustration of the architecture is presented in \Cref{fig:sbi_approaches}. Each station’s three-component seismogram is first processed independently by a shared CNN, which extracts station-specific temporal features while enforcing equivariance across the seismic network. To stabilise training and preserve physically meaningful amplitude information, waveforms are normalised per trace and augmented with an explicit log-amplitude feature \citep{munchmeyer2021transformer}. The resulting feature sequences are combined with sinusoidal temporal embeddings and station-location embeddings derived from the array geometry, enabling the model to reason jointly about waveform content, timing, and spatial configuration.

Aggregation across stations and time is performed using an axial transformer \citep{ho2020axial}. Rather than applying full self-attention over all station tokens as in \citet{munchmeyer2021transformer}, the model factorises attention into separate station-wise and time-wise operations, preserving the temporal structure within the per-station CNN-extracted features. The attention mechanism in the transformer architecture is invariant under permutations of stations, and  can be trained with a varying, arbitrary number of stations. It is therefore a highly generalisable feature extractor, that could in principle be applied to varying station configurations. 

As stated in the section above, this network is trained end-to-end with a NDE head network, which models the posterior distribution. This relies on a compressed summary of the seismic observations, $\mathbf{t}$, to model perform density modelling $\log q_{\phi}(\mathbf{m} \mid \mathbf{t})$. We therefore perform aggregation of the transformer embeddings via learned query tokens, and mean pool them into a single fixed-dimensional embedding. This is passed through a small feedforward neural network to produce the final compressed summary $\mathbf{t}$. The exact implementation can be found in the provided software. 


\subsection{Incorporating 3-D Earth structure uncertainty } 
\label{sec:2d_structure_approach}

Although we consider 1D Earth model uncertainty that affects all stations in the same manner, in many practical applications, it is often necessary to handle 3-D Earth structure uncertainty in moment tensor inversions. To first order, the lateral heterogeneities induce station-specific timeshifts in the waveforms \citep[as used for e.g. teleseismic station corrections in the ISC-EHB catalogue;][]{engdahl1998global}.  Prior work often pre-processes the observations by aligning them with the best-fitting synthetics, often by maximising per-station correlations. Ideally, though, we should incorporate this 3-D Earth structure uncertainty in moment tensor solutions themselves. One approach treats station-specific time-shifts as a proxy for \textit{all} unresolved Earth structure uncertainties and includes them as extra parameters in the inversion \citep{hu2023seismic, hu2025bayesian}. These  ``nuisance'' parameters can then be marginalised over for the desired focal mechanism (following \Cref{eq:marginal_posterior_1}). However, as noted in \Cref{sec:theory_intro}, inference over the joint distribution of source parameters and nuisance parameters increases the dimensionality of the problem, resulting in significantly increased computational costs for traditional MCMC-based inference procedures (for instance, \citealt{hu2025bayesian} runs over $5\times10^6$ forward model evaluations per inversion).  

The approach taken in \citet{hu2023seismic} is useful in a likelihood-based analysis because attempting to pre-marginalise the likelihood over station-specific time-shifts (as done for 1-D Earth structure uncertainty in Equations \ref{eq:marginal_likelihood} - \ref{eq:final_gaussian}) can further degrade the accuracy of the inversion. However, in theory SBI allows us to add these station-specific time shifts as extra nuisance parameters that we simulate in the prior predictive model of \Cref{eq:prior_predictive}. These can then be marginalised \textit{along} with the 1-D Earth structure uncertainty with the flexible empirical density estimation formulation. 

In practice, our optimal score compression formulation in \Cref{sec:optimal_score_compression} relies on a relatively accurate Gaussian likelihood, which makes this challenging. However, the deep learning-based compression can straightforwardly handle arbitrary station-specific time-shifts as an extra source of uncertainty. For the real data examples in \Cref{sec:real_data}, we therefore modify the prior predictive model for data generation to include time-shift nuisance parameters: 
\begin{equation}
\mathbf{m} \sim p(\mathbf{m}), \quad
\Omega \sim p(\Omega), \quad
\mathbf{t} \sim p(\mathbf{t}), \quad
\mathbf{D} \sim p(\mathbf{D} \mid \mathbf{m}, \Omega, \mathbf{t}),
\label{eq:prior_predictive_timeshift}
\end{equation}
where each $t_i$ is a station-specific time-shift parameter drawn from an assumed prior distribution $p(\mathbf{t})$. These nuisance parameters account for uncertainties in absolute timing and are marginalised over implicitly during training.

\section{Evaluation}
\label{sec:evaluation}
\subsection{Covariance goodness-of-fit}

Given the Gaussian approximation adopted in \Cref{eq:final_gaussian}, it is important to assess whether the inferred covariance structure provides an adequate statistical description of the theory errors. If the Gaussian assumption holds, the $\chi^2$ statistic
\begin{equation}
\chi^2
= (\mathbf{D} - g(\mathbf{m}, \Omega))^\mathrm{T}
\mathbf{C}^{-1}
(\mathbf{D} - g(\mathbf{m}, \Omega))
\label{eq:chi2_def}
\end{equation}
follows a $\chi^2$ distribution with $\mathrm{n}_\mathrm{dof} = \dim(\mathbf{D})$ degrees of freedom, or equivalently a reduced statistic $\chi^2_\mathrm{red} = \chi^2 /\mathrm{n}_\mathrm{dof}$. This offers a simple and computationally inexpensive diagnostic of the approximation quality.

In practice, $\chi^2$ values can be computed for an ensemble of synthetic observations from the prior predictive model. The empirical distribution of $\chi^2$ values may then be compared against the analytic expectation. Then, any  systematic discrepancies between the observed and expected $\chi^2_\mathrm{red}$ distribution, such as skewed or heavy-tailed behaviour, indicate a breakdown of the Gaussian covariance approximation \citep{tarantola2005inverse}. This comparison can be performed visually using the reduced-$\chi^2$ distribution, or quantitatively via the Kolmogorov–Smirnov (KS) statistic. The KS statistic measures the maximum absolute difference between the empirical and theoretical cumulative distributions,
\begin{equation}
D_\mathrm{KS}
= \max \left| F_\mathrm{emp} - F_{\chi^2_\textrm{red}} \right|.
\label{eq:ks_stat}
\end{equation}
\subsection{Moment tensor inversion quality}

To assess whether a given moment tensor inversion approach produces posteriors consistent with the true source parameters, we quantify both bias and uncertainty in the inferred solutions. For each focal mechanism parameter, we compute the posterior mean and measure the absolute bias relative to the known true value. In addition, we compute the posterior standard deviation for each parameter, which provides a measure of how tightly each approach constrains each focal mechanism parameter.

More discriminatory are calibration tests. These check whether the inferred posteriors are statistically consistent with the true solution over many repeated artificial inversions. This enables a practitioner to identify whether an inversion procedure is systematically biased or fails to capture the true uncertainty in the posterior. Approaches for evaluating posterior calibration have received a lot of attention in the field of SBI \citep{talts2018validating,hermans2021,lueckmann2021benchmarking}, since calibration is an essential pre-condition for reliable and trust-worthy posterior inference. 

Estimating empirical coverage in high-dimensional parameter spaces can be computationally demanding, as many existing approaches require explicit density estimation in order to construct credible regions around the true parameters for quantile evaluation \citep[e.g.][]{hermans2021}. \citet{lemos2023sampling} introduced an alternative method, termed Tests of Accuracy with Random Points (TARP), which enables substantially more efficient coverage estimation. Rather than explicitly computing credible regions, TARP evaluates the fraction of posterior samples lying between a randomly drawn reference point and the true model parameters. Repeating this procedure across multiple inversions yields an empirical distribution of credibility levels, which can be directly compared to the expected uniform distribution under perfect calibration \citep[see][for a more detailed explanation, with examples for moment tensor inversion]{saoulis2025full}. In this work, we employ TARP to assess whether both Gaussian-likelihood inversions and SBI-based methods produce well-calibrated posterior distributions.

\begin{figure*}
 \centering
    \includegraphics[width=\textwidth]{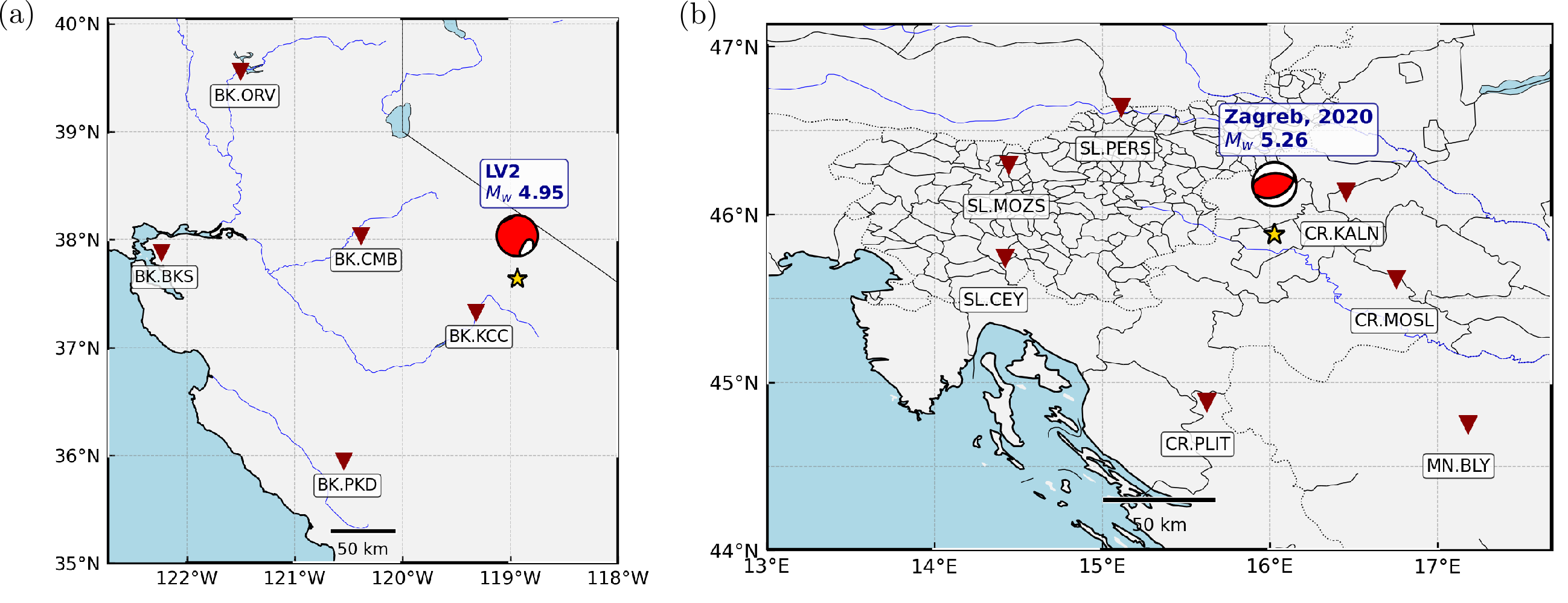}
   \caption{Earthquake location (gold star), focal mechanism (red beachball), and station configuration (brown triangles) for the two events studied in this manuscript. Panel a) shows the LV2 volcanic event in California originally studied in \citet{dreger2000dilational}, with focal mechanism solution from \citet{phạm2021toward}. We also use this source-receiver configuration for the synthetic experiments in the main text. Panel b) shows the 2020 tectonic event near Zagreb, Croatia with focal mechanism solution from \citet{hu2025bayesian}. }
   \label{fig:event_network_maps}
\end{figure*}

\section{Synthetic experiments}
\label{sec:synthetic_inversions}

\subsection{Experimental details}
\label{sec:experimental_details}
\subsubsection{Forward modelling}

For our synthetic experiments, we use the source–receiver geometry shown in \Cref{fig:event_network_maps}. This configuration corresponds to the network geometry of the Long Valley Caldera event studied later in this manuscript (LV2; see \Cref{sec:LV2}). We perform a suite of synthetic tests with moderate magnitude events, $ M_\mathrm{W}\sim 5$, where theory errors dominate over data errors.

We follow the experimental and processing setup of \citet{phạm2021toward}. All Green's functions are computed using the Computer Programs in Seismology code \citep{herrmann2013computer}. Unless otherwise stated, we use a source depth of 5 km and all synthetics are band-passed filtered between 20-50 s. All data are sampled at 1 Hz, and we select a time window of 200 s after the origin time for all stations. We use the SoCal 1-D Earth model as a fiducial Earth model $\Omega^*$ \citep{dreger1990broadband}, and follow \citet{phạm2021toward} in parametrising the Earth model prior $p(\Omega)$ in terms of a perturbation parameter $\kappa$. Starting from $\Omega^*$, we generate prior samples by independently applying fractional perturbations to the $V_\mathrm{p}$, $V_\mathrm{s}$, and layer depths in each layer. The parameter $\kappa$ specifies the mean percentage amplitude of these perturbations, such that larger $\kappa$ corresponds to greater deviation from the fiducial model.

For each experiment, we compute $N=300$ sets of Greens functions for different $\Omega^{(n)} \sim p(\Omega)$. We may then compute local estimates $\mathbf{C}_t(\mathbf{m})$ using the Monte Carlo sampling procedure specified in \Cref{eq:mc_theory_cov}. 

For all experiments, we parameterise the seismic source using the symmetric moment tensor $\mathbf{m} = (M_{xx}, M_{yy}, M_{zz}, M_{xy}, M_{xz}, M_{yz})$, and adopt a uniform prior over its six independent components, $p(\mathbf{m}) = \mathcal{U}[-4\times10^{16}, 4\times10^{16}]$. All density modelling and sampling algorithms are performed in the moment tensor component basis. For evaluation and visualisation, we use the \citet{tape2015uniform} parametrisation to represent the isotropic and compensated linear vector dipole (CLVD) components $\delta$ and $\gamma$, along with the moment magnitude measure $M_w$ and best fitting double-couple source orientation angles strike, dip and rake. 

The sensitivity kernels are computed using 5-point derivative stencils. We perform iterative least squares to find an estimate of $\mathbf{m}_\textrm{MLE}$ for \textit{every event} for the Gaussian likelihood approach and score compression-based SBI approach. This is achieved by repeatedly estimating the local sensitivity kernels $\mathbf{G}(\mathbf{m})$ and covariance $\mathbf{C}_t(\mathbf{m})$, and applying the MLE estimation formula in \Cref{eq:optimal_score_mle}. We perform 8 iterations, and select the iteration $\mathbf{m}_\textrm{MLE}$ that minimises the $\chi^2$ (misfit) of the event. In practice, this was generally not the final iteration as we found the optimisation problem was not perfectly convex or stable. 

\subsubsection{Gaussian likelihood sampling}

We draw comparisons between the results of moment tensor inversions using the approximate Gaussian likelihood against those using SBI. We compute a local estimate of the Gaussian covariance $\mathbf{C}(\mathbf{m_\textrm{MLE}})$ as described in \Cref{eq:data_theory_covariance} (or \Cref{eq:regularised_final_covariance} for real data), and use MCMC to draw samples from the posterior. We use \texttt{emcee} \citep{foreman2013emcee} to implement a simple, embarrassingly parallel Metropolis-Hastings sampling strategy. We manually optimised the step-size of the burn-in and final sampling stage for each experimental setup to ensure that the posterior solutions converged. We parallelise this over 30 cores (Intel Xeon Gold 6230 @2.10 GHz) and run approximately $100,000$ forward model evaluations for each Gaussian likelihood-based inversion. Final posterior samples are computed with a per-walker burn-in fraction of 0.4. 

\subsubsection{SBI implementation details}

For the score compression approach, we use the prior truncation procedure in Equation \ref{eq:prior_fisher} with $N=15$, finding this struck a good balance between ensuring well-calibrated but discriminative posteriors for all experiments \citep[see][who found similar results]{saoulis2025full}. We run $8,000$ forward model evaluations from the prior predictive model \Cref{eq:prior_predictive}, and use the resulting compressed dataset $\mathcal{D}=\{\mathbf{m}, \mathbf{t}\}$ to train the NDE to perform posterior estimation (with $10\%$ of data held back as a validation set for model selection). We use a MAF architecture with 5 transformation layers for the NDE, finding it trained quickly on CPUs and was sufficiently flexible to accurately model the target posteriors. Training was performed until a stopping criteria was reached on the validation set. The total duration of MLE estimation of $\mathbf{m}$, dataset generation and training took around $15$ minutes on average for each inversion, using parallelisation over 30 cores. 

For the deep learning-based compression approach, we performed a short manual optimisation to improve the performance of the architecture. The per-station 1-D CNN processes 3 component, 200 s duration seismic traces.  The network consists of a stack of 6 1-D convolutional layers with uniform kernel sizes, with early strided convolutions used to progressively downsample the temporal dimension while increasing the feature dimensionality. This design yields per-station features with a temporal resolution of around 30 time samples and 128 feature channels per station. We then add sinuisoidal embeddings that encode the station location for each station feature block, and time embeddings encoding the time index of each of the 128-dimensional feature channels. 

We build the axial transformer with multiple layers consisting of multi-head attention (4 heads) with model dimension of 256. The transformer learns 8 query tokens, which query the processed seismic features over 4 attention layers. After the 4 layers, these queries are mean pooled and processed by a 2 layer feedforward network to produce a final summary embedding $\mathbf{t}$ with dimension 128. This summary is passed to a rational-quadratic neural spline flow (RQ-NSF) with 5 transformation layers to perform density estimation. We found that this NDE architecture significantly improved performance over the MAF used for the score compression approach, likely as a result of facilitating improved gradient flow back through the feature extractor network. 

The deep learning-based inference model was trained using $100,000$ forward model evaluations from the prior predictive model in Equation \ref{eq:prior_predictive}, with $10\%$ held out as validation data. Training was performed for 300 epochs with a batch size of 128, and the lowest validation loss model was selected. Dataset generation took approximately 10 minutes and training took approximately 12 hours on a single NVIDIA RTX A6000 GPU. Note that, once trained, this model was globally applicable for any moment tensor $\mathbf{m}$, unlike the other two approaches. 

All models were trained using the Adam optimiser. For the simple MAF NDE for score compression, we used a learning rate of $1\times10^{-4}$. For the deep learning model, we used a linear learning rate ramp-up to $1\times10^{-4}$ combined with a cosine decay schedule to $1\times10^{-5}$, finding that annealing the learning rate toward the end of training led to slightly improved models. We used a small weight decay of $1\times10^{-4}$ and a dropout rate of $10\%$, and found that models generally did not overfit during training. All the precise architectural details can also be found in the provided software. 

\subsection{Probing the validity of the Gaussian assumption}

\begin{figure}
 \centering
    \includegraphics[width=0.48\textwidth]{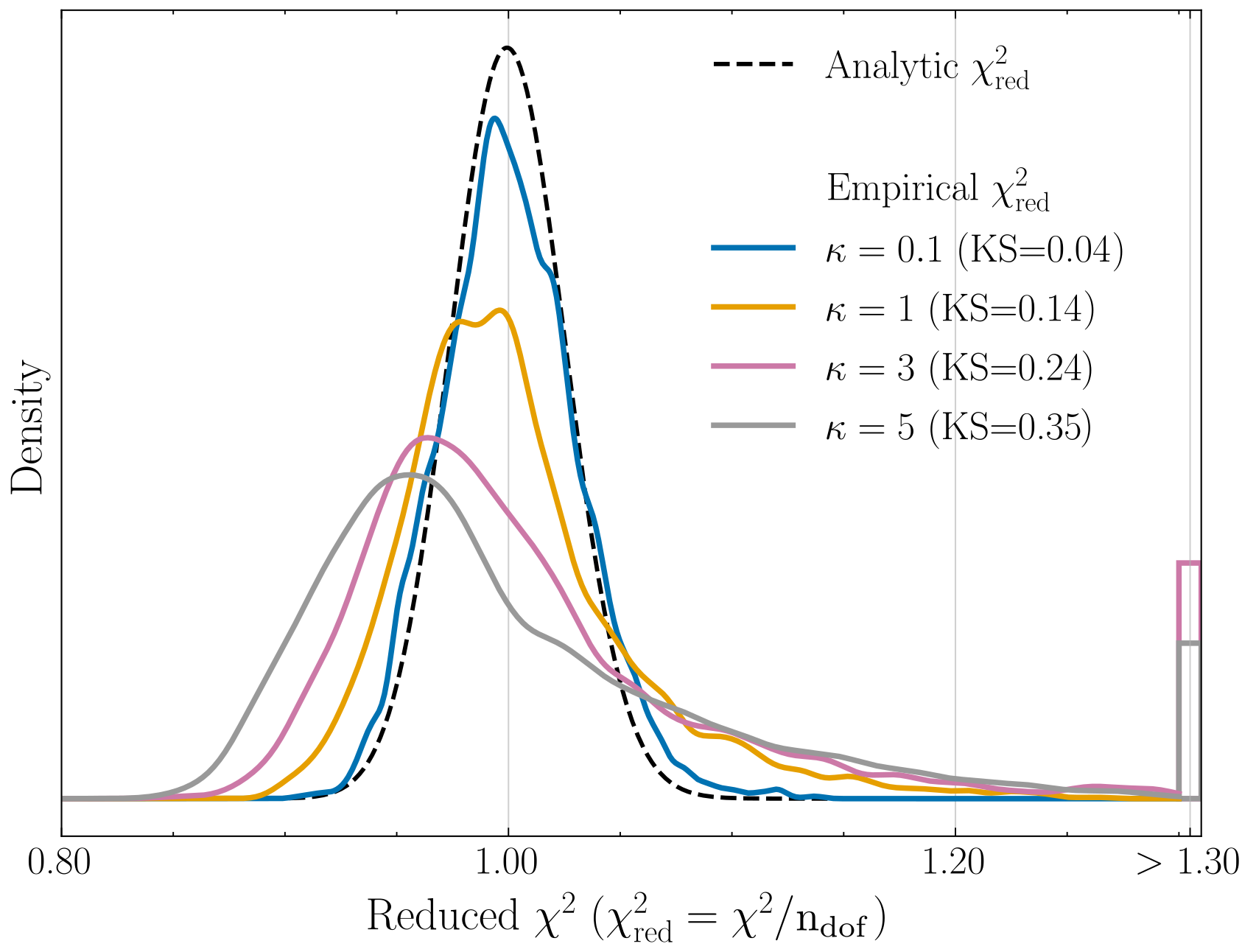}
   \caption{A comparison between the analytically expected $\chi^2_\mathrm{red}$ statistic distribution against their empirical distributions under varying levels of Earth structure uncertainty ($\kappa \in [0.1,1,3,5]$). The degree of agreement between the analytic and empirical distributions probes how well the Gaussian likelihood models the observed variability in the data observations $\mathbf{D}$. We quantify this with the KS statistic in \Cref{eq:ks_stat}. We find that even under very minor Earth structure perturbations ($\sim1\%$) a Gaussian likelihood becomes a very poor approximation. }
   \label{fig:chi2_comparison_figure}
\end{figure}

We begin by using the $\chi^2$ test from \Cref{eq:chi2_def} to evaluate the accuracy of a Gaussian approximation of the theory errors, encoded by the Gaussian covariance $\mathbf{C}_t(\mathbf{m})$. This provides a straightforward way of evaluating the accuracy of the Gaussian assumption without needing to perform inference. We estimate the theory covariance $\mathbf{C}_t(\mathbf{m^*})$ for a representative moment tensor $\mathbf{m^*}$ under varying levels of Earth structure uncertainty, $\kappa \in [0.1,1,3,5]$ and compute the $\chi^2_\mathrm{red}$ distribution for an ensemble of 2000 random events using the LV2 source-receiver geometry in \Cref{fig:event_network_maps}a.  

\Cref{fig:chi2_comparison_figure} shows that under even minor Earth structure perturbations, on the order of $1\%$ of the velocity model, the Gaussian assumption of a local estimate of  $\mathbf{C}_t(\mathbf{m^*})$ breaks down. This provides evidence that under minor Earth structure uncertainty, Gaussian approximations of the theory errors may lead to biased posterior inference.

\begin{figure}
 \centering
    \includegraphics[width=0.48\textwidth]{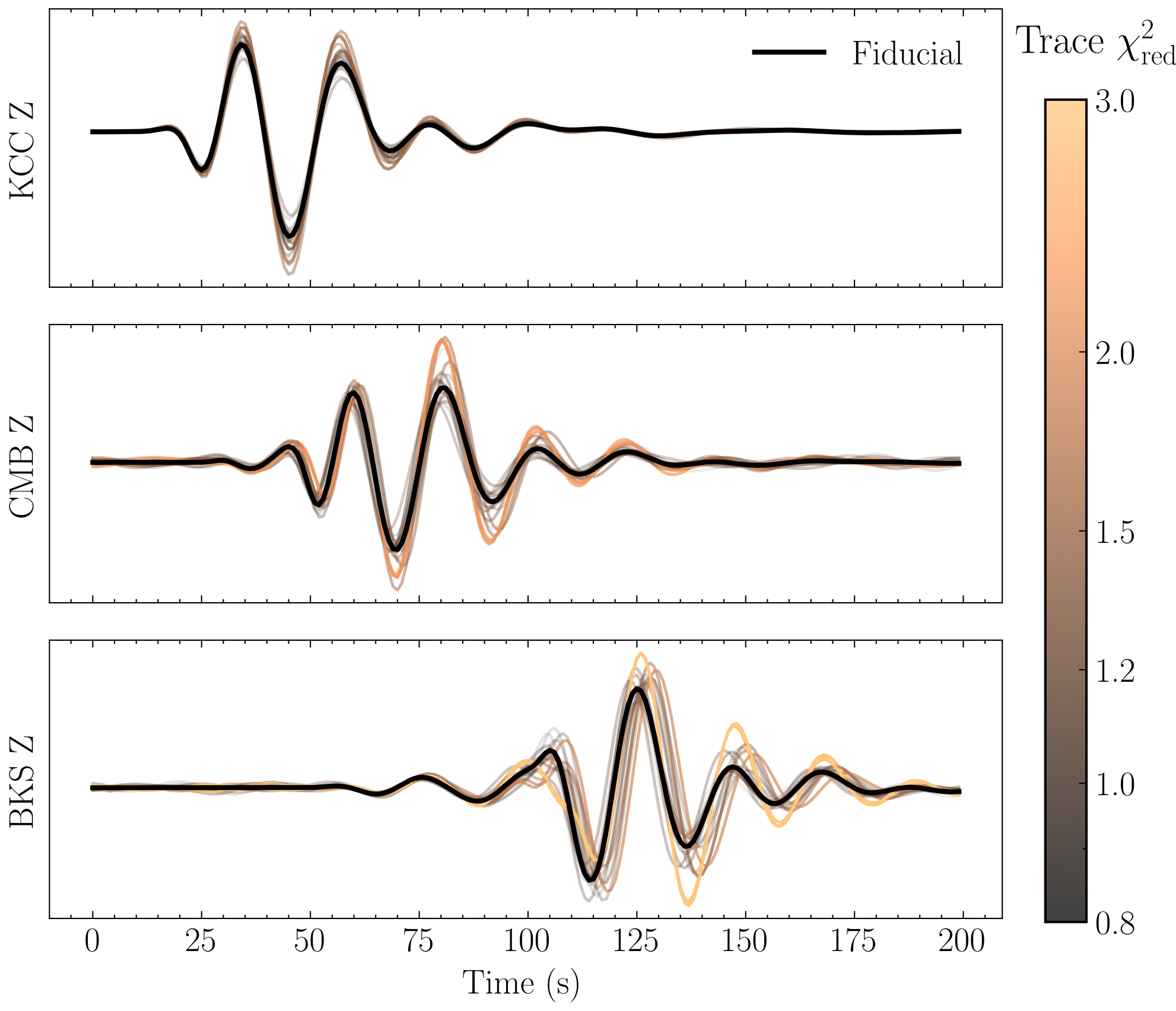}
   \caption{Examples of the per-trace $\chi^2_\mathrm{red}$ statistic under Earth structure perturbations for $\kappa =3\%$. Higher values of $\chi^2_\mathrm{red}$ are denoted by a brighter-copper colour, with significant departure from those expected under a Gaussian likelihood. Even under modest perturbations in the Earth structure, there are significant non-Gaussian effects in the observations.}
   \label{fig:traces_chi2_example}
\end{figure}

We present some examples of the effect of velocity structure perturbations in \Cref{fig:traces_chi2_example}, alongside the per-trace $\chi^2_\mathrm{red}$ statistic. We find that, intuitively, stations at greater epicentral distances are more sensitive to  velocity model uncertainties. The resulting time-shifts are not always perfectly modelled by the Gaussian covariance $\mathbf{C}_t$, leading to occasional traces with higher $\chi^2_\mathrm{red}$. However, traces with large changes to waveform amplitude (and low time-shift relative to the fiducial observation) appear to cause the most severe modelling issues. One potential explanation is that the covariance estimate $\mathbf{C}_t(\mathbf{m})$ is dominated by the variance induced by time-shifts. This leads to higher modelled variance in regions of the waveform with higher gradients (i.e. when particle velocity $\mathrm{d}x/\mathrm{dt}$ is large), and lower variance in regions where the gradient approaches zero (i.e. the turning points of the waveform). This behaviour is demonstrated in \Cref{fig:covariances}(b). The modelled covariance, which is a first-order approximation that does not model higher order statistics, performs poorly on rarer traces with large amplitude differences and low time-shifts from the fiducial. 

In summary, we find that even under minor velocity model uncertainty, a Gaussian covariance provides a poor statistical fit of the observed variability in the observations. However, the degree to which this matters in a practical sense is best tested through actual moment tensor inversions.

\begin{figure*}
 \centering
    \includegraphics[width=0.98\textwidth]{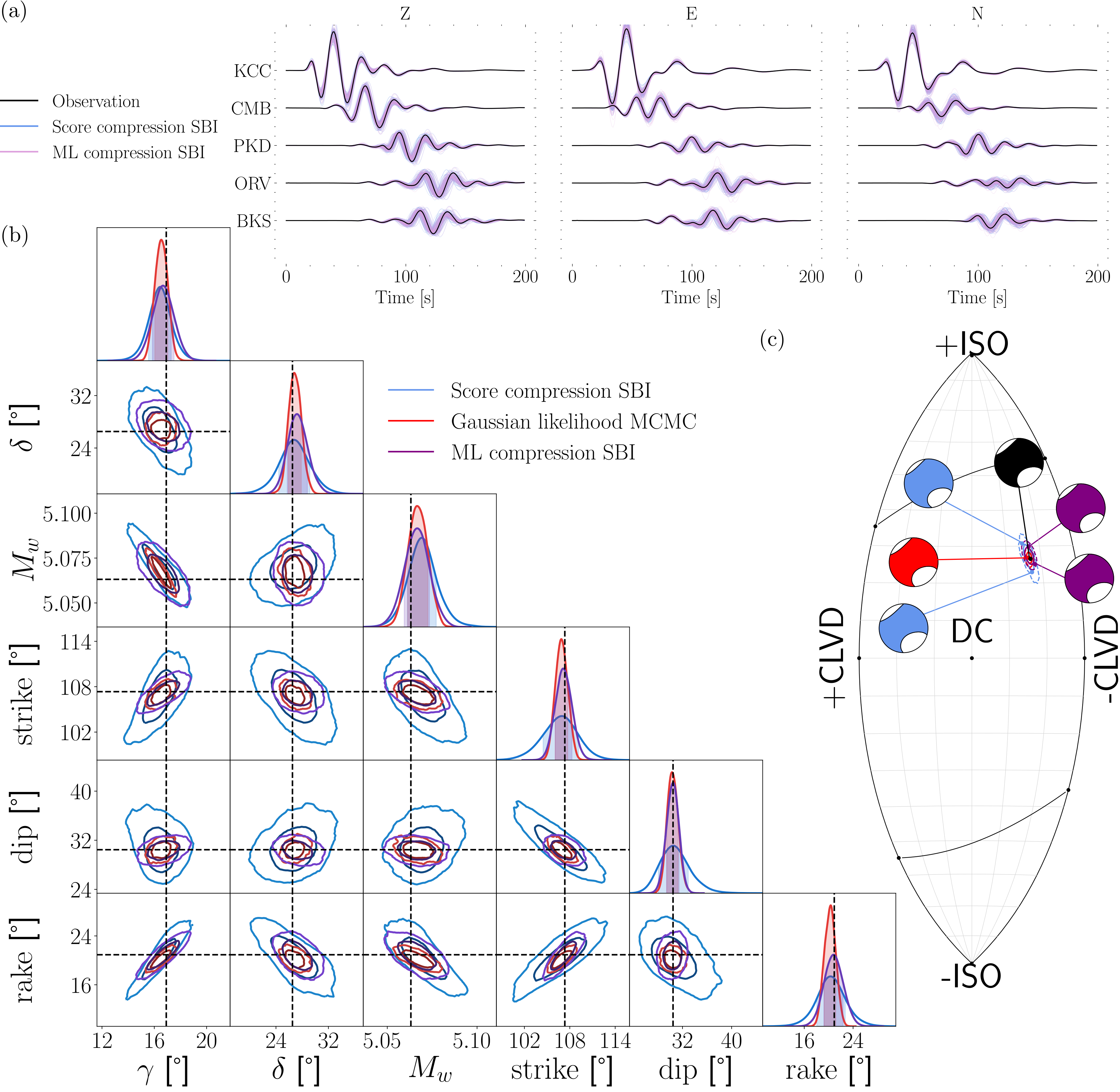}
   \caption{Posterior inference on an artificial example comparing the results using a Gaussian likelihood (red) against the two SBI frameworks: score compression with shallow density estimation (blue) and deep learning based compression and density estimation (purple). Panel (a) shows the observation and posterior predictive checks for the two SBI approaches.  The focal mechanism parameters $\gamma, \delta, M_W$ and best fitting double-couple plane orientation posterior are shown in (b), as well as the location on the lune plot in (c) (the top left pair plot in (b) shows the same lune coordinates). The $\pm[1,2]\sigma$ contours are shown for each posterior, and the true artificial solution is marked by the black dashed line and black beachball in (c). }
   \label{fig:artificial_good_examples}
\end{figure*}

\subsection{Synthetic inversions}
\subsubsection{Case study 1: Gaussian likelihood vs. SBI under high uncertainty}
\label{sec:GL_vs_SBI_casestudy}

We run a large number of synthetic inversions under varying levels of Earth structure uncertainty. To begin with, we explore how each of the three approaches performs under moderate ($\kappa=5\%$) uncertainty in the Earth velocity model. 

\begin{figure*}
 \centering
    \includegraphics[width=\textwidth]{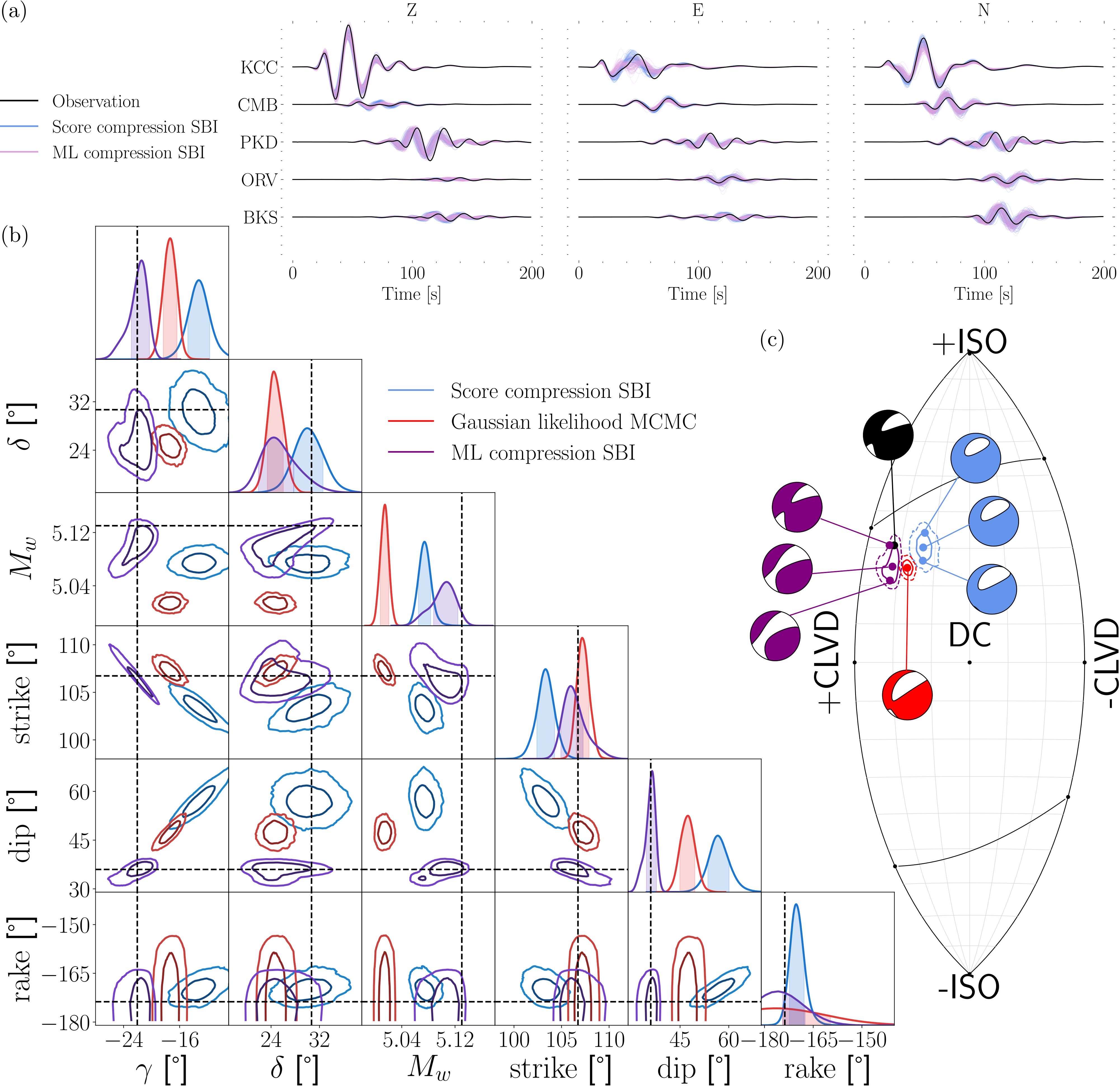}
   \caption{Posterior inference on a more problematic artificial example, where the Gaussian likelihood fails to handle the randomly perturbed 1-D velocity structure. Panel (a) shows the observation and posterior predictive checks for the two SBI approaches.  The focal mechanism posteriors are shown in (b), as well as the location on the lune plot in (c) (the top left pair plot in (b) shows the same lune coordinates), with the artificial solution marked in black. Plot details same as in Figure 6. }
   \label{fig:artificial_challenging_examples}
\end{figure*}

In Figs. \ref{fig:artificial_good_examples} and \ref{fig:artificial_challenging_examples}, we show two examples of moment tensor inversion with each of the three approaches: Gaussian likelihood-based MCMC, SBI with optimal score compression and SBI with a deep-learning based compression model. \Cref{fig:artificial_good_examples} provides an example where the Gaussian likelihood approximation is sufficiently accurate to provide a good solution, leading to posterior contours that are consistent with the artificial solution. The Gaussian likelihood approach yields posteriors with the lowest uncertainty, which we interpret below. The ML-based compression approach yields significantly tighter posterior contours than the score compression algorithm. The deep learning compression architecture is therefore capable of extracting more information from an observation than the linearised, gradient-based score compression, indicating that there are non-negligible non-linearities in the forward problem.

However, \Cref{fig:artificial_challenging_examples} presents a synthetic example where the Gaussian likelihood assumption leads to failure for both the Gaussian likelihood MCMC approach, and the score compression SBI approach. Both of these approaches yield moment tensor solutions which are inconsistent with the artificial event. In this instance, the iterative least squares procedure for estimating a local theory covariance $\mathbf{C}_t(\mathbf{m^*)}$ fails to converge since the Gaussian likelihood is not sufficiently accurate. Since both approaches rely on an accurate $\mathbf{m}^*$ (the score compression SBI approach through the prior truncation procedure in \Cref{eq:prior_fisher}), a severely biased estimate of $\mathbf{m}^*$ causes these approaches to fail. On the other hand, the ML-based compression algorithm has no such limitations, and yields a posterior that is consistent with the artificial event. 

We systematically evaluate how each moment tensor inversion approach performs. We perform 400 artificial moment tensor inversions, with each approach, drawing moment tensors from the same uniform prior. We then perform coverage testing on each of the resulting posterior sets using TARP, as described in \Cref{sec:evaluation}.

\begin{figure}
 \centering
    \includegraphics[width=0.48\textwidth]{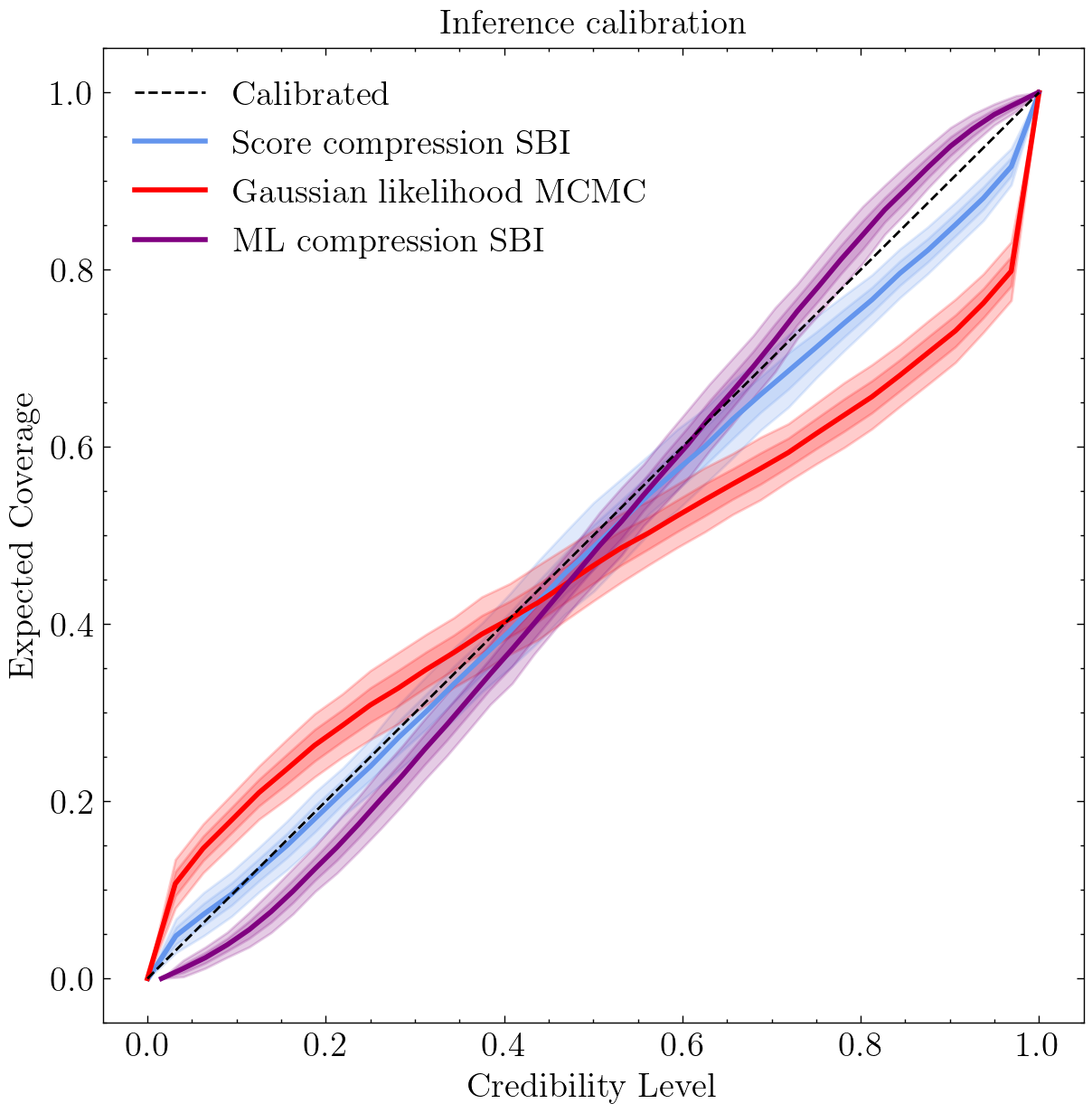}
   \caption{Calibration statistics for each of the three inference methods over 400 artificial inversions with randomly sampled input source models and perturbed 1-D Earth models. The ideal calibration is denoted by the black dashed line on the diagonal. The Gaussian likelihood approach yields overconfident ($\sim50\%$) and slightly biased posterior contours, resulting from inaccuracies in the approximate likelihood. The score compression SBI approach yields much more faithful posterior contours, with a minor degree of bias likely caused by the prior truncation procedure (discussed in the main text). The ML-based compression SBI approach is slightly conservative ($\sim10\%$) but nonetheless trustworthy and unbiased.  The $\pm[1,2]\sigma$ uncertainty regions are estimated from bootstrapping the coverage results 50 times.}
   \label{fig:calibration_comparison}
\end{figure}

The results of this coverage testing are presented in \Cref{fig:calibration_comparison}. \Cref{fig:calibration_comparison} demonstrates that only the SBI approaches can produce posteriors that are consistent with the true moment tensor solutions, even under high Earth structure uncertainty. The main issue uncovered by this test is that the Gaussian likelihood approach yields severely overconfident, and mildly biased, moment tensor posteriors. On the other hand, the score compression-based SBI approach produces much better calibrated posteriors with only a small degree of bias. 

We can interpret these results with the help of Figs. \ref{fig:artificial_good_examples} and \ref{fig:artificial_challenging_examples}. The Gaussian likelihood approach yields much lower uncertainties, but these often fail to cover the true solution, leading to overconfidence. On the other hand, the score compression-based SBI approach produces much greater uncertainties (as in \Cref{fig:artificial_good_examples}) which are generally consistent with the true solution. However, both approaches demonstrate inference bias: \Cref{fig:artificial_challenging_examples} provides an example of how inaccuracies in the Gaussian likelihood can introduce this bias. 

The ML-based compression SBI approach exhibits a mild degree of under-confidence, producing conservative uncertainty estimates that tend to overestimate its own uncertainty (by around $10\%$). This is not ideal since it indicates that the model could produce tighter constraints on the moment tensor: i.e., the deep-learning based compression algorithm is extracting more information from the observations that are not fully taken advantage of by the NDE head. From a practical perspective, however, this under-confidence is far preferable to overconfidence, in the sense that at least one can be sure the modelled posterior covers the true solution.  There is a wide body of literature that has worked towards ensuring SBI approaches yield conservative or calibrated posteriors, rather than over-confident posteriors \citep{delaunoy2022towards, hermans2021}.

We also systematically evaluate the modelled uncertainties and biases of each approach over the 400 artificial inversions. The results for each approach are shown in \Cref{table:uncertainties_and_bias}. The Gaussian likelihood approach produces low uncertainties, to some degree driven by its over-confidence. We also find that a Gaussian likelihood produces systematic posterior biases, particularly for $M_\mathrm{W}$ where the magnitude is significantly under-estimated for a large number of events. The biases on the focal mechanism parameters $\gamma$ and $\delta$ are minor, however, and we did not find much evidence of particular mechanisms being over- or under-represented.

\begin{table*}
\centering
\begin{tabular}{l l l c c c c c c}
\hline
Method & Compression & Metric & $\gamma$ & $\delta$ & $ M_w (10^{-2})$ & strike & dip & rake \\
\hline
\multirow{4}{*}{SBI}
 & \multirow{2}{*}{Optimal score} & $\sigma$ & 1.51 & 2.10 & 0.96 & 2.56 & 1.91 & 2.94 \\
 &  & Bias & -0.17 & 0.49 & -0.23 & 0.05 & -0.24 & 0.29  \\
 & \multirow{2}{*}{Deep learning} & $\sigma$ & 0.71 & 0.84 & 0.37 & 1.25 & 0.95 & 1.52 \\
 &  & Bias & -0.06 & 0.06 & -0.01 & -0.23 & 0.06 & -0.26 \\

\hline
\multirow{2}{*}{Gaussian}
 & \multirow{2}{*}{ -} & $\sigma$ & 0.66 & 0.87 & 0.45 & 0.79 & 0.71 & 1.04 \\
 &  & Bias & -0.15 & 0.15 & -0.32 & 0.07 & -0.18 & 0.53  \\
\hline
\end{tabular}
\caption{Average posterior standard deviations ($\sigma$) and biases for each inference method over the 400 artificial inversions with Earth structure uncertainty of $\kappa=5\%$ (see also Figure \ref{fig:artificial_good_examples}). Bias is computed per inversion by computing the difference between the mean of the modelled posterior and the true source parameters. $M_w$ values are reported in units of $10^{-2}$, all other values in units of degrees. }
\label{table:uncertainties_and_bias}
\end{table*}

Between the SBI approaches, we find that the deep-learning based compression produces significantly tighter (around a factor of $\times 2$) constraints on the source parameters. This advantage stems from the fact that the deep-learning model is capable of extracting more information from the observations than score compression. In addition, since the deep learning-based compression algorithm does not rely on an inaccurate Gaussian likelihood, it produces posteriors with lower systematic bias, particularly for the focal mechanism parameters $\gamma, \delta$, and  $M_\mathrm{W}$.

\subsubsection{Case study 2: Gaussian likelihood inversion quality under different experimental configurations}

We next explore how the Gaussian likelihood assumption fares under varying experimental configurations. We test the following scenarios with input source models randomly sampled from the prior:
\begin{enumerate}
    \item differing levels of Earth model uncertainty,
    \item introducing shorter period data in the filtering window,
    \item inversions with a more balanced station configuration.
\end{enumerate}

\begin{figure*}
 \centering
    \includegraphics[width=\textwidth]{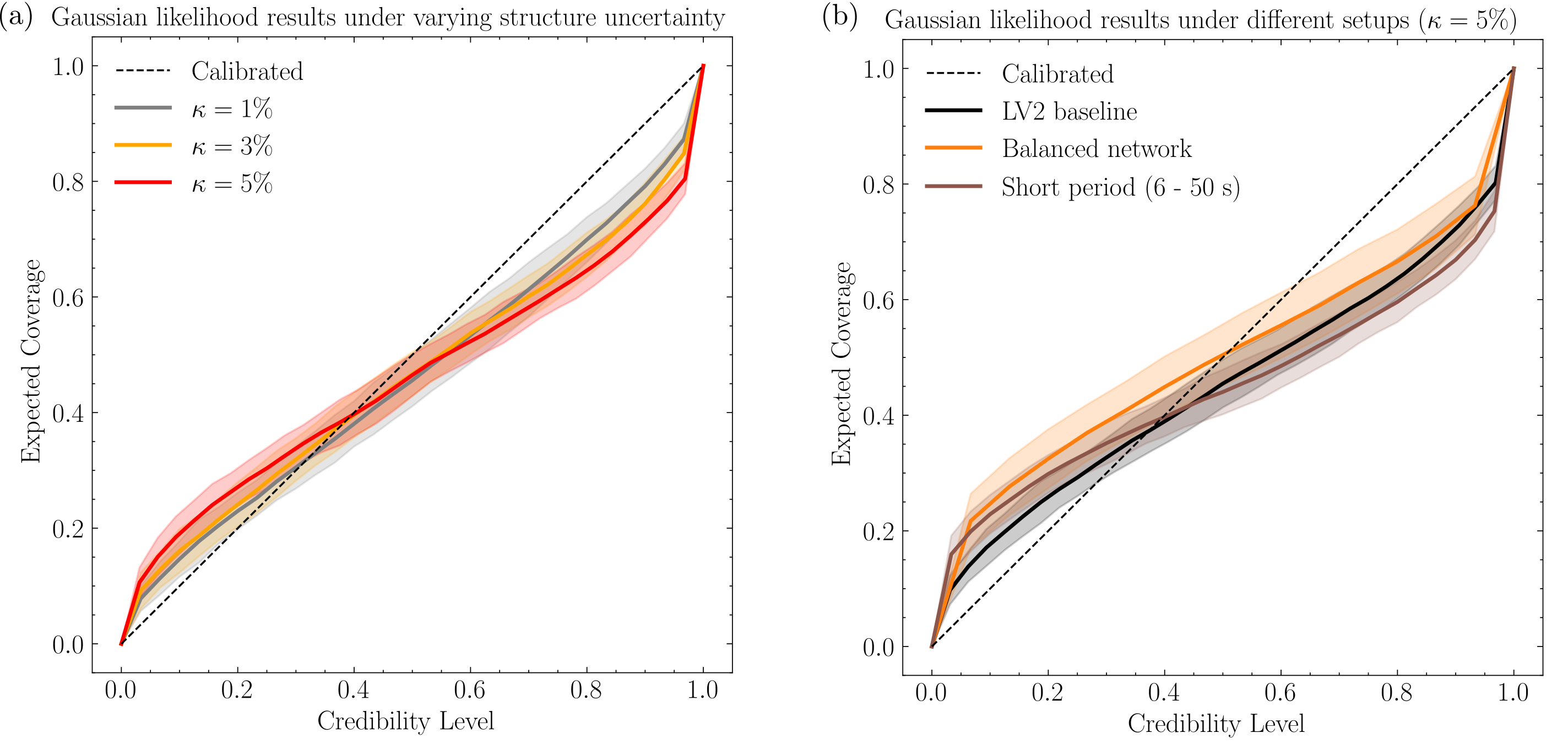}
   \caption{The calibration performance of the Gaussian likelihood inversions under varied experimental configurations. Panel (a) shows the results under varying levels of Earth structure uncertainty, parametrised by fractional perturbations $\kappa$. Panel (b) explores two different experimental configurations: one with a balanced network with full azimuthal coverage, and another that includes shorter periods in the inversion. }
   \label{fig:GL_calibration_curves}
\end{figure*}

We begin by exploring model calibration under three levels of Earth structure uncertainty, $\kappa \in[1,3,5]\%$ from scenario (i). Fig. \Cref{fig:GL_calibration_curves} shows the results of 300 artificial inversions under each of these uncertainty scenarios. We find that while even under mild uncertainty ($\kappa=1\%$), a Gaussian likelihood yields significantly overconfident posteriors that underestimate the moment tensor component uncertainty by around 30\%. This increases to a $50\%$ underestimation by $\kappa=5\%$. In addition, we find the Gaussian likelihood approach leads to systematic bias in the moment tensor solutions across all levels of Earth structure uncertainty. This is shown by the asymmetry of the calibration curve, reflecting the biases in the focal mechanism uncovered in \Cref{sec:GL_vs_SBI_casestudy}. 

We next consider shorter period data by filtering between 6-50s from scenario (ii). We expect that this should exacerbate modelling errors due to stronger sensitivity to smaller scale structures and, more importantly, lead to more non-Gaussian effects in the theory errors.  \Cref{fig:GL_calibration_curves} demonstrates that the Gaussian likelihood assumption is significantly degraded in this context, leading to poorer calibration relative to the (already problematic) baseline 20 - 50 s period band. We find that the Gaussian likelihood assumption now underestimates its moment tensor component uncertainties by around a factor of $\times2$, and there is a persistent and significant bias. We show a typical example of posterior inference in Fig. S\ref{SI:fig_2}.

Finally, we repeat this procedure for scenario (iii), performing repeated inversions using a balanced network with good azimuthal coverage around the event, with the exact configuration shown in Fig. S\ref{SI:fig_1}. \Cref{fig:GL_calibration_curves} finds a similar degree of overconfidence to the baseline case (again around $50\%$ overconfident). This indicates that good station coverage does not mitigate against a poorly specified likelihood in terms of producing well-calibrated moment tensor uncertainties. However, it is important to note that in such scenarios, the moment tensor is recovered with higher precision. Better station coverage is therefore still desirable, since incorrect uncertainty estimation may not affect the overall interpretation of the focal mechanism solution. We can nonetheless infer that the limitations with the likelihood-based approach uncovered in this section cannot be attributed to poor station coverage. 

By comparison, we find that the SBI approaches produce reliable posteriors under scenarios (i) and (iii), with calibration curves for the SBI approaches shown in Figs. S\ref{SI:fig_3} \& S\ref{SI:fig_4}. However, including higher frequency information in the waveforms, as done in scenario (ii), exacerbates the bias in the score compression SBI approach. This is likely for the same reasons discussed in \Cref{sec:GL_vs_SBI_casestudy} (i.e., biased MLE estimates of $\mathbf{m}$ due to an incorrect Gaussian likelihood), though calibration is still much better than using the Gaussian likelihood directly. The ML-based compression approach remains reliable, unbiased and slightly conservative for most of the scenarios tested, though we find its performance also degrades on the short period data. We provide some comment on this in Fig. S\ref{SI:fig_4}. We also test to what extent model misspecification affects the ML-based compression approach in Fig. S\ref{SI:fig_4}, and find that while it induces predictable miscalibration, the method does not catastrophically fail. 

We present an in-depth analysis showing the posterior accuracies over the 300 synthetic inversions for several of the scenarios tested in Figs. S\ref{SI:fig_5}, S\ref{SI:fig_6} \& S\ref{SI:fig_7}. We also repeat the posterior width and bias analysis of \Cref{table:uncertainties_and_bias} for the $\kappa=3\%$ uncertainty and shorter period inversions in Tables S\ref{SI:table_1} \& S\ref{SI:table_2}.

\subsubsection{Case study 3: shallow, isotropic events}

Shallow isotropic sources have proven theoretically challenging to fully resolve using regional surface waves, even when accounting for uncertainty in the Earth structure \citep{kvrivzova2013resolvability,hu2023seismic, chiang2025bayesian, pham2025global}. This is because fundamentally different source mechanisms, such as isotropic and vertical CLVD, could produce nearly identical regional surface waves \citep{kawakatsu1996observability,ford2012event,hu2023seismic}. The ambiguity could result in much larger solution uncertainty for non-double-couple source types than for seismic sources at greater depths. Here we test SBI's ability to retrieve truthworthy uncertainty estimates in the challenging setting. To do so, we study a synthetic scenario using the same station configuration as in the LV2 event (\Cref{fig:event_network_maps}) and consider highly isotropic sources with a source depth of 500 m. We set the Earth structure uncertainty to $\kappa=5\%$, and test each of the three approaches on 10 randomly sampled isotropic events. 

\begin{figure*}
 \centering
    \includegraphics[width=\textwidth]{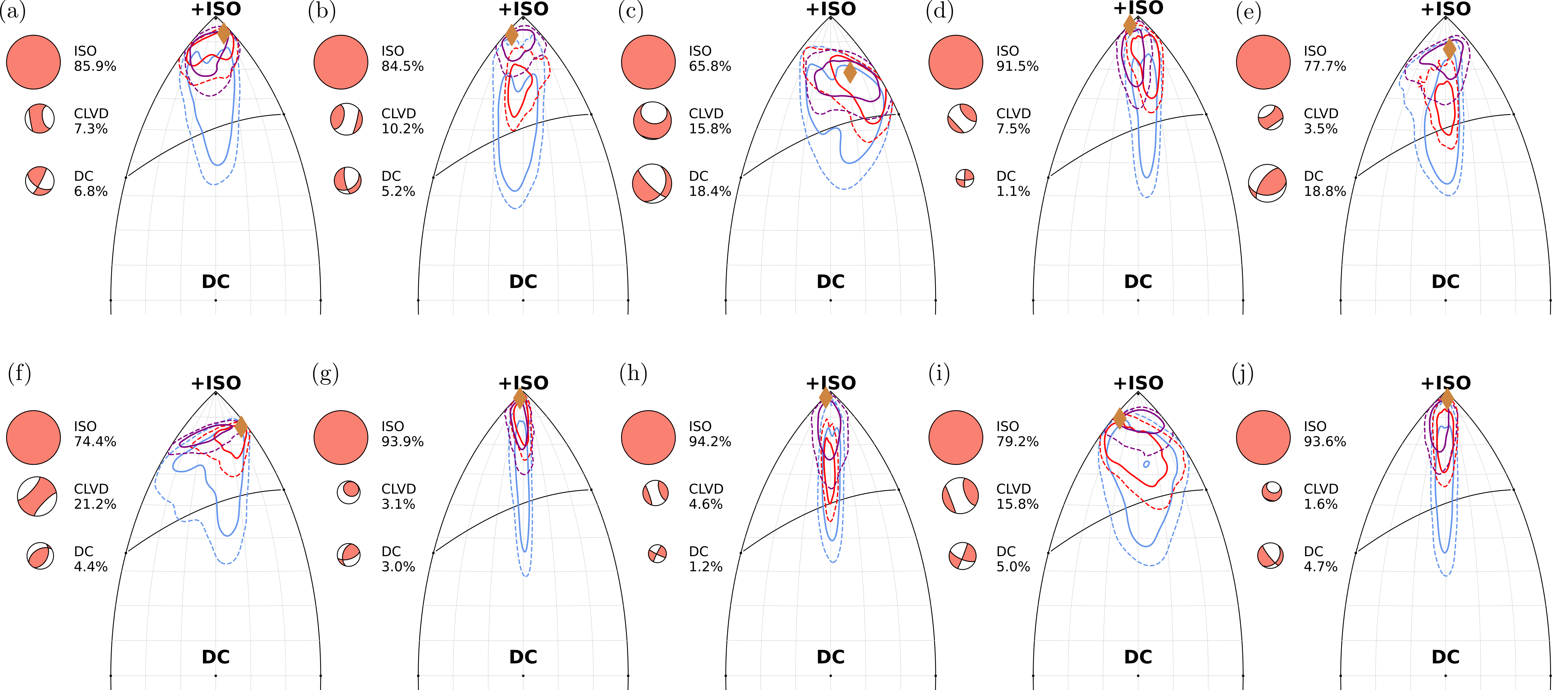}
   \caption{Inference results with each approach on 10 random artificial shallow isotropic sources. We show the isotropic half of the lune plot for each inversion, with the true solution marked by an orange diamond. The focal mechanism decomposition of the true source into its fractional isotropic, CLVD and double-couple components is shown to the left of each lune. The posterior solutions from the Gaussian likelihood approach (red), the score-compression SBI approach (blue) and the ML-based compression SBI approach (purple) are shown as contours denoting the $\pm[1,2]$ sigma regions. }
   \label{fig:shallow_isotropic_sources}
\end{figure*}

The results of this analysis are shown in \Cref{fig:shallow_isotropic_sources}. We focus only on the lune plot in order to evaluate how well each approach resolves the focal mechanism components of each event. We find several instances where the Gaussian likelihood-based approach (and, by extension, the score compression SBI approach) significantly underestimates the isotropic component of the event. These can be seen most clearly for lunes (b), (e), (h), and (i), where the maximum a posteriori point for both a Gaussian likelihood-based analysis and the score compression approach substantially underestimates the isotropic component. In contrast, the ML-based compression approach to SBI produces posterior contours that more faithfully recover the isotropic component of the event. 

It is worth noting that the uniform prior of the moment tensor components assumed here leads to very low prior density toward the pure isotropic sources. This is true for the Tape parametrisation as well \citep{tape2015uniform}. This induces a preference for smaller isotropic components, particularly when the observational uncertainties are poorly modelled by the likelihood. Prior volume effects may therefore explain the strong isotropic---double-couple trade-off visible for many inversions in \Cref{fig:shallow_isotropic_sources}. Nonetheless, we may reasonably conclude that the more accurate (implicit) likelihood modelling performed by the ML-based approach ensures improved posterior solutions, regardless of the prior chosen. These preliminary results, not yet fully explored in this study, are promising for future real data applications in the context of non-proliferation seismology \citep{ford2012event,chiang2018moment}. 

\subsection{A note on computational costs}

As described in \Cref{sec:experimental_details}, the ML-based compression algorithm had a significant drawback relative to the score compression approach: for each source-receiver configuration, it cost $12$ GPU hours to train the deep learning model, in conjunction with the larger number of forward simulations required for training. 

However, once trained, the ML-based approach is highly efficient, and can perform posterior inference on hundreds of events within seconds. This advantage was felt directly in the synthetic tests in  \Cref{sec:synthetic_inversions}, where for each experimental configuration the ML-based compression training and inference time was fixed at 12 hours. However, for the other two approaches the local covariance  $\mathbf{m}_\textrm{MLE}$ estimation and all inference overheads had to be repeated for every event. For the 300 synthetic inversions required to produce reliable coverage statistics, this amounted to 3-4 days of compute using 30 CPUs for each experimental configuration. For each experiment this totaled $\sim25\times10^6$ forward model evaluations for the Gaussian likelihood approach and $\sim3.5\times10^6$ for the score compression approach; 1 to 2 orders of magnitude more than the $0.1\times10^6$ forward model evaluations for the ML-based approach. 

It would be quite restrictive to need to re-train a deep-learning compression model for each event under study. Fortunately, while we do not investigate this here, prior studies using similar architectures have demonstrated the ability to generalise across varying source locations and source-receiver geometries (e.g., \citealt{munchmeyer2021transformer, hourcade2025pegsgraph}). A single ML model could therefore be trained to perform inversions in a given region even under varying source-receiver geometries. In principle, this could enable extremely efficient Bayesian source mechanism determination for whole earthquake catalogues, with several order of magnitudes reduction in computational cost relative to standard methods. 

\section{Application to real data}
\label{sec:real_data}

We repeat the inversion procedures for the two real events in \Cref{fig:event_network_maps}. All data are processed as in the synthetic experiments, using a 20 - 50 s period filtering band and a 200 s window following the origin time of the event. The per-station component noise variance is estimated by using the window prior to the event, and then used to rescale the data covariance matrices $\mathbf{C}_d$. 

For the Gaussian likelihood and the score compression SBI approaches we use prior moment tensor solutions to estimate station-specific time-shifts and align the observations with the synthetics. We then performed iterative least squares for a best fitting moment tensor estimate $\mathbf{m}_\textrm{MLE}$. Finding that this did not yield perfectly stable or recoverable best fitting solutions, we also performed a single round of MCMC, which mildly improved (but did not ensure) the stability of the inversion process. As this did not occur during the synthetic tests, we attribute these issues to further unmodelled theory errors. 

For the ML-based compression approach, we perform no alignment and implicitly marginalise over station specific time-shifts using the approach detailed in \Cref{sec:2d_structure_approach}. We use a simple, heuristic prior over time-shifts that absorbs errors in the absolute origin time with a Gaussian distribution, and per-station corrections that are uniformly distributed:
\begin{equation}
t_0 \sim \mathcal{N}(0, \sigma_\textrm{ot}^2),
\qquad
\delta_i \sim \mathcal{U}(-\Delta_\textrm{sc}, \Delta_\textrm{sc}),
\qquad
t_i = t_0 + \delta_i .
\end{equation}
We use $p(\mathbf{t}; \sigma_\textrm{ot} =2\:\textrm{s},\Delta_\textrm{sc}=3\:\textrm{s} )$, which is empirically motivated by the best fitting time-shifts found in prior regional studies \citep{phạm2021toward, hu2025bayesian}. We did not perform any detailed analysis regarding this choice, but found that the ML-based compression solutions were not significantly impacted by small changes to $\sigma_\textrm{ot}$ and $\Delta_\textrm{sc}$.

\subsection{Long Valley Caldera $M_W\:4.9$, 11/22/1997}

\label{sec:LV2}

\begin{figure*}
 \centering
    \includegraphics[width=\textwidth]{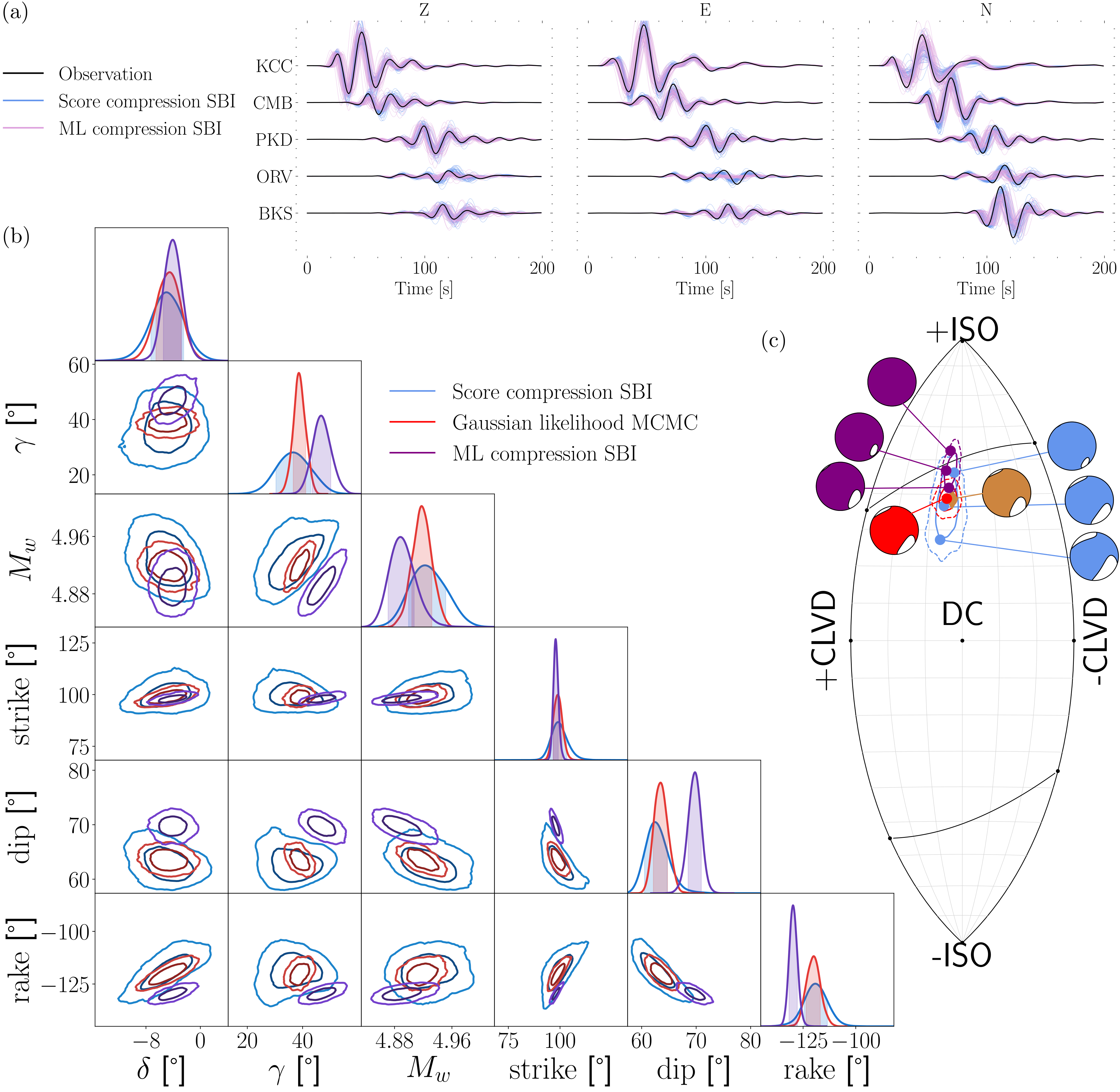}
   \caption{Posterior inference on an the LV2 volcanic event in Southern California. Panels (a), (b) and (c) are as before in \Cref{fig:artificial_good_examples}, with the gold coloured beachball in (c) showing the LV2 focal mechanism solution from \citet{phạm2021toward}.}
   \label{fig:LV2_real_event_inversion}
\end{figure*}

We reanalyse a well-studied event at the Long Valley Caldera in California, USA at 17:20:35 on 11/22/1997, referred to as LV2 in \citet{minson2008stable}. Prior studies have indicated a significant non-double-couple component likely related to volcanic processes at the caldera \citep{dreger2000dilational, phạm2021toward}. We use the SoCal 1-D Earth model as in the synthetic examples, and use $\kappa=5\%$ as the fractional uncertainty in the layer depths and layer velocities $V_p$ and $V_s$. We use a theory error regularisation factor of $\epsilon=0.005$ in \Cref{eq:regularised_final_covariance}. We present the results of each of the three approaches: a Gaussian likelihood based inversion, SBI with score compression, and SBI with ML-based compression in \Cref{fig:LV2_real_event_inversion}. We also show the focal mechanism solution found in \citet{phạm2021toward}, who introduced the Gaussian likelihood methodology we reproduce here in \Cref{fig:LV2_real_event_inversion}c. 

As in the synthetic examples, we find close agreement between the Gaussian likelihood and score compression-based SBI approach for the maximum a posteriori solution of the focal mechanism. Again, however, the SBI approach yields substantially higher uncertainties, which our results in \Cref{sec:synthetic_inversions} suggest are a better characterisation of the true uncertainty. On the other hand, the ML-based approach yields slightly different moment tensor solutions, with a fractionally higher isotropic component than the other two approaches. We found this higher isotropic component persisted across ML training runs, as well as when time-shifts were manually corrected as in the two other approaches (i.e. by training a model with $p(\mathbf{t}; \sigma_\textrm{ot} =0\:\textrm{s},\Delta_\textrm{sc}=0\:\textrm{s} )$ and aligning the observation with the best fitting synthetics). All solutions yield similar trade-offs between component pairs, even when they slightly disagree on the best solution or uncertainty widths (note e.g. a strong positive tradeoff between the magnitude $M_w$ and isotropic component $\gamma$).  

We show a random subset of 10,000 posterior predictive checks in \Cref{fig:LV2_real_event_inversion}a. These posterior predictive checks show that both SBI approaches yield plausible fits to the data. For some stations (e.g. KCC), the ML-based compression approach yields better fits to the data, while for others (e.g. ORV) the score compression approach seems to fit better. We verify quantitatively in S\ref{SI:table_3} that the ML-based moment tensor ensemble produces synthetics that are competitive with or improve over the the other two approaches in terms of station-component power content, envelope misfit, and $\chi^2$ fit, which all provide somewhat independent probes of waveform fit. We interpret this as providing evidence that all approaches provide reasonable agreement with the data.

\begin{figure*}
 \centering
    \includegraphics[width=\textwidth]{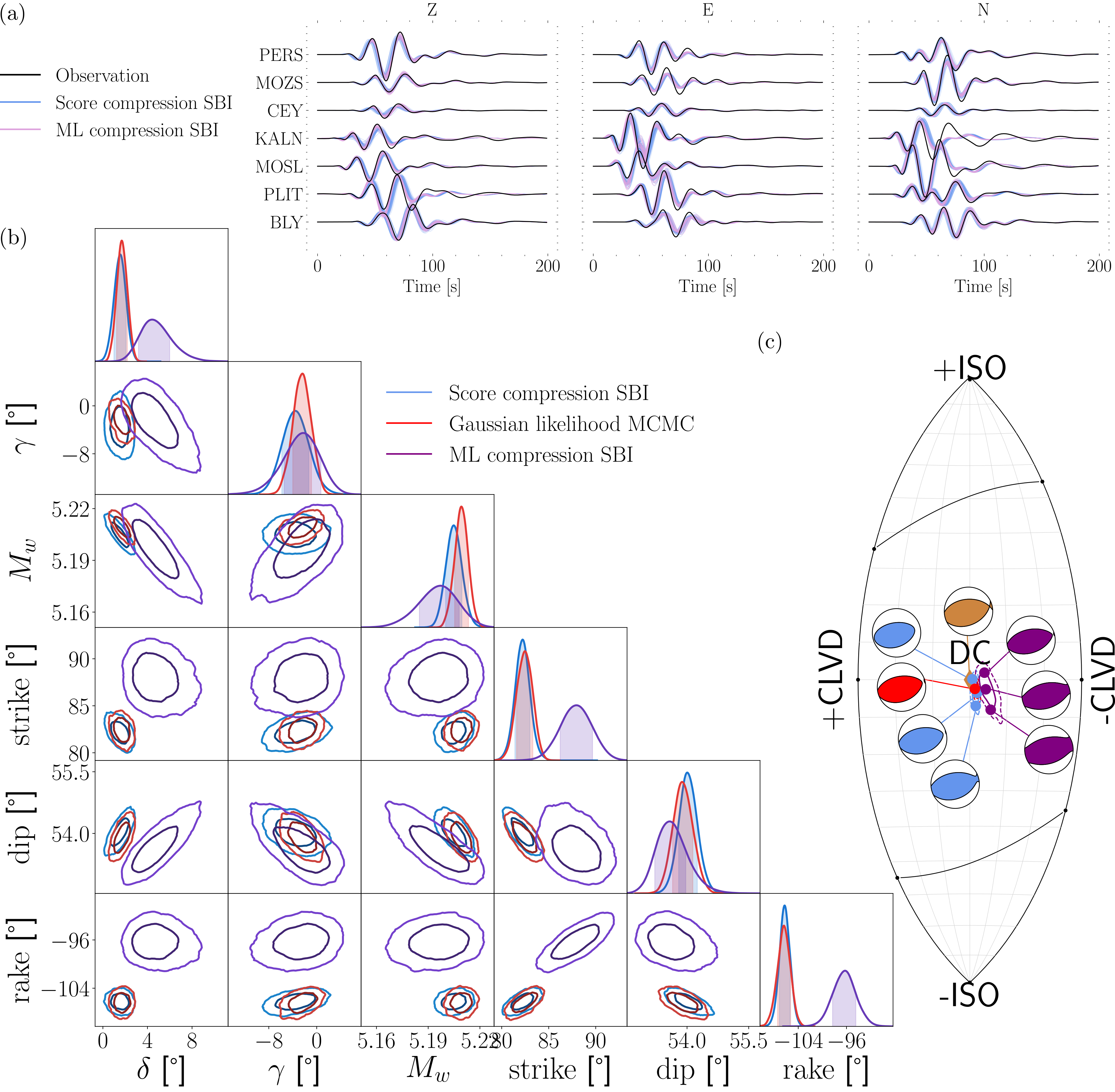}
   \caption{Posterior inference on the Zagreb 2020 event. Panel (a) shows the posterior predictive checks for the two SBI approaches. For visualisation purposes, for the ML-based compression approach we plot a random subset of the 10\% lowest $\chi^2$ synthetics (to avoid showing the broad prior over station time shifts $p(\mathbf{t})$). Panels (b) and (c) are as before, with the gold coloured beachball in (c) showing the Zagreb 2020 mainshock (deviatoric) solution from \citet{hu2025bayesian}. }
   \label{fig:croatia_real_event_inversion}
\end{figure*}

\subsection{Zagreb, Croatia, $M_W\:5.3$ 2020/03/22}

The 2020 Zagreb earthquake had a significant human toll, causing one fatality, many injuries, and reported economic losses exceeding several billion euros \citep{novak2020zagreb, atalic2021mw5}. The focal mechanism of the earthquake has been analysed in several prior works \citep[e.g.,][]{markuvsic2020zagreb, hu2025bayesian}. \citet{hu2025bayesian} used a station-specific time-shift approach to treat unresolved Earth structure uncertainty, and retrieved a double-couple source-type mechanism largely consistent with the tectonic background in the region. 

To aid interpretation, we use stations utilised in their analysis, and use the same 2-D composite model (made up of three different 1-D Earth models depending on the station location) used in their work. We present details of the composite model in Fig. S\ref{SI:fig_8}, and defer to prior work for a more detailed exposition of the tectonic and geophysical background of the region \citep[e.g.,][]{herak2009seismicity, stipvcevic2011crustal, stipvcevic2020crustal,hu2025bayesian}. We apply $\kappa=3\%$ perturbations to the velocity models, motivated by the approximate uncertainty levels discovered in \citet{stipvcevic2020crustal}. These are applied to each of the 1-D Earth models independently. This leads to a problem of greater complexity (and realism) than the LV2 event forward modelling, since structure perturbations are now only azimuthally (rather than globally) correlated.  

The results of all three approaches are presented in \Cref{fig:croatia_real_event_inversion}. We find close agreement of a double-couple mechanism with the Gaussian likelihood approach and the score compression SBI approach with \citet{hu2025bayesian}, though there are minor differences in fault orientation and magnitude. We found some instability in the exact maximum likelihood solution $\mathbf{m}_\textrm{MLE}$ even with a large theory regularisation term $\epsilon=0.05$, indicating significant unmodelled errors. 

Although broadly compatible with the other two approaches and with the solution of \citet{hu2025bayesian}, the ML-based compression approach shows evidence of a minor CLVD component to the event, and prefers different strike and rake orientations (with discrepancies between  $5-10\degree$). The ML-based compression approach also has larger uncertainties than in the previous examples, which we hypothesise results from the more challenging to model Earth structure uncertainties, with independent 1-D Earth models causing uncorrelated errors at different stations. We verified that training with a larger dataset of simulations reduced these uncertainties, so conclude this effect is related to increased inherent difficulties in modelling the composite model uncertainties. 

The ML-based compression yields a plausible solution --- it is consistent with a very low ($\sim 2\degree$) CLVD component within the 95\% confidence interval. It is also possible that nearby faults with differing orientations may have been activated during or contributed to the main shock, introducing a small modelled non-double-couple component. More thorough investigation is warranted to better understand this fractional CLVD component. While the ML-based approach performs a more comprehensive treatment of the uncertainties (i.e., by marginalising over station-specific time-shifts and not making any assumptions about the form of the errors), it is not possible to rule out that unmodelled effects may introduce bias in the resulting solution. As such, we encourage some caution in further interpretation of this result.

Again, a quantitative analysis of the posterior predictive checks in Table S\ref{SI:table_4} suggests that all approaches provide comparable fits to the data. However, we find that the posterior predictive checks in \Cref{fig:croatia_real_event_inversion}a are in some sense under-dispersive and do not fully cover the observations. This is an indication of model misspecification; i.e., the theory errors induced by our 1-D Earth structure perturbations underestimate or do not capture the true modelling uncertainty. 

\section{Discussion} 
\label{sec:discussion}

This work has focused on uncertainties arising from 1-D Earth structure, with azimuthal variations treated in a simplified manner through station-specific time shifts. An important next step is to rigorously assess how the different inversion approaches perform on more realistic data. The score-compression SBI approach is closely tied to a Gaussian likelihood, whose behaviour and robustness on real data is better understood. Nonetheless, 3-D Earth structure effects, particularly at lower periods, will likely exacerbate the Gaussian likelihood mismodelling issues uncovered in this work.  

The reliability of the fully data-driven deep learning compression remains  unclear. In particular, unmodelled effects such as 3-D Earth structure and site response may lead to more severe out-of-distribution errors, which could underlie the minor moment tensor solution differences uncovered in \Cref{sec:real_data} \citep{van2020uncertainty, schmitt2023detecting, pierre2026mitigating}. A more comprehensive evaluation is therefore warranted, for example by testing the deep learning approach on 3-D synthetics that also include realistic observational contaminants.

Future work should attempt to model more realistic Earth structure uncertainties, such as 3-D heterogenieties \citep[as in e.g.,][]{phạm2024towards} and anisotropy. These could be derived from tomographic model ensembles \citep{chiang2025bayesian}, for which SBI is better suited to deal with than likelihood-based analysis (as explained in \Cref{sec:theory_intro}). Beyond Earth structure, there are many more modelling phenomena that SBI could be applied to: source time functions \citep{vallee2011scardec, stahler2014fully}, source  location uncertainties \citep[and receiver uncertainties in the case of ocean bottom seismometers;][]{stachnik2012determination, trabattoni2020orienting, lindner2023bayesian}, composite or finite source models, or potentially even errors in the forward modelling itself. All these effects could be treated either as model parameters to be inverted for or ``nuisance'' parameters to marginalise over. 

We explored two extremes of compression for SBI: a generic compression algorithm applied on the raw full waveforms, and a fully ML-based data-driven approach. These both had limitations: the former approach was brittle under minor non-linearities, while the latter requires designing and training a tailored deep learning architecture. In practice, the field of SBI has heavily relied on ``two-step'' compression approaches \citep{Alsing2018, alsing2019fast, Cranmer2020, gatti2024dark}, which first extracts more stable physically motivated summaries from the observations before compressing them further. For full-waveform seismic observations, there are a wide range of options to consider: power spectra, time-frequency representations or wavelet coefficients \citep{cesca2006amplitude,fichtner2008theoretical,vavryvcuk2012moment}; peak amplitude, waveform envelope or coda statistics \citep{nakahara2008seismogram,eulenfeld2022fast}; arrival times, first-motion polarities and S to P amplitude ratios (\citealp{hardebeck2002new, skoumal2024skhash}, see also \citealp{song2025foconet} for a relevant ML-based example). Indeed, our purely data-driven approach could no doubt also benefit from some hybrid: for instance, including physically motivated summaries \citep{makinen2024hybrid, jeffrey2025dark} or explicit simulator feedback (e.g., \citealt{graikos2022diffusion,chung2023diffusion, holzschuh2024flow}; see \citealp{thuerey2021physics} for a review of methods), to incorporate information from the forward model during inference.

This work utilised the neural posterior estimation (NPE) framework, which trains a model of the posterior. This allows for rapid posterior sampling, but is somewhat restrictive in that it bakes in the prior used for dataset generation. Neural likelihood estimation (NLE) effectively learns probabilistic emulator of the forward model (including all uncertainty effects), and once trained can be used to probe the effects of the prior on moment tensor solutions \citep{papamakarios2019sequential}. Likelihoods across independent measurements can also be multiplied, which could enable the application of SBI to joint inversions of full-waveform data with first-motion polarities \citep{hamidbeygi2023bayesian, chiang2025bayesian} or InSAR data \citep{delouis2002joint, weston2011global}.

\section{Conclusions}
\label{sec:conclusions}

This manuscript has demonstrated that accurate Bayesian moment tensor inversions under model uncertainty rely on well-characterised likelihoods. Standard Gaussian likelihood approximations introduce bias and overconfidence in the solutions. Simulation-based inference (SBI) produces better models of the likelihood using machine learning, enabling accurate and calibrated moment tensor inversions under challenging sources of uncertainty. 

We showed that Gaussian likelihood approximations break down under very minor velocity model uncertainty. Predictably, as the degree of velocity structure uncertainty increases from $1\%$ to $5\%$ more pathologies are introduced in the Gaussian likelihood-based moment tensor solutions. These problems are exacerbated when treating higher frequency data. We also showed that Gaussian likelihood inversions of shallow isotropic sources may significantly underestimate the isotropic component of the source. 

We introduced two frameworks for empirical likelihood modelling with SBI. One leveraged the generic and lightweight score compression algorithm, which uses a Taylor expansion of the forward model to reduce the dimensionality of full-waveform seismic observations. Score compression-based SBI produced much better calibrated moment tensor posterior solutions than a Gaussian likelihood approach, albeit with very occasional biases. The second framework used a fully data-driven approach, using a deep learning model to compress the observations and model the posterior. The latter framework is significantly more flexible and can straightforwardly incorporate more complex sources of uncertainty. Importantly, the flexibility and efficiency of deep learning-based SBI has the potential to enable rapid and robust Bayesian moment tensor inversion at the catalogue level.

\begin{acknowledgments}
We thank Iva Dasović and Krešimir Kuk at the Croatian Seismograph Network, as well as Jinyin Hu, for providing expertise and access to the data for the Zagreb 2020 earthquake. We are grateful to Davide Piras, Alessio Spurio Mancini, and Benjamin Joachimi for helpful discussions in preparation for this work. AAS was supported by the STFC UCL Centre for Doctoral Training in Data Intensive Science (grant ST/W00674X/1) and by departmental and industry contributions. AAS was also supported by the A. G. Leventis Foundation educational grant scheme. T.-S. P. acknowledges financial support from the Australian Research Council through a Discovery Early Career Research Award (DE230100025).

\end{acknowledgments}

\begin{dataavailability}
\texttt{Python} code used to produce the results of this study is available at \href{https://github.com/asaoulis/seismo-sbi}{https://github.com/asaoulis/seismo-sbi}. We used several widely available Python packages for processing the data and training the neural networks:
\texttt{obspy} \citep{beyreuther2010obspy}, \texttt{PyTorch} \citep{paszke2019pytorch}, \texttt{PyTorch Lightning} \citep{lightning}, \texttt{sbi} \citep{tejero2020sbi}, and \texttt{nflows} \citep{nflows}. 
\end{dataavailability}

\bibliographystyle{gji}
\bibliography{bibliography}

\begin{thebibliography}{141}
\expandafter\ifx\csname natexlab\endcsname\relax\def\natexlab#1{#1}\fi

\bibitem[Abercrombie \& Ekstr{\"o}m(2001)]{abercrombie2001earthquake}
Abercrombie, R.~E. \& Ekstr{\"o}m, G., 2001.
\newblock Earthquake slip on oceanic transform faults, {\it Nature\/}, {\bf 410}(6824), 74--77.

\bibitem[Akuhara et~al.(2025)Akuhara, Shinohara, Yamada, Azuma, Hino, Obana, Takahashi, Fujie, Kodaira, Murai, et~al.]{akuhara2025non}
Akuhara, T., Shinohara, M., Yamada, T., Azuma, R., Hino, R., Obana, K., Takahashi, T., Fujie, G., Kodaira, S., Murai, Y., et~al., 2025.
\newblock Non-double-couple components of the 2024 noto earthquake aftershocks: influence on focal mechanism estimation, {\it Earth, Planets and Space\/}, {\bf 77}(1), 145.

\bibitem[Alsing \& Wandelt(2018)]{Alsing2018}
Alsing, J. \& Wandelt, B., 2018.
\newblock Generalized massive optimal data compression, {\it MNRAS\/}, {\bf 476}, 60--64, doi: 10.1093/mnrasl/sly029.

\bibitem[Alsing et~al.(2017)Alsing, Wandelt, \& Feeney]{Alsing2017}
Alsing, J., Wandelt, B., \& Feeney, S., 2017.
\newblock Massive optimal data compression and density estimation for scalable, likelihood-free inference in cosmology, {\it MNRAS\/}, {\bf 000}, 1--14.

\bibitem[Alsing et~al.(2019)Alsing, Charnock, Feeney, \& Wandelt]{alsing2019fast}
Alsing, J., Charnock, T., Feeney, S., \& Wandelt, B., 2019.
\newblock Fast likelihood-free cosmology with neural density estimators and active learning, {\it Monthly Notices of the Royal Astronomical Society\/}, {\bf 488}(3), 4440--4458.

\bibitem[Atali{\'c} et~al.(2021)Atali{\'c}, Uro{\v{s}}, {\v{S}}avor~Novak, Dem{\v{s}}i{\'c}, \& Nastev]{atalic2021mw5}
Atali{\'c}, J., Uro{\v{s}}, M., {\v{S}}avor~Novak, M., Dem{\v{s}}i{\'c}, M., \& Nastev, M., 2021.
\newblock The mw5. 4 zagreb (croatia) earthquake of march 22, 2020: impacts and response, {\it Bulletin of Earthquake Engineering\/}, {\bf 19}(9), 3461--3489.

\bibitem[{Atlas Collaboration}(2025)]{atlas2025implementation}
{Atlas Collaboration}, 2025.
\newblock An implementation of neural simulation-based inference for parameter estimation in atlas, {\it Reports on Progress in Physics\/}, {\bf 88}(6), 067801.

\bibitem[Beyreuther et~al.(2010)Beyreuther, Barsch, Krischer, Megies, Behr, \& Wassermann]{beyreuther2010obspy}
Beyreuther, M., Barsch, R., Krischer, L., Megies, T., Behr, Y., \& Wassermann, J., 2010.
\newblock Obspy: A python toolbox for seismology, {\it Seismological Research Letters\/}, {\bf 81}(3), 530--533.

\bibitem[Blom et~al.(2023)Blom, Hardalupas, \& Rawlinson]{blom2023mitigating}
Blom, N., Hardalupas, P.-S., \& Rawlinson, N., 2023.
\newblock Mitigating the effect of errors in source parameters on seismic (waveform) tomography, {\it Geophysical Journal International\/}, {\bf 232}(2), 810--828.

\bibitem[Cesca et~al.(2006)Cesca, Buforn, \& Dahm]{cesca2006amplitude}
Cesca, S., Buforn, E., \& Dahm, T., 2006.
\newblock Amplitude spectra moment tensor inversion of shallow earthquakes in spain, {\it Geophysical Journal International\/}, {\bf 166}(2), 839--854.

\bibitem[Charnock et~al.(2018)Charnock, Lavaux, \& Wandelt]{charnock2018automatic}
Charnock, T., Lavaux, G., \& Wandelt, B.~D., 2018.
\newblock Automatic physical inference with information maximizing neural networks, {\it Physical Review D\/}, {\bf 97}(8), 083004.

\bibitem[Chiang et~al.(2018)Chiang, Ichinose, Dreger, Ford, Matzel, Myers, \& Walter]{chiang2018moment}
Chiang, A., Ichinose, G.~A., Dreger, D.~S., Ford, S.~R., Matzel, E.~M., Myers, S.~C., \& Walter, W., 2018.
\newblock Moment tensor source-type analysis for the democratic people’s republic of korea--declared nuclear explosions (2006--2017) and 3 september 2017 collapse event, {\it Seismological Research Letters\/}, {\bf 89}(6), 2152--2165.

\bibitem[Chiang et~al.(2025)Chiang, Ford, Pasyanos, \& Simmons]{chiang2025bayesian}
Chiang, A., Ford, S.~R., Pasyanos, M.~E., \& Simmons, N.~A., 2025.
\newblock Bayesian inference for the seismic moment tensor using regional waveforms and teleseismic-p polarities with a data-derived distribution of velocity models and source locations, {\it Bulletin of the Seismological Society of America\/}.

\bibitem[Chung et~al.(2023)Chung, Kim, Mccann, Klasky, \& Ye]{chung2023diffusion}
Chung, H., Kim, J., Mccann, M.~T., Klasky, M.~L., \& Ye, J.~C., 2023.
\newblock Diffusion posterior sampling for general noisy inverse problems, in {\em The Eleventh International Conference on Learning Representations\/}.

\bibitem[Cranmer et~al.(2020)Cranmer, Brehmer, \& Louppe]{Cranmer2020}
Cranmer, K., Brehmer, J., \& Louppe, G., 2020.
\newblock The frontier of simulation-based inference, {\it Proceedings of the National Academy of Sciences of the United States of America\/}, {\bf 117}, 30055--30062, doi: 10.1073/PNAS.1912789117.

\bibitem[Dax et~al.(2025)Dax, Green, Gair, Gupte, P{\"u}rrer, Raymond, Wildberger, Macke, Buonanno, \& Sch{\"o}lkopf]{dax2025real}
Dax, M., Green, S.~R., Gair, J., Gupte, N., P{\"u}rrer, M., Raymond, V., Wildberger, J., Macke, J.~H., Buonanno, A., \& Sch{\"o}lkopf, B., 2025.
\newblock Real-time inference for binary neutron star mergers using machine learning, {\it Nature\/}, {\bf 639}(8053), 49--53.

\bibitem[Deistler et~al.(2022)Deistler, Goncalves, \& Macke]{deistler2022truncated}
Deistler, M., Goncalves, P.~J., \& Macke, J.~H., 2022.
\newblock Truncated proposals for scalable and hassle-free simulation-based inference, in {\em Advances in Neural Information Processing Systems\/}, vol.~35, pp. 23135--23149, Curran Associates, Inc.

\bibitem[Deistler et~al.(2025)Deistler, Boelts, Steinbach, Moss, Moreau, Gloeckler, Rodriguez, Linhart, Lappalainen, Miller, Goncalves, Lueckmann, Schröder, \& Macke]{deistler2025sbi}
Deistler, M., Boelts, J., Steinbach, P., Moss, G., Moreau, T., Gloeckler, M., Rodriguez, P. L.~C., Linhart, J., Lappalainen, J.~K., Miller, B.~K., Goncalves, P.~J., Lueckmann, J.-M., Schröder, C., \& Macke, J.~H., 2025.
\newblock Simulation-based inference: A practical guide, {\it arXiv\/}.

\bibitem[Delaunoy et~al.(2022)Delaunoy, Hermans, Rozet, Wehenkel, \& Louppe]{delaunoy2022towards}
Delaunoy, A., Hermans, J., Rozet, F., Wehenkel, A., \& Louppe, G., 2022.
\newblock Towards reliable simulation-based inference with balanced neural ratio estimation, {\it Advances in Neural Information Processing Systems\/}, {\bf 35}, 20025--20037.

\bibitem[Delouis et~al.(2002)Delouis, Giardini, Lundgren, \& Salichon]{delouis2002joint}
Delouis, B., Giardini, D., Lundgren, P., \& Salichon, J., 2002.
\newblock Joint inversion of insar, gps, teleseismic, and strong-motion data for the spatial and temporal distribution of earthquake slip: Application to the 1999 izmit mainshock, {\it Bulletin of the Seismological Society of America\/}, {\bf 92}(1), 278--299.

\bibitem[Dreger \& Helmberger(1990)]{dreger1990broadband}
Dreger, D.~S. \& Helmberger, D.~V., 1990.
\newblock Broadband modeling of local earthquakes, {\it Bulletin of the Seismological Society of America\/}, {\bf 80}(5), 1162--1179.

\bibitem[Dreger et~al.(2000)Dreger, Tkalcic, \& Johnston]{dreger2000dilational}
Dreger, D.~S., Tkalcic, H., \& Johnston, M., 2000.
\newblock Dilational processes accompanying earthquakes in the long valley caldera, {\it Science\/}, {\bf 288}(5463), 122--125.

\bibitem[Duputel et~al.(2012)Duputel, Rivera, Fukahata, \& Kanamori]{duputel2012uncertainty}
Duputel, Z., Rivera, L., Fukahata, Y., \& Kanamori, H., 2012.
\newblock Uncertainty estimations for seismic source inversions, {\it Geophysical Journal International\/}, {\bf 190}(2), 1243--1256.

\bibitem[Duputel et~al.(2014)Duputel, Agram, Simons, Minson, \& Beck]{duputel2014accounting}
Duputel, Z., Agram, P.~S., Simons, M., Minson, S.~E., \& Beck, J.~L., 2014.
\newblock Accounting for prediction uncertainty when inferring subsurface fault slip, {\it Geophysical journal international\/}, {\bf 197}(1), 464--482.

\bibitem[Duputel et~al.(2015)Duputel, Jiang, Jolivet, Simons, Rivera, Ampuero, Riel, Owen, Moore, Samsonov, et~al.]{duputel2015iquique}
Duputel, Z., Jiang, J., Jolivet, R., Simons, M., Rivera, L., Ampuero, J.-P., Riel, B., Owen, S.~E., Moore, A.~W., Samsonov, S.~V., et~al., 2015.
\newblock The iquique earthquake sequence of april 2014: Bayesian modeling accounting for prediction uncertainty, {\it Geophysical Research Letters\/}, {\bf 42}(19), 7949--7957.

\bibitem[Durkan et~al.(2019)Durkan, Bekasov, Murray, \& Papamakarios]{durkan2019neural}
Durkan, C., Bekasov, A., Murray, I., \& Papamakarios, G., 2019.
\newblock Neural spline flows, in {\em Advances in Neural Information Processing Systems\/}, vol.~32, Curran Associates, Inc.

\bibitem[Durkan et~al.(2020{\natexlab{a}})Durkan, Bekasov, Murray, \& Papamakarios]{nflows}
Durkan, C., Bekasov, A., Murray, I., \& Papamakarios, G., 2020{\natexlab{a}}.
\newblock {nflows}: normalizing flows in {PyTorch}, doi: 10.5281/zenodo.4296287.

\bibitem[Durkan et~al.(2020{\natexlab{b}})Durkan, Murray, \& Papamakarios]{durkan2020contrastive}
Durkan, C., Murray, I., \& Papamakarios, G., 2020{\natexlab{b}}.
\newblock On contrastive learning for likelihood-free inference, in {\em Proceedings of the 37th International Conference on Machine Learning\/}, vol. 119 of {\bf Proceedings of Machine Learning Research}, pp. 2771--2781, PMLR.

\bibitem[Dziewonski \& Woodhouse(1983)]{dziewonski1983experiment}
Dziewonski, A.~M. \& Woodhouse, J.~H., 1983.
\newblock An experiment in systematic study of global seismicity: Centroid-moment tensor solutions for 201 moderate and large earthquakes of 1981, {\it Journal of Geophysical Research: Solid Earth\/}, {\bf 88}(B4), 3247--3271.

\bibitem[Engdahl et~al.(1998)Engdahl, van~der Hilst, \& Buland]{engdahl1998global}
Engdahl, E.~R., van~der Hilst, R., \& Buland, R., 1998.
\newblock Global teleseismic earthquake relocation with improved travel times and procedures for depth determination, {\it Bulletin of the Seismological Society of America\/}, {\bf 88}(3), 722--743.

\bibitem[Eulenfeld et~al.(2022)Eulenfeld, Dahm, Heimann, \& Wegler]{eulenfeld2022fast}
Eulenfeld, T., Dahm, T., Heimann, S., \& Wegler, U., 2022.
\newblock Fast and robust earthquake source spectra and moment magnitudes from envelope inversion, {\it Bulletin of the Seismological Society of America\/}, {\bf 112}(2), 878--893.

\bibitem[Falcon(2019)]{lightning}
Falcon, W., 2019.
\newblock Pytorch lightning.

\bibitem[Ferreira et~al.(2011)Ferreira, Weston, \& Funning]{ferreira2011global}
Ferreira, A., Weston, J., \& Funning, G., 2011.
\newblock Global compilation of interferometric synthetic aperture radar earthquake source models: 2. effects of 3-d earth structure, {\it Journal of Geophysical Research: Solid Earth\/}, {\bf 116}(B8).

\bibitem[Ferreira \& Woodhouse(2006)]{ferreira2006long}
Ferreira, A.~M. \& Woodhouse, J.~H., 2006.
\newblock Long-period seismic source inversions using global tomographic models, {\it Geophysical Journal International\/}, {\bf 166}(3), 1178--1192.

\bibitem[Fichtner et~al.(2008)Fichtner, Kennett, Igel, \& Bunge]{fichtner2008theoretical}
Fichtner, A., Kennett, B.~L., Igel, H., \& Bunge, H.-P., 2008.
\newblock Theoretical background for continental-and global-scale full-waveform inversion in the time--frequency domain, {\it Geophysical Journal International\/}, {\bf 175}(2), 665--685.

\bibitem[Ford et~al.(2012)Ford, Walter, \& Dreger]{ford2012event}
Ford, S.~R., Walter, W.~R., \& Dreger, D.~S., 2012.
\newblock Event discrimination using regional moment tensors with teleseismic-p constraints, {\it Bulletin of the Seismological Society of America\/}, {\bf 102}(2), 867--872.

\bibitem[Foreman-Mackey et~al.(2013)Foreman-Mackey, Hogg, Lang, \& Goodman]{foreman2013emcee}
Foreman-Mackey, D., Hogg, D.~W., Lang, D., \& Goodman, J., 2013.
\newblock emcee: the mcmc hammer, {\it Publications of the Astronomical Society of the Pacific\/}, {\bf 125}(925), 306.

\bibitem[Gatti et~al.(2024)Gatti, Jeffrey, Whiteway, Williamson, Jain, Ajani, Anbajagane, Giannini, Zhou, Porredon, et~al.]{gatti2024dark}
Gatti, M., Jeffrey, N., Whiteway, L., Williamson, J., Jain, B., Ajani, V., Anbajagane, D., Giannini, G., Zhou, C., Porredon, A., et~al., 2024.
\newblock Dark energy survey year 3 results: Simulation-based cosmological inference with wavelet harmonics, scattering transforms, and moments of weak lensing mass maps. validation on simulations, {\it Physical Review D\/}, {\bf 109}(6), 063534.

\bibitem[Gerardi et~al.(2024)Gerardi, Cuceu, Joachimi, Nadathur, \& Font-Ribera]{gerardi2024optimal}
Gerardi, F., Cuceu, A., Joachimi, B., Nadathur, S., \& Font-Ribera, A., 2024.
\newblock Optimal data compression for lyman-$\alpha$ forest cosmology, {\it Monthly Notices of the Royal Astronomical Society\/}, {\bf 528}(2), 2667--2678.

\bibitem[Gloeckler et~al.(2024)Gloeckler, Deistler, Weilbach, Wood, \& Macke]{gloeckler2024all}
Gloeckler, M., Deistler, M., Weilbach, C.~D., Wood, F., \& Macke, J.~H., 2024.
\newblock All-in-one simulation-based inference, in {\em International Conference on Machine Learning\/}, pp. 15735--15766, PMLR.

\bibitem[Gon{\c{c}}alves et~al.(2020)Gon{\c{c}}alves, Lueckmann, Deistler, Nonnenmacher, {\"O}cal, Bassetto, Chintaluri, Podlaski, Haddad, Vogels, et~al.]{gonccalves2020training}
Gon{\c{c}}alves, P.~J., Lueckmann, J.-M., Deistler, M., Nonnenmacher, M., {\"O}cal, K., Bassetto, G., Chintaluri, C., Podlaski, W.~F., Haddad, S.~A., Vogels, T.~P., et~al., 2020.
\newblock Training deep neural density estimators to identify mechanistic models of neural dynamics, {\it elife\/}, {\bf 9}, e56261.

\bibitem[Graikos et~al.(2022)Graikos, Malkin, Jojic, \& Samaras]{graikos2022diffusion}
Graikos, A., Malkin, N., Jojic, N., \& Samaras, D., 2022.
\newblock Diffusion models as plug-and-play priors, {\it Advances in Neural Information Processing Systems\/}, {\bf 35}, 14715--14728.

\bibitem[Greenberg et~al.(2019)Greenberg, Nonnenmacher, \& Macke]{greenberg2019automatic}
Greenberg, D.~S., Nonnenmacher, M., \& Macke, J.~H., 2019.
\newblock Automatic posterior transformation for likelihood-free inference.

\bibitem[Hallo \& Gallovi{\v{c}}(2016)]{hallo2016fast}
Hallo, M. \& Gallovi{\v{c}}, F., 2016.
\newblock Fast and cheap approximation of green function uncertainty for waveform-based earthquake source inversions, {\it Geophysical Journal International\/}, {\bf 207}(2), 1012--1029.

\bibitem[Hamidbeygi et~al.(2023)Hamidbeygi, Vasyura-Bathke, Dettmer, Eaton, \& Dosso]{hamidbeygi2023bayesian}
Hamidbeygi, M., Vasyura-Bathke, H., Dettmer, J., Eaton, D.~W., \& Dosso, S.~E., 2023.
\newblock Bayesian estimation of non-linear centroid moment tensors using multiple seismic data sets, {\it Geophysical Journal International\/}, {\bf 235}(3), 2948--2961.

\bibitem[Hardebeck \& Shearer(2002)]{hardebeck2002new}
Hardebeck, J.~L. \& Shearer, P.~M., 2002.
\newblock A new method for determining first-motion focal mechanisms, {\it Bulletin of the Seismological Society of America\/}, {\bf 92}(6), 2264--2276.

\bibitem[Hejrani et~al.(2017)Hejrani, Tkal{\v{c}}i{\'c}, \& Fichtner]{hejrani2017centroid}
Hejrani, B., Tkal{\v{c}}i{\'c}, H., \& Fichtner, A., 2017.
\newblock Centroid moment tensor catalogue using a 3-d continental scale earth model: Application to earthquakes in papua new guinea and the solomon islands, {\it Journal of Geophysical Research: Solid Earth\/}, {\bf 122}(7), 5517--5543.

\bibitem[Herak et~al.(2009)Herak, Herak, \& Tomljenovi{\'c}]{herak2009seismicity}
Herak, D., Herak, M., \& Tomljenovi{\'c}, B., 2009.
\newblock Seismicity and earthquake focal mechanisms in north-western croatia, {\it Tectonophysics\/}, {\bf 465}(1-4), 212--220.

\bibitem[Hermans et~al.(2020)Hermans, Begy, \& Louppe]{hermans20nlre}
Hermans, J., Begy, V., \& Louppe, G., 2020.
\newblock Likelihood-free {MCMC} with amortized approximate ratio estimators, in {\em Proceedings of the 37th International Conference on Machine Learning\/}, vol. 119 of {\bf Proceedings of Machine Learning Research}, pp. 4239--4248, PMLR.

\bibitem[Hermans et~al.(2021)Hermans, Delaunoy, Rozet, Wehenkel, \& Louppe]{hermans2021}
Hermans, J., Delaunoy, A., Rozet, F., Wehenkel, A., \& Louppe, G., 2021.
\newblock Averting a crisis in simulation-based inference, {\it stat\/}, {\bf 1050}, 14.

\bibitem[Herrmann(2013)]{herrmann2013computer}
Herrmann, R.~B., 2013.
\newblock Computer programs in seismology: An evolving tool for instruction and research, {\it Seismological Research Letters\/}, {\bf 84}(6), 1081--1088.

\bibitem[Hingee et~al.(2011)Hingee, Tkal{\v{c}}i{\'c}, Fichtner, \& Sambridge]{hingee2011seismic}
Hingee, M., Tkal{\v{c}}i{\'c}, H., Fichtner, A., \& Sambridge, M., 2011.
\newblock Seismic moment tensor inversion using a 3-d structural model: applications for the australian region, {\it Geophysical Journal International\/}, {\bf 184}(2), 949--964.

\bibitem[Hj{\"o}rleifsd{\'o}ttir \& Ekstr{\"o}m(2010)]{hjorleifsdottir2010effects}
Hj{\"o}rleifsd{\'o}ttir, V. \& Ekstr{\"o}m, G., 2010.
\newblock Effects of three-dimensional earth structure on cmt earthquake parameters, {\it Physics of the Earth and Planetary Interiors\/}, {\bf 179}(3-4), 178--190.

\bibitem[Ho et~al.(2020)Ho, Kalchbrenner, Weissenborn, \& Salimans]{ho2020axial}
Ho, J., Kalchbrenner, N., Weissenborn, D., \& Salimans, T., 2020.
\newblock Axial attention in multidimensional transformers.

\bibitem[Holzschuh \& Thuerey(2024)]{holzschuh2024flow}
Holzschuh, B. \& Thuerey, N., 2024.
\newblock Flow matching for posterior inference with simulator feedback, {\it arXiv preprint arXiv:2410.22573\/}.

\bibitem[Hourcade et~al.(2025)Hourcade, Juhel, \& Bletery]{hourcade2025pegsgraph}
Hourcade, C., Juhel, K., \& Bletery, Q., 2025.
\newblock Pegsgraph: A graph neural network for fast earthquake characterization based on prompt elastogravity signals, {\it Journal of Geophysical Research: Machine Learning and Computation\/}, {\bf 2}(1), e2024JH000360.

\bibitem[Hu et~al.(2023)Hu, Phạm, \& Tkal{\v{c}}i{\'c}]{hu2023seismic}
Hu, J., Phạm, T.-S., \& Tkal{\v{c}}i{\'c}, H., 2023.
\newblock Seismic moment tensor inversion with theory errors from 2-d earth structure: implications for the 2009--2017 dprk nuclear blasts, {\it Geophysical Journal International\/}, {\bf 235}(3), 2035--2054.

\bibitem[Hu et~al.(2025)Hu, Tkalčić, Phạm, Herak, Dasovi{\'c}, \& Br{\v{c}}i{\'c}]{hu2025bayesian}
Hu, J., Tkalčić, H., Phạm, T.-S., Herak, M., Dasovi{\'c}, I., \& Br{\v{c}}i{\'c}, M.~M., 2025.
\newblock Bayesian reassessment of seismic moment tensors and their uncertainties in the adriatic sea region, {\it Seismica\/}, {\bf 4}(2).

\bibitem[Huang \& Zhang(2026)]{huang2026msep}
Huang, X. \& Zhang, Y., 2026.
\newblock Msep-tformer: a multitask source estimation parameter transformer network for earthquake monitoring, {\it Geophysical Journal International\/}, {\bf 244}(1), ggaf453.

\bibitem[Jeffrey et~al.(2021)Jeffrey, Alsing, \& Lanusse]{jeffrey2021likelihood}
Jeffrey, N., Alsing, J., \& Lanusse, F., 2021.
\newblock Likelihood-free inference with neural compression of des sv weak lensing map statistics, {\it Monthly Notices of the Royal Astronomical Society\/}, {\bf 501}(1), 954--969.

\bibitem[Jeffrey et~al.(2025)Jeffrey, Whiteway, Gatti, Williamson, Alsing, Porredon, Prat, Doux, Jain, Chang, et~al.]{jeffrey2025dark}
Jeffrey, N., Whiteway, L., Gatti, M., Williamson, J., Alsing, J., Porredon, A., Prat, J., Doux, C., Jain, B., Chang, C., et~al., 2025.
\newblock Dark energy survey year 3 results: likelihood-free, simulation-based w cdm inference with neural compression of weak-lensing map statistics, {\it Monthly Notices of the Royal Astronomical Society\/}, {\bf 536}(2), 1303--1322.

\bibitem[Kawakatsu(1996)]{kawakatsu1996observability}
Kawakatsu, H., 1996.
\newblock Observability of the isotropic component of a moment tensor, {\it Geophysical Journal International\/}, {\bf 126}(2), 525--544.

\bibitem[Kolb \& Leki{\'c}(2014)]{kolb2014receiver}
Kolb, J. \& Leki{\'c}, V., 2014.
\newblock Receiver function deconvolution using transdimensional hierarchical bayesian inference, {\it Geophysical Journal International\/}, {\bf 197}(3), 1719--1735.

\bibitem[K{\v{r}}{\'\i}{\v{z}}ov{\'a} et~al.(2013)K{\v{r}}{\'\i}{\v{z}}ov{\'a}, Zahradn{\'\i}k, \& Kiratzi]{kvrivzova2013resolvability}
K{\v{r}}{\'\i}{\v{z}}ov{\'a}, D., Zahradn{\'\i}k, J., \& Kiratzi, A., 2013.
\newblock Resolvability of isotropic component in regional seismic moment tensor inversion, {\it Bulletin of the Seismological Society of America\/}, {\bf 103}(4), 2460--2473.

\bibitem[Kubo et~al.(2024)Kubo, Naoi, \& Kano]{kubo2024recent}
Kubo, H., Naoi, M., \& Kano, M., 2024.
\newblock Recent advances in earthquake seismology using machine learning, {\it Earth, Planets and Space\/}, {\bf 76}(1), 36.

\bibitem[Lanzieri et~al.(2025)Lanzieri, Zeghal, Makinen, Boucaud, Starck, \& Lanusse]{lanzieri2025optimal}
Lanzieri, D., Zeghal, J., Makinen, T.~L., Boucaud, A., Starck, J.-L., \& Lanusse, F., 2025.
\newblock Optimal neural summarization for full-field weak lensing cosmological implicit inference, {\it Astronomy \& Astrophysics\/}, {\bf 697}, A162.

\bibitem[Ledoit \& Wolf(2004)]{ledoit2004well}
Ledoit, O. \& Wolf, M., 2004.
\newblock A well-conditioned estimator for large-dimensional covariance matrices, {\it Journal of multivariate analysis\/}, {\bf 88}(2), 365--411.

\bibitem[Lehman et~al.(2025)Lehman, Krippendorf, Weller, \& Dolag]{lehman2025learning}
Lehman, K., Krippendorf, S., Weller, J., \& Dolag, K., 2025.
\newblock Learning optimal summary statistics of galaxy catalogs with sbi, {\it Journal of Cosmology and Astroparticle Physics\/}, {\bf 2025}(12), 032.

\bibitem[Lemos et~al.(2023)Lemos, Coogan, Hezaveh, \& Perreault-Levasseur]{lemos2023sampling}
Lemos, P., Coogan, A., Hezaveh, Y., \& Perreault-Levasseur, L., 2023.
\newblock Sampling-based accuracy testing of posterior estimators for general inference, {\it arXiv preprint arXiv:2302.03026\/}.

\bibitem[Lindner et~al.(2023)Lindner, Rietbrock, Bie, Goes, Collier, Rychert, Harmon, Hicks, Henstock, \& Group]{lindner2023bayesian}
Lindner, M., Rietbrock, A., Bie, L., Goes, S., Collier, J., Rychert, C., Harmon, N., Hicks, S.~P., Henstock, T., \& Group, V.~W., 2023.
\newblock Bayesian regional moment tensor from ocean bottom seismograms recorded in the lesser antilles: Implications for regional stress field, {\it Geophysical Journal International\/}, {\bf 233}(2), 1036--1054.

\bibitem[Lueckmann et~al.(2019)Lueckmann, Bassetto, Karaletsos, \& Macke]{lueckmann2019likelihood}
Lueckmann, J.-M., Bassetto, G., Karaletsos, T., \& Macke, J.~H., 2019.
\newblock Likelihood-free inference with emulator networks, in {\em Symposium on Advances in Approximate Bayesian Inference\/}, pp. 32--53, PMLR.

\bibitem[Lueckmann et~al.(2021)Lueckmann, Boelts, Greenberg, Goncalves, \& Macke]{lueckmann2021benchmarking}
Lueckmann, J.-M., Boelts, J., Greenberg, D., Goncalves, P., \& Macke, J., 2021.
\newblock Benchmarking simulation-based inference, in {\em International conference on artificial intelligence and statistics\/}, pp. 343--351, PMLR.

\bibitem[Makinen et~al.(2024)Makinen, Sui, Wandelt, Porqueres, \& Heavens]{makinen2024hybrid}
Makinen, T.~L., Sui, C., Wandelt, B.~D., Porqueres, N., \& Heavens, A., 2024.
\newblock Hybrid summary statistics, {\it arXiv preprint arXiv:2410.07548\/}.

\bibitem[Marku{\v{s}}i{\'c} et~al.(2020)Marku{\v{s}}i{\'c}, Stanko, Korbar, Beli{\'c}, Penava, \& Kordi{\'c}]{markuvsic2020zagreb}
Marku{\v{s}}i{\'c}, S., Stanko, D., Korbar, T., Beli{\'c}, N., Penava, D., \& Kordi{\'c}, B., 2020.
\newblock The zagreb (croatia) m5. 5 earthquake on 22 march 2020, {\it Geosciences\/}, {\bf 10}(7), 252.

\bibitem[McBrearty \& Beroza(2022)]{mcbrearty2022earthquake}
McBrearty, I.~W. \& Beroza, G.~C., 2022.
\newblock Earthquake location and magnitude estimation with graph neural networks, in {\em 2022 IEEE international conference on image processing (ICIP)\/}, pp. 3858--3862, IEEE.

\bibitem[McBrearty \& Beroza(2023)]{mcbrearty2023earthquake}
McBrearty, I.~W. \& Beroza, G.~C., 2023.
\newblock Earthquake phase association with graph neural networks, {\it Bulletin of the Seismological Society of America\/}, {\bf 113}(2), 524--547.

\bibitem[Minson et~al.(2013)Minson, Simons, \& Beck]{minson2013bayesian}
Minson, S., Simons, M., \& Beck, J., 2013.
\newblock Bayesian inversion for finite fault earthquake source models i—theory and algorithm, {\it Geophysical Journal International\/}, {\bf 194}(3), 1701--1726.

\bibitem[Minson \& Dreger(2008)]{minson2008stable}
Minson, S.~E. \& Dreger, D.~S., 2008.
\newblock Stable inversions for complete moment tensors, {\it Geophysical Journal International\/}, {\bf 174}(2), 585--592.

\bibitem[Minson et~al.(2014)Minson, Simons, Beck, Ortega, Jiang, Owen, Moore, Inbal, \& Sladen]{minson2014bayesian}
Minson, S.~E., Simons, M., Beck, J., Ortega, F., Jiang, J., Owen, S., Moore, A., Inbal, A., \& Sladen, A., 2014.
\newblock Bayesian inversion for finite fault earthquake source models--ii: the 2011 great tohoku-oki, japan earthquake, {\it Geophysical Journal International\/}, {\bf 198}(2), 922--940.

\bibitem[Mousavi \& Beroza(2022)]{mousavi2022deep}
Mousavi, S.~M. \& Beroza, G.~C., 2022.
\newblock Deep-learning seismology, {\it Science\/}, {\bf 377}(6607), eabm4470.

\bibitem[Mousavi et~al.(2020)Mousavi, Ellsworth, Zhu, Chuang, \& Beroza]{mousavi2020earthquake}
Mousavi, S.~M., Ellsworth, W.~L., Zhu, W., Chuang, L.~Y., \& Beroza, G.~C., 2020.
\newblock Earthquake transformer—an attentive deep-learning model for simultaneous earthquake detection and phase picking, {\it Nature communications\/}, {\bf 11}(1), 3952.

\bibitem[M{\"u}nchmeyer et~al.(2021)M{\"u}nchmeyer, Bindi, Leser, \& Tilmann]{munchmeyer2021transformer}
M{\"u}nchmeyer, J., Bindi, D., Leser, U., \& Tilmann, F., 2021.
\newblock The transformer earthquake alerting model: A new versatile approach to earthquake early warning, {\it Geophysical Journal International\/}, {\bf 225}(1), 646--656.

\bibitem[Nakahara(2008)]{nakahara2008seismogram}
Nakahara, H., 2008.
\newblock Seismogram envelope inversion for high-frequency seismic energy radiation from moderate-to-large earthquakes, {\it Advances in Geophysics\/}, {\bf 50}, 401--426.

\bibitem[Novak et~al.(2020)Novak, Uros, Atalic, Herak, Demsic, Banicek, Lazarevic, Bijelic, Crnogorac, \& Todoric]{novak2020zagreb}
Novak, M.~S., Uros, M., Atalic, J., Herak, M., Demsic, M., Banicek, M., Lazarevic, D., Bijelic, N., Crnogorac, M., \& Todoric, M., 2020.
\newblock Zagreb earthquake of 22 march 2020-preliminary report on seismologic aspects and damage to buildings.

\bibitem[Papamakarios \& Murray(2016)]{papamakarios2016fast}
Papamakarios, G. \& Murray, I., 2016.
\newblock Fast $\epsilon$-free inference of simulation models with bayesian conditional density estimation, in {\em Advances in Neural Information Processing Systems\/}, vol.~29, Curran Associates, Inc.

\bibitem[Papamakarios et~al.(2017)Papamakarios, Pavlakou, \& Murray]{papamakarios2017masked}
Papamakarios, G., Pavlakou, T., \& Murray, I., 2017.
\newblock Masked autoregressive flow for density estimation, in {\em Advances in Neural Information Processing Systems\/}, vol.~30, Curran Associates, Inc.

\bibitem[Papamakarios et~al.(2019)Papamakarios, Sterratt, \& Murray]{papamakarios2019sequential}
Papamakarios, G., Sterratt, D., \& Murray, I., 2019.
\newblock Sequential neural likelihood: Fast likelihood-free inference with autoregressive flows, in {\em The 22nd international conference on artificial intelligence and statistics\/}, pp. 837--848, PMLR.

\bibitem[Papamakarios et~al.(2021)Papamakarios, Nalisnick, Rezende, Mohamed, \& Lakshminarayanan]{papamakarios2021normalizing}
Papamakarios, G., Nalisnick, E., Rezende, D.~J., Mohamed, S., \& Lakshminarayanan, B., 2021.
\newblock Normalizing flows for probabilistic modeling and inference, {\it J. Mach. Learn. Res.\/}, {\bf 22}(1).

\bibitem[Park et~al.(2025)Park, Gatti, \& Jain]{park2025dimensionality}
Park, M., Gatti, M., \& Jain, B., 2025.
\newblock Dimensionality reduction techniques for statistical inference in cosmology, {\it Physical Review D\/}, {\bf 111}(6), 063523.

\bibitem[Paszke et~al.(2019)Paszke, Gross, Massa, Lerer, Bradbury, Chanan, Killeen, Lin, Gimelshein, Antiga, et~al.]{paszke2019pytorch}
Paszke, A., Gross, S., Massa, F., Lerer, A., Bradbury, J., Chanan, G., Killeen, T., Lin, Z., Gimelshein, N., Antiga, L., et~al., 2019.
\newblock Pytorch: An imperative style, high-performance deep learning library, {\it Advances in neural information processing systems\/}, {\bf 32}.

\bibitem[Petersen et~al.(2021)Petersen, Cesca, Heimann, Niemz, Dahm, K{\"u}hn, Kummerow, \& Plenefisch]{petersen2021regional}
Petersen, G.~M., Cesca, S., Heimann, S., Niemz, P., Dahm, T., K{\"u}hn, D., Kummerow, J., \& Plenefisch, T., 2021.
\newblock Regional centroid moment tensor inversion of small to moderate earthquakes in the alps using the dense alparray seismic network: challenges and seismotectonic insights, {\it Solid earth\/}, {\bf 12}(6), 1233--1257.

\bibitem[Phạm \& Tkal{\v{c}}i{\'c}(2021)]{phạm2021toward}
Phạm, T.-S. \& Tkal{\v{c}}i{\'c}, H., 2021.
\newblock Toward improving point-source moment-tensor inference by incorporating 1d earth model's uncertainty: Implications for the long valley caldera earthquakes, {\it Journal of Geophysical Research: Solid Earth\/}, {\bf 126}(11), e2021JB022477.

\bibitem[Phạm et~al.(2024)Phạm, Tkal{\v{c}}i{\'c}, Hu, \& Kim]{phạm2024towards}
Phạm, T.-S., Tkal{\v{c}}i{\'c}, H., Hu, J., \& Kim, S., 2024.
\newblock Towards a new standard for seismic moment tensor inversion containing 3-d earth structure uncertainty, {\it Geophysical Journal International\/}, {\bf 238}(3), 1840--1853.

\bibitem[Phạm et~al.(2025)Phạm, Tkal{\v{c}}i{\'c}, Hu, \& Wei]{pham2025global}
Phạm, T.-S., Tkal{\v{c}}i{\'c}, H., Hu, J., \& Wei, Z., 2025.
\newblock On the global centroid moment tensor achievements and the next generation earthquake catalogs, {\it Physics of the Earth and Planetary Interiors\/}, p. 107490.

\bibitem[Pierre et~al.(2026)Pierre, R{\'e}galdo-Saint~Blancard, Hahn, \& Eickenberg]{pierre2026mitigating}
Pierre, S., R{\'e}galdo-Saint~Blancard, B., Hahn, C., \& Eickenberg, M., 2026.
\newblock Mitigating model misspecification in simulation-based inference for galaxy clustering, {\it Physical Review D\/}, {\bf 113}(4), 043536.

\bibitem[Poppeliers \& Preston(2021)]{poppeliers2021effects}
Poppeliers, C. \& Preston, L., 2021.
\newblock The effects of earth model uncertainty on the inversion of seismic data for seismic source functions, {\it Geophysical Journal International\/}, {\bf 224}(1), 100--120.

\bibitem[Poppeliers \& Preston(2022)]{poppeliers2022efficient}
Poppeliers, C. \& Preston, L., 2022.
\newblock An efficient method to propagate model uncertainty when inverting seismic data for time domain seismic moment tensors, {\it Geophysical Journal International\/}, {\bf 231}(2), 1221--1232.

\bibitem[Prelogovi{\'c} \& Mesinger(2024)]{prelogovic2024informative}
Prelogovi{\'c}, D. \& Mesinger, A., 2024.
\newblock How informative are summaries of the cosmic 21 cm signal?, {\it Astronomy \& Astrophysics\/}, {\bf 688}, A199.

\bibitem[R{\"o}sler et~al.(2021)R{\"o}sler, Stein, \& Spencer]{rosler2021uncertainties}
R{\"o}sler, B., Stein, S., \& Spencer, B.~D., 2021.
\newblock Uncertainties in seismic moment tensors inferred from differences between global catalogs, {\it Seismological Research Letters\/}, {\bf 92}(6), 3698--3711.

\bibitem[R{\"o}sler et~al.(2024{\natexlab{a}})R{\"o}sler, Spencer, \& Stein]{rosler2024global}
R{\"o}sler, B., Spencer, B.~D., \& Stein, S., 2024{\natexlab{a}}.
\newblock Which global moment tensor catalog provides the most precise non-double-couple components?, {\it Seismological Research Letters\/}, {\bf 95}(4), 2444--2451.

\bibitem[R{\"o}sler et~al.(2024{\natexlab{b}})R{\"o}sler, Stein, Ringler, \& Vack{\'a}{\v{r}}]{rosler2024apparent}
R{\"o}sler, B., Stein, S., Ringler, A., \& Vack{\'a}{\v{r}}, J., 2024{\natexlab{b}}.
\newblock Apparent non-double-couple components as artifacts of moment tensor inversion, {\it Seismica\/}, {\bf 3}(1).

\bibitem[Saoulis et~al.(2025)Saoulis, Piras, Spurio~Mancini, Joachimi, \& Ferreira]{saoulis2025full}
Saoulis, A., Piras, D., Spurio~Mancini, A., Joachimi, B., \& Ferreira, A., 2025.
\newblock Full-waveform earthquake source inversion using simulation-based inference, {\it Geophysical Journal International\/}, {\bf 241}(3), 1740--1761.

\bibitem[Sawade et~al.(2022)Sawade, Beller, Lei, \& Tromp]{sawade2022global}
Sawade, L., Beller, S., Lei, W., \& Tromp, J., 2022.
\newblock Global centroid moment tensor solutions in a heterogeneous earth: The cmt3d catalogue, {\it Geophysical Journal International\/}, {\bf 231}(3), 1727--1738.

\bibitem[Schmitt et~al.(2023)Schmitt, B{\"u}rkner, K{\"o}the, \& Radev]{schmitt2023detecting}
Schmitt, M., B{\"u}rkner, P.-C., K{\"o}the, U., \& Radev, S.~T., 2023.
\newblock Detecting model misspecification in amortized bayesian inference with neural networks, in {\em Dagm german conference on pattern recognition\/}, pp. 541--557, Springer.

\bibitem[Si et~al.(2024{\natexlab{a}})Si, Wu, Li, Wang, \& Zhu]{si2024all}
Si, X., Wu, X., Li, Z., Wang, S., \& Zhu, J., 2024{\natexlab{a}}.
\newblock An all-in-one seismic phase picking, location, and association network for multi-task multi-station earthquake monitoring, {\it Communications Earth \& Environment\/}, {\bf 5}(1), 22.

\bibitem[Si et~al.(2024{\natexlab{b}})Si, Wu, Sheng, Zhu, \& Li]{si2024seisclip}
Si, X., Wu, X., Sheng, H., Zhu, J., \& Li, Z., 2024{\natexlab{b}}.
\newblock Seisclip: A seismology foundation model pre-trained by multimodal data for multipurpose seismic feature extraction, {\it IEEE Transactions on Geoscience and Remote Sensing\/}, {\bf 62}, 1--13.

\bibitem[Simut{\.e} et~al.(2023)Simut{\.e}, Boehm, Krischer, Gokhberg, Vall{\'e}e, \& Fichtner]{simute2023bayesian}
Simut{\.e}, S., Boehm, C., Krischer, L., Gokhberg, A., Vall{\'e}e, M., \& Fichtner, A., 2023.
\newblock Bayesian seismic source inversion with a 3-d earth model of the japanese islands, {\it Journal of Geophysical Research: Solid Earth\/}, {\bf 128}(1), e2022JB024231.

\bibitem[Skoumal et~al.(2024)Skoumal, Hardebeck, \& Shearer]{skoumal2024skhash}
Skoumal, R.~J., Hardebeck, J.~L., \& Shearer, P.~M., 2024.
\newblock Skhash: A python package for computing earthquake focal mechanisms, {\it Seismological Research Letters\/}, {\bf 95}(4), 2519--2526.

\bibitem[Sokos \& Zahradn{\'\i}k(2013)]{sokos2013evaluating}
Sokos, E. \& Zahradn{\'\i}k, J., 2013.
\newblock Evaluating centroid-moment-tensor uncertainty in the new version of isola software, {\it Seismological Research Letters\/}, {\bf 84}(4), 656--665.

\bibitem[Song et~al.(2022)Song, Kim, Rhie, \& Park]{song2022moment}
Song, J.-H., Kim, S., Rhie, J., \& Park, D., 2022.
\newblock Moment tensor solutions for earthquakes in the southern korean peninsula using three-dimensional seismic waveform simulations, {\it Frontiers in Earth Science\/}, {\bf 10}, 945022.

\bibitem[Song et~al.(2025)Song, Men-Andrin, Ellsworth, \& Beroza]{song2025foconet}
Song, X., Men-Andrin, M., Ellsworth, W.~L., \& Beroza, G.~C., 2025.
\newblock Foconet: transformer-based focal-mechanism determination, {\it Authorea Preprints\/}.

\bibitem[Stachnik et~al.(2012)Stachnik, Sheehan, Zietlow, Yang, Collins, \& Ferris]{stachnik2012determination}
Stachnik, J., Sheehan, A.~F., Zietlow, D., Yang, Z., Collins, J., \& Ferris, A., 2012.
\newblock Determination of new zealand ocean bottom seismometer orientation via rayleigh-wave polarization, {\it Seismological Research Letters\/}, {\bf 83}(4), 704--713.

\bibitem[St{\"a}hler \& Sigloch(2014)]{stahler2014fully}
St{\"a}hler, S.~C. \& Sigloch, K., 2014.
\newblock Fully probabilistic seismic source inversion--part 1: Efficient parameterisation, {\it Solid Earth\/}, {\bf 5}(2), 1055--1069.

\bibitem[St{\"a}hler \& Sigloch(2016)]{stahler2016fully}
St{\"a}hler, S.~C. \& Sigloch, K., 2016.
\newblock Fully probabilistic seismic source inversion--part 2: Modelling errors and station covariances, {\it Solid Earth\/}, {\bf 7}(6), 1521--1536.

\bibitem[Stip{\v{c}}evi{\'c} et~al.(2011)Stip{\v{c}}evi{\'c}, Tkal{\v{c}}i{\'c}, Herak, Marku{\v{s}}i{\'c}, \& Herak]{stipvcevic2011crustal}
Stip{\v{c}}evi{\'c}, J., Tkal{\v{c}}i{\'c}, H., Herak, M., Marku{\v{s}}i{\'c}, S., \& Herak, D., 2011.
\newblock Crustal and uppermost mantle structure beneath the external dinarides, croatia, determined from teleseismic receiver functions, {\it Geophysical journal international\/}, {\bf 185}(3), 1103--1119.

\bibitem[Stip{\v{c}}evi{\'c} et~al.(2020)Stip{\v{c}}evi{\'c}, Herak, Molinari, Dasovi{\'c}, Tkal{\v{c}}i{\'c}, \& Gosar]{stipvcevic2020crustal}
Stip{\v{c}}evi{\'c}, J., Herak, M., Molinari, I., Dasovi{\'c}, I., Tkal{\v{c}}i{\'c}, H., \& Gosar, A., 2020.
\newblock Crustal thickness beneath the dinarides and surrounding areas from receiver functions, {\it Tectonics\/}, {\bf 39}(3), e2019TC005872.

\bibitem[Stockman et~al.(2024)Stockman, Lawson, \& Werner]{stockman2024sb}
Stockman, S., Lawson, D.~J., \& Werner, M.~J., 2024.
\newblock Sb-etas: using simulation based inference for scalable, likelihood-free inference for the etas model of earthquake occurrences, {\it Statistics and Computing\/}, {\bf 34}(5), 174.

\bibitem[Sun et~al.(2023)Sun, Ross, Zhu, \& Azizzadenesheli]{sun2023phase}
Sun, H., Ross, Z.~E., Zhu, W., \& Azizzadenesheli, K., 2023.
\newblock Phase neural operator for multi-station picking of seismic arrivals, {\it Geophysical Research Letters\/}, {\bf 50}(24), e2023GL106434.

\bibitem[Takemura et~al.(2020)Takemura, Okuwaki, Kubota, Shiomi, Kimura, \& Noda]{takemura2020centroid}
Takemura, S., Okuwaki, R., Kubota, T., Shiomi, K., Kimura, T., \& Noda, A., 2020.
\newblock Centroid moment tensor inversions of offshore earthquakes using a three-dimensional velocity structure model: slip distributions on the plate boundary along the nankai trough, {\it Geophysical Journal International\/}, {\bf 222}(2), 1109--1125.

\bibitem[Talts et~al.(2018)Talts, Betancourt, Simpson, Vehtari, \& Gelman]{talts2018validating}
Talts, S., Betancourt, M., Simpson, D., Vehtari, A., \& Gelman, A., 2018.
\newblock Validating bayesian inference algorithms with simulation-based calibration, {\it arXiv preprint arXiv:1804.06788\/}.

\bibitem[Tape \& Tape(2015)]{tape2015uniform}
Tape, W. \& Tape, C., 2015.
\newblock A uniform parametrization of moment tensors, {\it Geophysical Journal International\/}, {\bf 202}(3), 2074--2081.

\bibitem[Tarantola(2005)]{tarantola2005inverse}
Tarantola, A., 2005.
\newblock {\it Inverse problem theory and methods for model parameter estimation\/}, SIAM.

\bibitem[Tarantola et~al.(1982)Tarantola, Valette, et~al.]{tarantola1982inverse}
Tarantola, A., Valette, B., et~al., 1982.
\newblock Inverse problems= quest for information, {\it Journal of geophysics\/}, {\bf 50}(1), 159--170.

\bibitem[Tejero-Cantero et~al.(2020)Tejero-Cantero, Boelts, Deistler, Lueckmann, Durkan, Gon{\c{c}}alves, Greenberg, \& Macke]{tejero2020sbi}
Tejero-Cantero, A., Boelts, J., Deistler, M., Lueckmann, J.-M., Durkan, C., Gon{\c{c}}alves, P., Greenberg, D., \& Macke, J., 2020.
\newblock sbi: A toolkit for simulation-based inference, {\it Journal of Open Source Software\/}, {\bf 5}(52), 2505.

\bibitem[Thuerey et~al.(2021)Thuerey, Holl, Mueller, Schnell, Trost, \& Um]{thuerey2021physics}
Thuerey, N., Holl, P., Mueller, M., Schnell, P., Trost, F., \& Um, K., 2021.
\newblock Physics-based deep learning, {\it arXiv preprint arXiv:2109.05237\/}.

\bibitem[Thurin et~al.(2025)Thurin, Modrak, Tape, McPherson, Rodr{\'\i}guez-Cardozo, Kintner, Ding, Liu, \& Braunmiller]{thurin2025mtuq}
Thurin, J., Modrak, R., Tape, C., McPherson, A., Rodr{\'\i}guez-Cardozo, F., Kintner, J., Ding, L., Liu, Q., \& Braunmiller, J., 2025.
\newblock Mtuq: a framework for estimating moment tensors, point forces, and their uncertainties, {\it Geophysical Journal International\/}, {\bf 241}(2), 1373--1390.

\bibitem[Trabattoni et~al.(2020)Trabattoni, Barruol, Dreo, Boudraa, \& Fontaine]{trabattoni2020orienting}
Trabattoni, A., Barruol, G., Dreo, R., Boudraa, A., \& Fontaine, F., 2020.
\newblock Orienting and locating ocean-bottom seismometers from ship noise analysis, {\it Geophysical Journal International\/}, {\bf 220}(3), 1774--1790.

\bibitem[Vack{\'a}{\v{r}} et~al.(2017)Vack{\'a}{\v{r}}, Burj{\'a}nek, Gallovi{\v{c}}, Zahradn{\'\i}k, \& Clinton]{vackavr2017bayesian}
Vack{\'a}{\v{r}}, J., Burj{\'a}nek, J., Gallovi{\v{c}}, F., Zahradn{\'\i}k, J., \& Clinton, J., 2017.
\newblock Bayesian isola: new tool for automated centroid moment tensor inversion, {\it Geophysical Journal International\/}, {\bf 210}(2), 693--705.

\bibitem[Valentine \& Trampert(2012)]{valentine2012assessing}
Valentine, A.~P. \& Trampert, J., 2012.
\newblock Assessing the uncertainties on seismic source parameters: Towards realistic error estimates for centroid-moment-tensor determinations, {\it Physics of the Earth and Planetary Interiors\/}, {\bf 210}, 36--49.

\bibitem[Vall{\'e}e et~al.(2011)Vall{\'e}e, Charl{\'e}ty, Ferreira, Delouis, \& Vergoz]{vallee2011scardec}
Vall{\'e}e, M., Charl{\'e}ty, J., Ferreira, A.~M., Delouis, B., \& Vergoz, J., 2011.
\newblock Scardec: a new technique for the rapid determination of seismic moment magnitude, focal mechanism and source time functions for large earthquakes using body-wave deconvolution, {\it Geophysical Journal International\/}, {\bf 184}(1), 338--358.

\bibitem[Van~Amersfoort et~al.(2020)Van~Amersfoort, Smith, Teh, \& Gal]{van2020uncertainty}
Van~Amersfoort, J., Smith, L., Teh, Y.~W., \& Gal, Y., 2020.
\newblock Uncertainty estimation using a single deep deterministic neural network, in {\em International conference on machine learning\/}, pp. 9690--9700, PMLR.

\bibitem[Vasyura-Bathke et~al.(2021)Vasyura-Bathke, Dettmer, Dutta, Mai, \& Jonsson]{vasyura2021accounting}
Vasyura-Bathke, H., Dettmer, J., Dutta, R., Mai, P.~M., \& Jonsson, S., 2021.
\newblock Accounting for theory errors with empirical bayesian noise models in nonlinear centroid moment tensor estimation, {\it Geophysical Journal International\/}, {\bf 225}(2), 1412--1431.

\bibitem[Vavry{\v{c}}uk \& K{\"u}hn(2012)]{vavryvcuk2012moment}
Vavry{\v{c}}uk, V. \& K{\"u}hn, D., 2012.
\newblock Moment tensor inversion of waveforms: a two-step time-frequency approach, {\it Geophysical Journal International\/}, {\bf 190}(3), 1761--1776.

\bibitem[von Wietersheim-Kramsta et~al.(2025)von Wietersheim-Kramsta, Lin, Tessore, Joachimi, Loureiro, Reischke, \& Wright]{von2025kids}
von Wietersheim-Kramsta, M., Lin, K., Tessore, N., Joachimi, B., Loureiro, A., Reischke, R., \& Wright, A.~H., 2025.
\newblock Kids-sbi: Simulation-based inference analysis of kids-1000 cosmic shear, {\it Astronomy \& Astrophysics\/}, {\bf 694}, A223.

\bibitem[W{\'e}ber(2006)]{weber2006probabilistic}
W{\'e}ber, Z., 2006.
\newblock Probabilistic local waveform inversion for moment tensor and hypocentral location, {\it Geophysical Journal International\/}, {\bf 165}(2), 607--621.

\bibitem[Weston et~al.(2011)Weston, Ferreira, \& Funning]{weston2011global}
Weston, J., Ferreira, A., \& Funning, G., 2011.
\newblock Global compilation of interferometric synthetic aperture radar earthquake source models: 1. comparisons with seismic catalogs, {\it Journal of Geophysical Research: Solid Earth\/}, {\bf 116}(B8).

\bibitem[Yagi \& Fukahata(2011)]{yagi2011introduction}
Yagi, Y. \& Fukahata, Y., 2011.
\newblock Introduction of uncertainty of green's function into waveform inversion for seismic source processes, {\it Geophysical Journal International\/}, {\bf 186}(2), 711--720.

\bibitem[Zahradn{\'\i}k \& Sokos(2025)]{zahradnik2025isola2024}
Zahradn{\'\i}k, J. \& Sokos, E., 2025.
\newblock Isola2024: Assessing and understanding uncertainties of full moment tensors, {\it Seismological Research Letters\/}, {\bf 96}(4), 2647--2659.

\bibitem[Zammit-Mangion et~al.(2025)Zammit-Mangion, Sainsbury-Dale, \& Huser]{zammit2025neural}
Zammit-Mangion, A., Sainsbury-Dale, M., \& Huser, R., 2025.
\newblock Neural methods for amortized inference, {\it Annual Review of Statistics and Its Application\/}, {\bf 12}(1), 311--335.

\bibitem[Zhu \& Beroza(2019)]{zhu2019phasenet}
Zhu, W. \& Beroza, G.~C., 2019.
\newblock Phasenet: a deep-neural-network-based seismic arrival-time picking method, {\it Geophysical Journal International\/}, {\bf 216}(1), 261--273.

\bibitem[Zhu et~al.(2019)Zhu, Mousavi, \& Beroza]{zhu2019seismic}
Zhu, W., Mousavi, S.~M., \& Beroza, G.~C., 2019.
\newblock Seismic signal denoising and decomposition using deep neural networks, {\it IEEE Transactions on Geoscience and Remote Sensing\/}, {\bf 57}(11), 9476--9488.

\end{thebibliography}

\label{lastpage}

\section*{Supporting Information}

The Supporting Information for this study contains additional figures and explanations that provide further details on the methodology, results, and discussions presented in the main manuscript. Below is a brief description of each figure in the Supporting Information:

\refstepcounter{supsec}
\noindent
\textbf{Section S1:} A discussion of higher order terms in the optimal score compression scheme, which we neglect in our approach. \label{SI:sec_1}

\refstepcounter{supfig}
\noindent
\textbf{Figure S1:} Example of posterior inference and posterior predictive checks for an example shorter period $6-50$ s synthetic inversion. \label{SI:fig_2}

\refstepcounter{supfig}
\noindent
\textbf{Figure S2:} Diagram of the artificial balanced receiver configuration for one of the synthetic experiments. \label{SI:fig_1}

\refstepcounter{supfig}
\noindent
\textbf{Figure S3:} Calibration performance for the score compression SBI approaches for the various synthetic experiments addressed in \Cref{sec:synthetic_inversions}. \label{SI:fig_3}

\refstepcounter{supfig}
\noindent
\textbf{Figure S4:} Same as in Fig. S\ref{SI:fig_3}, but for the ML-based compression SBI approach. Also probes the effect of ML-based compression calibration under model misspecification. \label{SI:fig_4}

\refstepcounter{supfig}
\noindent
\textbf{Figure S5:} An analysis of the full distribution of per-parameter posterior accuracies for the 300 calibration test inversions with $\kappa=3\%$ for each of the three approaches.\label{SI:fig_5}

\refstepcounter{supfig}
\noindent
\textbf{Figure S6:} Same as  Fig. S\ref{SI:fig_5} but for $\kappa=5\%$. \label{SI:fig_6}

\refstepcounter{supfig}
\noindent
\textbf{Figure S7:}  Same as  Fig. S\ref{SI:fig_5} but for $\kappa=5\%$ and shorter period data between $6 - 50$ s. \label{SI:fig_7}

\refstepcounter{suptab}
\noindent
\textbf{Table S1:} The mean posterior standard deviation and bias per-parameter for each approach, as in \Cref{table:uncertainties_and_bias}, for $\kappa=3\%$. \label{SI:table_1}

\refstepcounter{suptab}
\noindent
\textbf{Table S2:} The mean posterior standard deviation and bias per-parameter for each approach, as in \Cref{table:uncertainties_and_bias}, for $\kappa=5\%$ and shorter period data between $6 - 50$ s. \label{SI:table_2}

\refstepcounter{supfig}
\noindent
\textbf{Figure S8:} The 2-D composite model for the Croatia event adapted from \citet{hu2025bayesian}, as described in the main text. \label{SI:fig_8}

\refstepcounter{suptab}
\noindent
\textbf{Table S3:} Results from a quantitative analysis of the posterior predictive checks for each of the three appraoches for the LV2 event. \label{SI:table_3}
\refstepcounter{suptab}
\noindent
\textbf{Table S4:} Same as Table\ref{SI:table_3} for the Croatia event. \label{SI:table_4}

\noindent
The full Supporting Information file is available online, linked with this article.

\end{document}


\maketitle

\section{Optimal score compression}
In the main text, we quoted the following result for optimal score compression:
\begin{linenomath*}
\begin{equation}
\mathbf{t} = \mathbf{m}_* + \mathbf{F}^{-1}_* \mathbf{G}^T_{\mathbf{m_*}} \mathbf{C}^{-1} (\mathbf{D} - \boldsymbol{\mu}_*), 
\label{eq:optimal_score_mle}
\end{equation}
\end{linenomath*}
where the compressed statistic $\mathbf{t}$ can be computed from a given observation $\mathbf{D}$ using the local sensitivity kernels $\mathbf{G}_{\mathbf{m_*}}$ and covariance estimate $\mathbf{C}_{\mathbf{m_*}}$ (note that $\mathbf{F}_* = \mathbf{G}^T_{\mathbf{m_*}} \mathbf{C}_{\mathbf{m_*}}^{-1}\mathbf{G}_{\mathbf{m_*}}$). However, if the covariance matrix $\mathbf{C}$ depends on model parameters $\mathbf{m}$, as it does for theory errors, the full result from \citet{Alsing2017} includes extra terms that rely on the gradients of the covariance $\nabla_\mathbf{m}\mathbf{C}$:
\begin{linenomath*}
\begin{equation}
\mathbf{t}
=
\mathbf{m}_*
+
\mathbf{F}_*^{-1}
\Bigg[
{\mathbf{G}_*}^{T}
\mathbf{C}_*^{-1}
(\mathbf{D} - \boldsymbol{\mu}_*)
+
\frac{1}{2}
(\mathbf{D} - \boldsymbol{\mu}_*)^{T}
\mathbf{C}_*^{-1}
\left(
\nabla \mathbf{C}_*
\right)
\mathbf{C}_*^{-1}
(\mathbf{D} - \boldsymbol{\mu}_*)
-
\frac{1}{2}
\operatorname{tr}
\!\left(
\mathbf{C}_*^{-1}
\nabla \mathbf{C}_*
\right)
\Bigg].
\label{eq:optimal_score_full}
\end{equation}
\end{linenomath*}
The first two terms correspond to \Cref{eq:optimal_score_mle}, while the latter two terms depend on non-zero gradients of the covariance $\nabla \mathbf{C}_*$. This correction is quadratic in the data and is therefore a higher-order correction term. In preparation for this work, we implemented and explored the results using the complete result for an optimal statistic $\mathbf{t}$. However, we found that this gave very poor performance, with the extra two terms completely dominating the MLE estimate $\mathbf{t}$. This was driven by very large gradient terms $\nabla_\mathbf{m} \mathbf{C}_*$; i.e. the theory covariance is very sensitive to perturbations to the moment tensor parameters. 

We therefore hypothesise that the strong model parameter dependency of the theory covariance $\mathbf{C}_t(\mathbf{m})$ causes the simple, linearised relationship assumed in the Taylor expansion to break down. For this reason, throughout this work we adopt the simplified optimal score statistic given by \Cref{eq:optimal_score_mle}, and neglect the covariance-gradient terms in \Cref{eq:optimal_score_full}. In practice, we find that this approximation yields significantly more stable behaviour while retaining the desired sensitivity to variations in the model parameters.

\section{Synthetic experiments}

\subsection{Short period inversion example}

\begin{figure*}
 \centering
    \includegraphics[width=\textwidth]{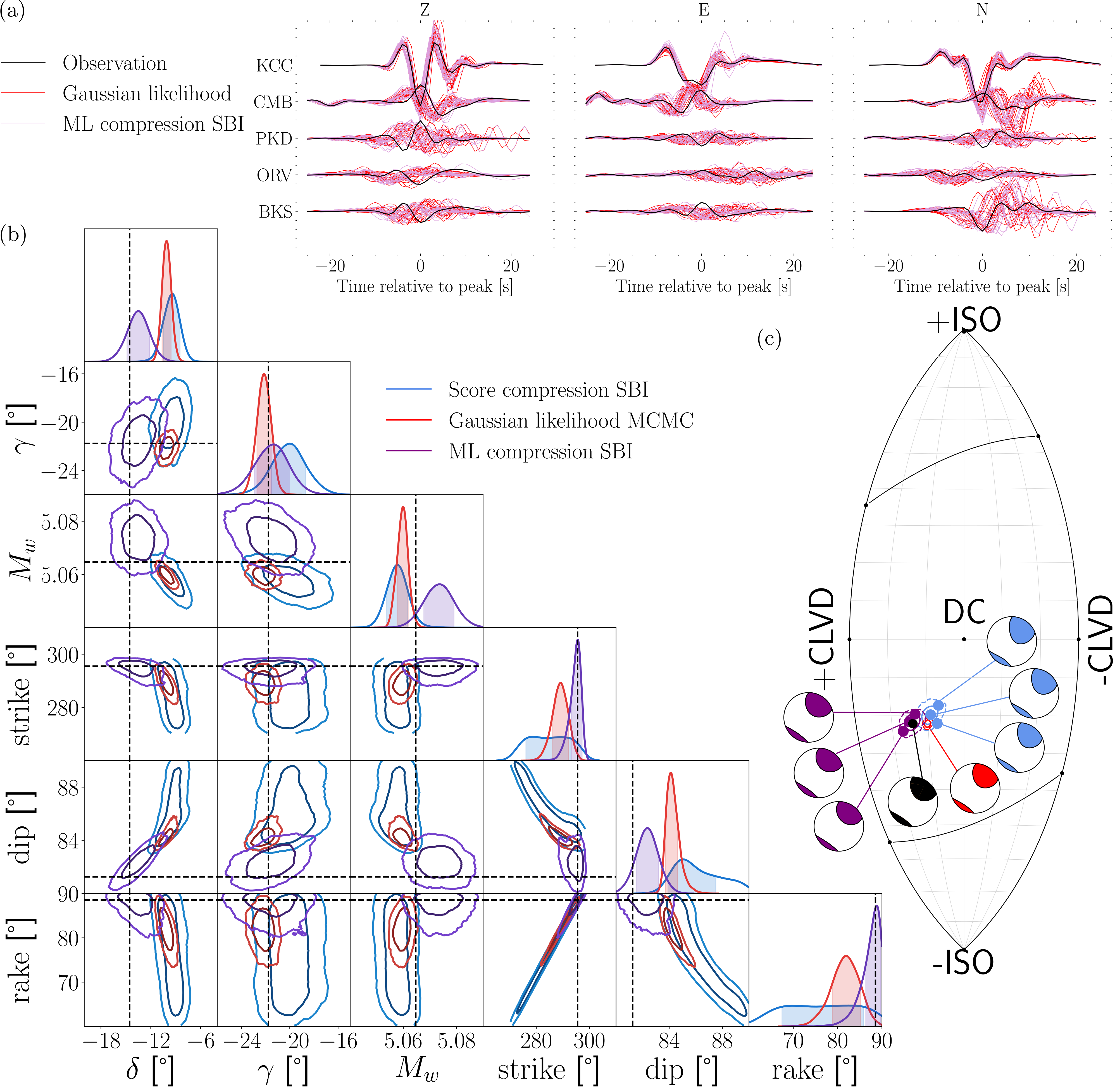}
   \caption{Full posterior inference on an artificial short period inversion comparing the results using a Gaussian likelihood (red) against the two SBI frameworks: score compression with shallow density estimation (blue) and deep learning based compression and density estimation (purple). Panel (a) shows a zoom in on the observation and posterior predictive checks for the two SBI approaches.  The focal mechanism parameters $\gamma, \delta, M_W$ and best fitting double-couple plane orientation posterior are shown in (b), as well as the location on the lune plot in (c) (the top left pair plot in (b) shows the same lune coordinates). The $\pm[1,2]\sigma$ contours are shown for each posterior, and the true artificial solution is marked by the black dashed line and black beachball in (c). }
   \label{fig:short_period_inversion}
\end{figure*}

In \Cref{fig:short_period_inversion} we present an example of the three inversion approaches for a shorter period filtering band between $6-50$ s, which we analysed in the main text. This illustrates the issues with the likelihood-based approaches to inversion, which can occasionally lead to biased moment tensor solutions and underestimation of uncertainty. 

The degree of bias shown in \Cref{fig:short_period_inversion} is not uncommon for the Gaussian likelihood approach; the $z$-score reaches $z>3$ for several parameters. Shown systematically for the 300 calibration test inversions in \Cref{fig:per_parameter_biases_short}, this occurs roughly $10-30\%$ of the time, depending on the parameter.

\subsection{Balanced network configuration}

\begin{figure*}
 \centering
    \includegraphics[width=0.7\textwidth]{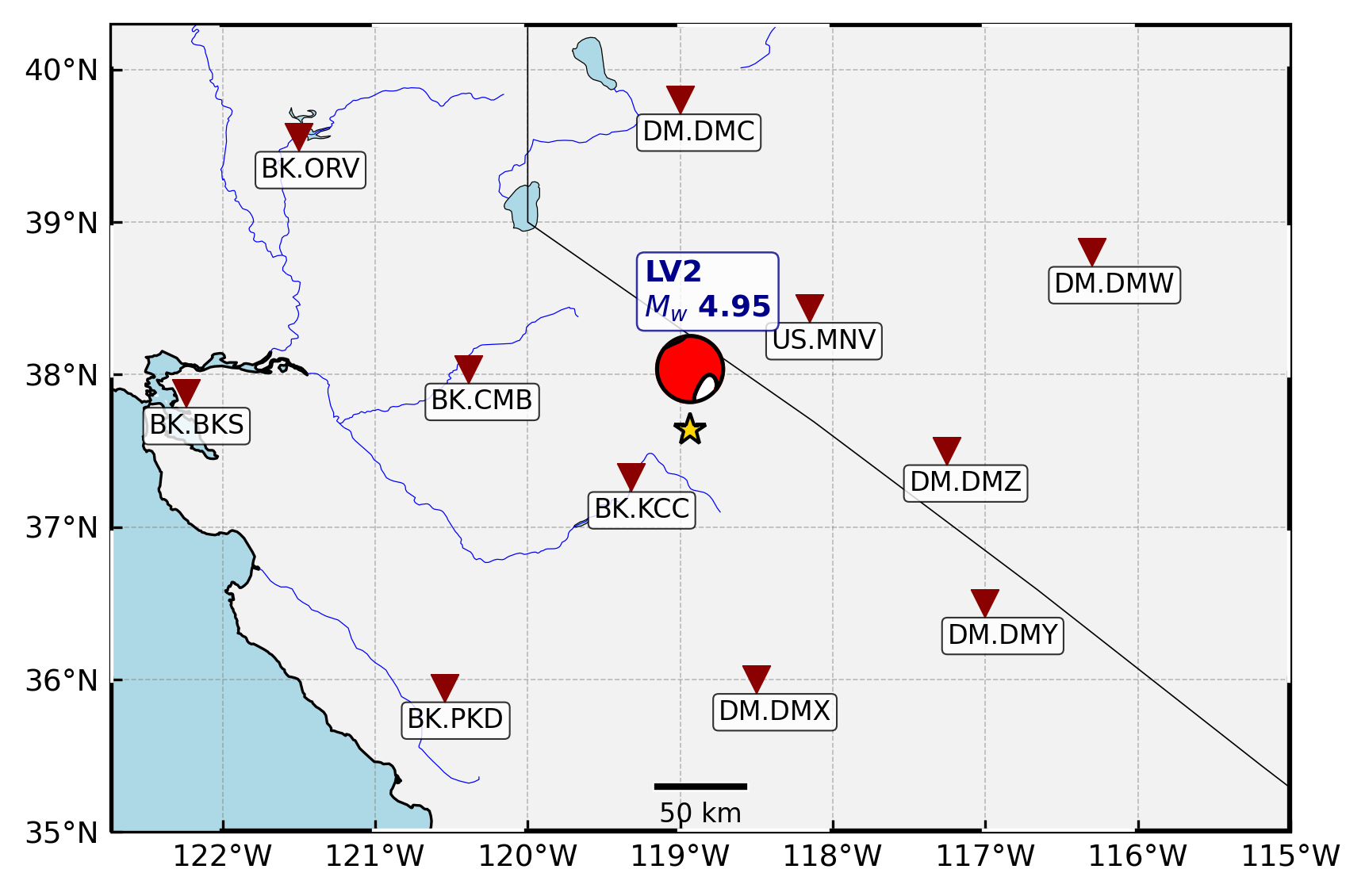}
   \caption{The (artificial) balanced network configuration for one of the synthetic tests in the main text. Stations US.MNV and dummy receivers with network code DM were added to this example, and were not used for the rest of the main text.}
   \label{fig:balanced_receivers}
\end{figure*}

In the main text, we explored how each inversion approach performed with a more azimuthally balanced network configuration. The network is shown in \Cref{fig:balanced_receivers}. The results in the main text demonstrated that even with very good station coverage, Gaussian likelihood assumptions still yield uncalibrated, and therefore misleading, posterior uncertainties. 

\subsection{Further calibration tests of the SBI approaches}

\begin{figure*}
 \centering
    \includegraphics[width=\textwidth]{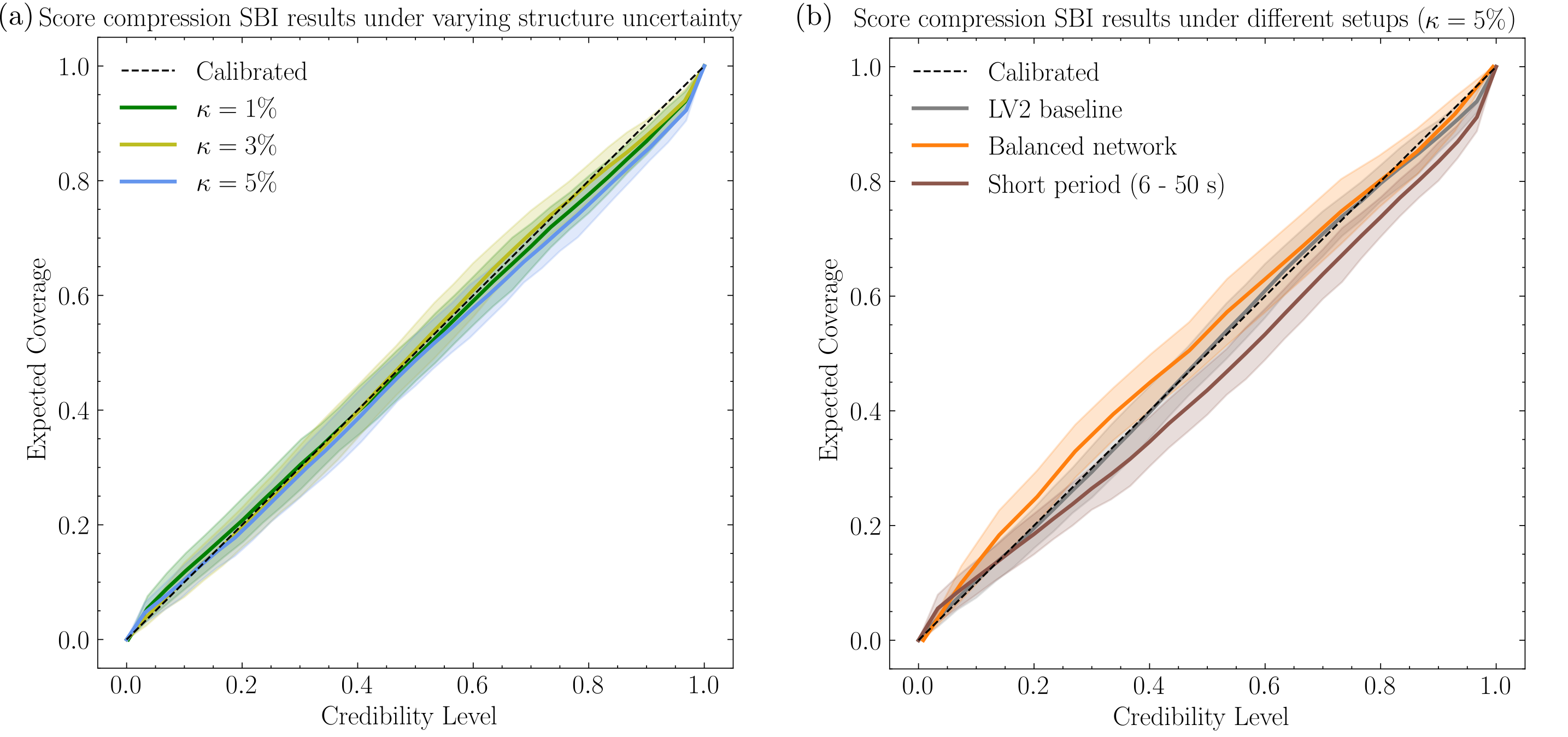}
   \caption{The calibration performance of the SBI with optimal score compression under varied experimental configurations. Panel (a) shows the results under varying levels of Earth structure uncertainty, parametrised by $\kappa$. Panel (b) explores two different experimental configurations: one with a balanced network with full azimuthal coverage, and another that includes shorter periods in the inversion. }
   \label{fig:SBI_calibration_curves}
\end{figure*}

\begin{figure*}
 \centering
    \includegraphics[width=\textwidth]{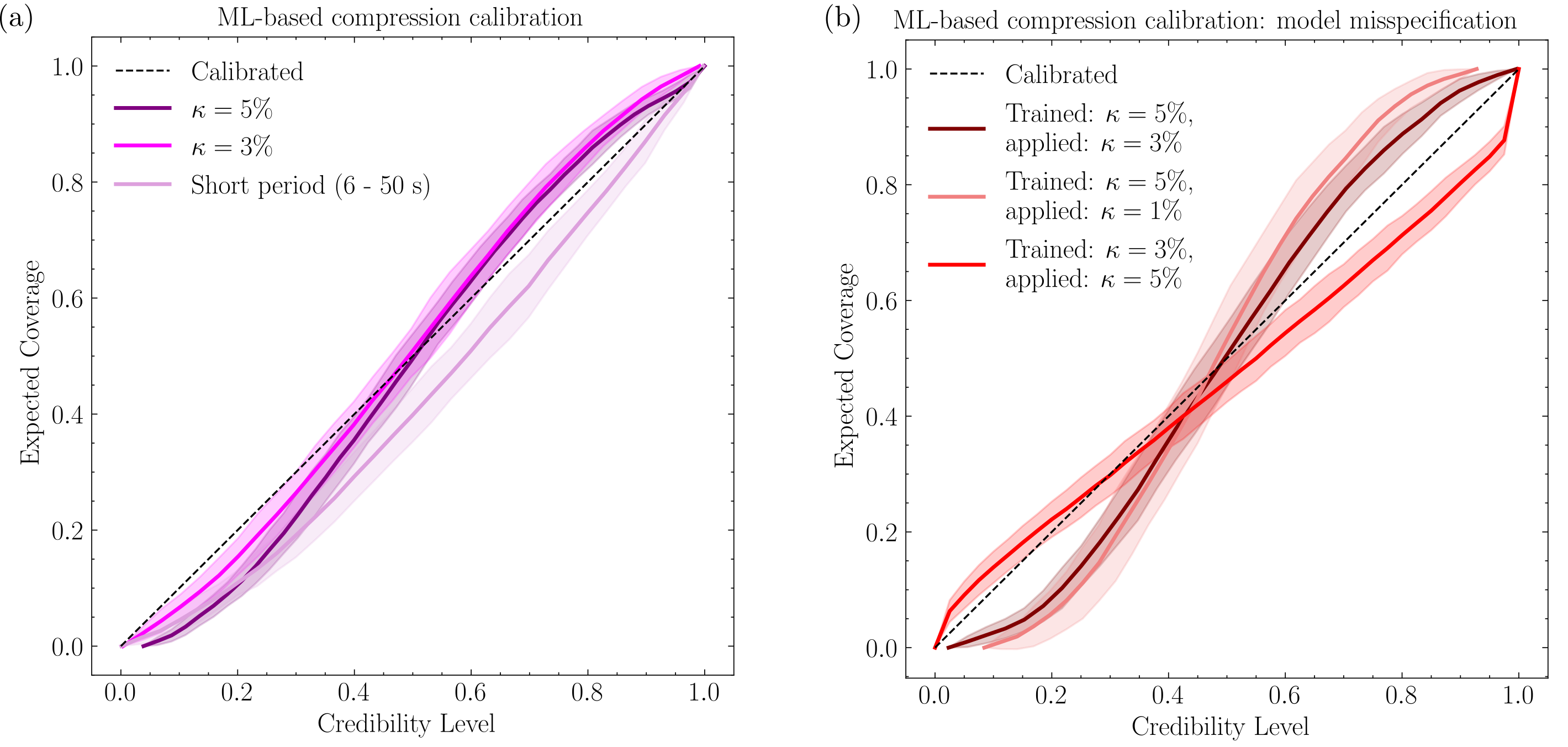}
   \caption{The calibration performance of the ML-based compression SBI approach under varied experimental configurations. Panel (a) shows the results under varying levels of Earth structure uncertainty, parametrised by $\kappa$ and differing filtering bands. Panel (b) explores model performance when exposed to greater or less Earth structure uncertainty than during initial training. }
   \label{fig:ML_calibration_curves}
\end{figure*}

In Figs. \ref{fig:SBI_calibration_curves} and \ref{fig:ML_calibration_curves} we show the calibration performance of the two SBI formalisms over various experiments. \Cref{fig:SBI_calibration_curves} repeats the varying experimental configurations from the main text using the score compression SBI approach. We find that across all experiments, the SBI approach is much better calibrated than a Gaussian likelihood-based inversion. We find minor bias for $\kappa=5\%$ as well as for the short period inversions, which we commented on in the main text. Nonetheless, the calibration statistics are acceptable and far improved over the likelihood-based alternative. 

\Cref{fig:ML_calibration_curves}(a) shows the performance of the ML-based compression SBI approach for the most challenging examples. For the varying Earth structure uncertainty levels $\kappa\in[3,5]\%$, we find a minor indication of underconfident (or conservative)
posteriors. We discuss this in the main text. However, for the short-period inversions we find an indication of bias. We found that training stability for the short period inversions was relatively poor, and the deep learning model began to overfit quite quickly. 
We attribute this to the more challenging modelling required to process the short period data, and flag this issue as an avenue for future work. In particular, improvements to model architecture or larger, more varied training datasets may be required to mitigate these issues. 

\Cref{fig:ML_calibration_curves}(b) shows a simple test of the effects of model misspecification on the ML-based compression approach. Specifically, we train the model on a given level of 1-D Earth structure uncertainty $\kappa_\textrm{trained}$, and apply it to artificial observations produced by a different level of uncertainty $\kappa_\textrm{applied}$. The results are relatively intuitive: models trained on a higher (lower) degree of Earth model uncertainty over(under)-estimate the uncertainty when applied to observations produced with lower (higher) Earth structure uncertainty. The encouraging result of these tests is that applying the model to slightly out-of-distribution data does not lead to model collapse or serious bias. However, as discussed in the main text, real observations will have very different imprints of model misspecification than those explored here, so more comprehensive testing is required in future work.

\subsection{Quantitative analysis of moment tensor solutions from each approach}

Here, we take a more in-depth look at the results of the 300 artificial inversions performed for each experimental configuration. We compute the $z$-score, which measures how far the true parameter lies from the posterior mean when scaled by the posterior uncertainty. 

For a given parameter $m$, the $z$-score is defined as
\begin{equation}
z = \frac{\mu_m - m_\mathrm{true}}{\sigma_m},
\label{eq:z_score}
\end{equation}
where $\mu$ and $\sigma$ denote the posterior mean and standard deviation, respectively. If the posterior is well calibrated, the $z$-scores computed across repeated synthetic experiments should be distributed according to a standard normal distribution, $\mathcal{N}(0,1)$. Deviations from this behaviour indicate either biased posterior estimates or misestimation of posterior uncertainties. While this assumes that the posterior is Gaussian (which is not the case in our examples), the $z$-score still provides a intuitive and interpretable diagnostic for the degree of discrepancy between the inferred moment tensor solutions and the true source. 

\begin{figure*}
 \centering
    \includegraphics[width=\textwidth]{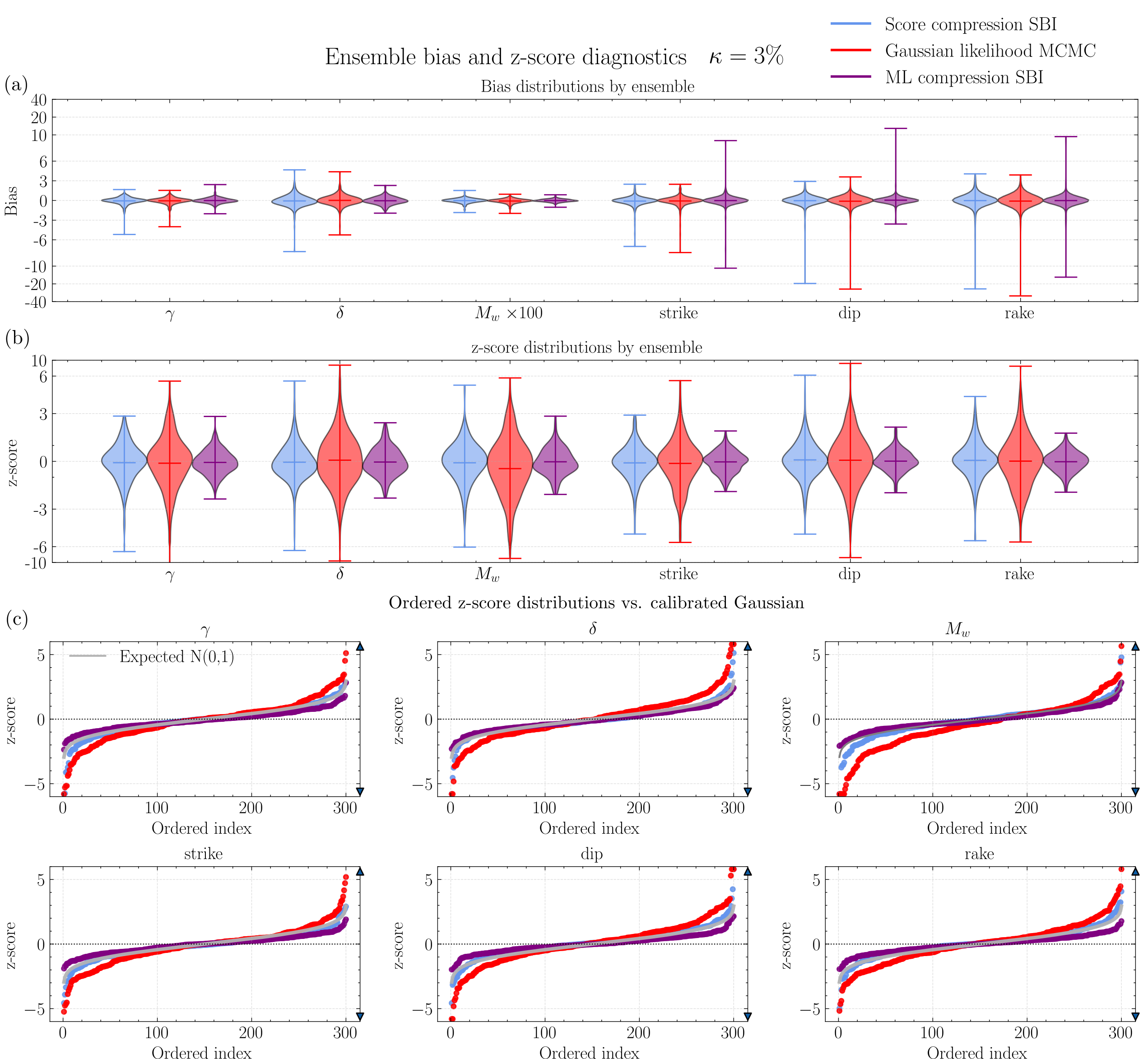}
   \caption{A detailed look at the posterior solutions for the $300$ artificial inversions used to perform coverage testing, for the $\kappa=3\%$ example. Panel (a) shows the per-parameter distributions of the bias between the posterior mean and the true solution. Panel (b) shows the same for the $z$-score distribution. Panel (c) shows the ordered per-parameter $z$-scores for all inversions against the expected distribution for well-calibrated (Gaussian) posteriors. }
   \label{fig:per_parameter_biases_3}
\end{figure*}

\begin{figure*}
 \centering
    \includegraphics[width=\textwidth]{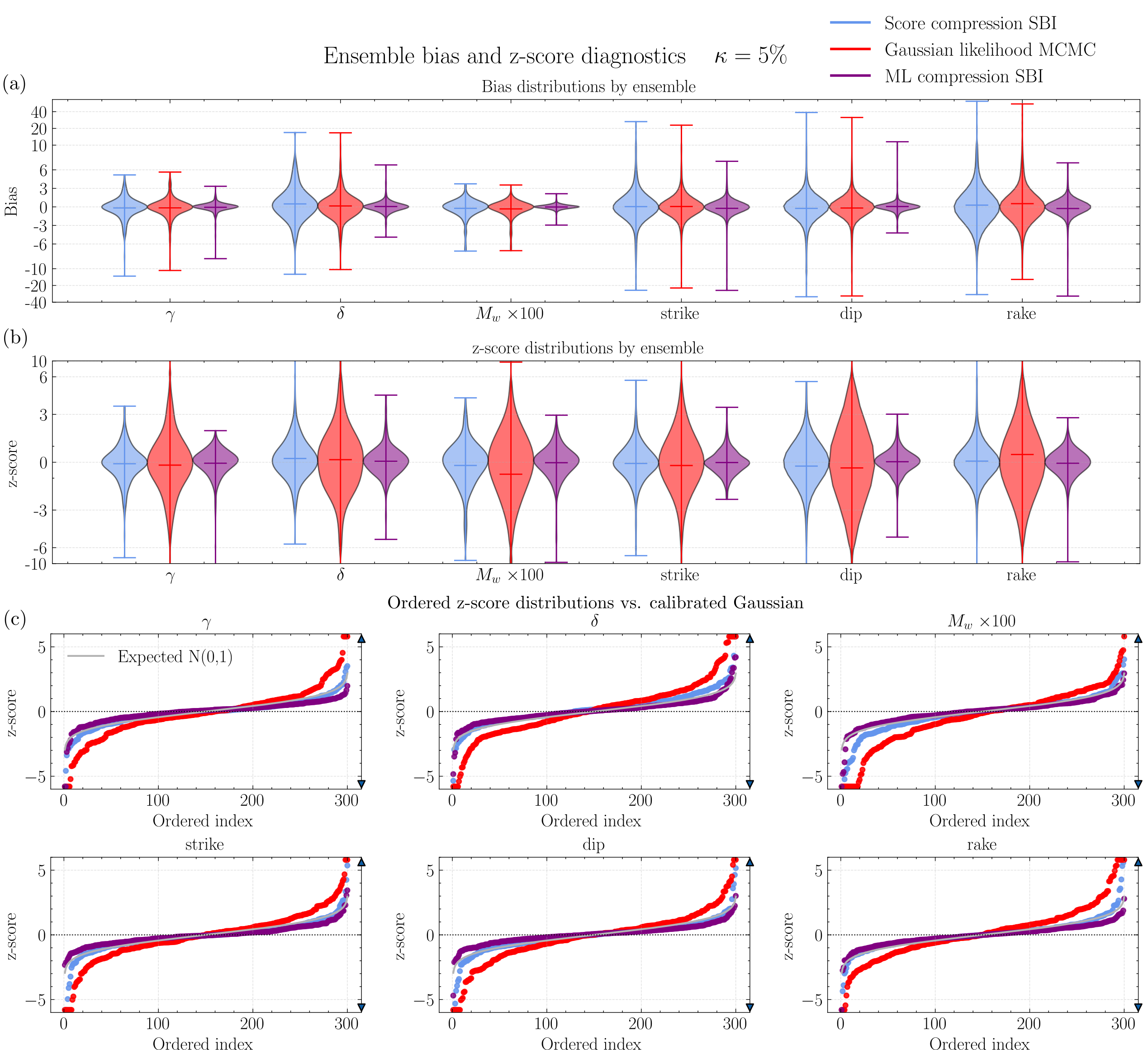}
   \caption{A detailed look at the posterior solutions for the $300$ artificial inversions used to perform coverage testing, for the $\kappa=5\%$ example. Panel (a) shows the per-parameter distributions of the bias between the posterior mean and the true solution. Panel (b) shows the same for the $z$-score distribution. Panel (c) shows the ordered per-parameter $z$-scores for all inversions against the expected distribution for well-calibrated (Gaussian) posteriors. }
   \label{fig:per_parameter_biases_5}
\end{figure*}

\begin{figure*}
 \centering
    \includegraphics[width=\textwidth]{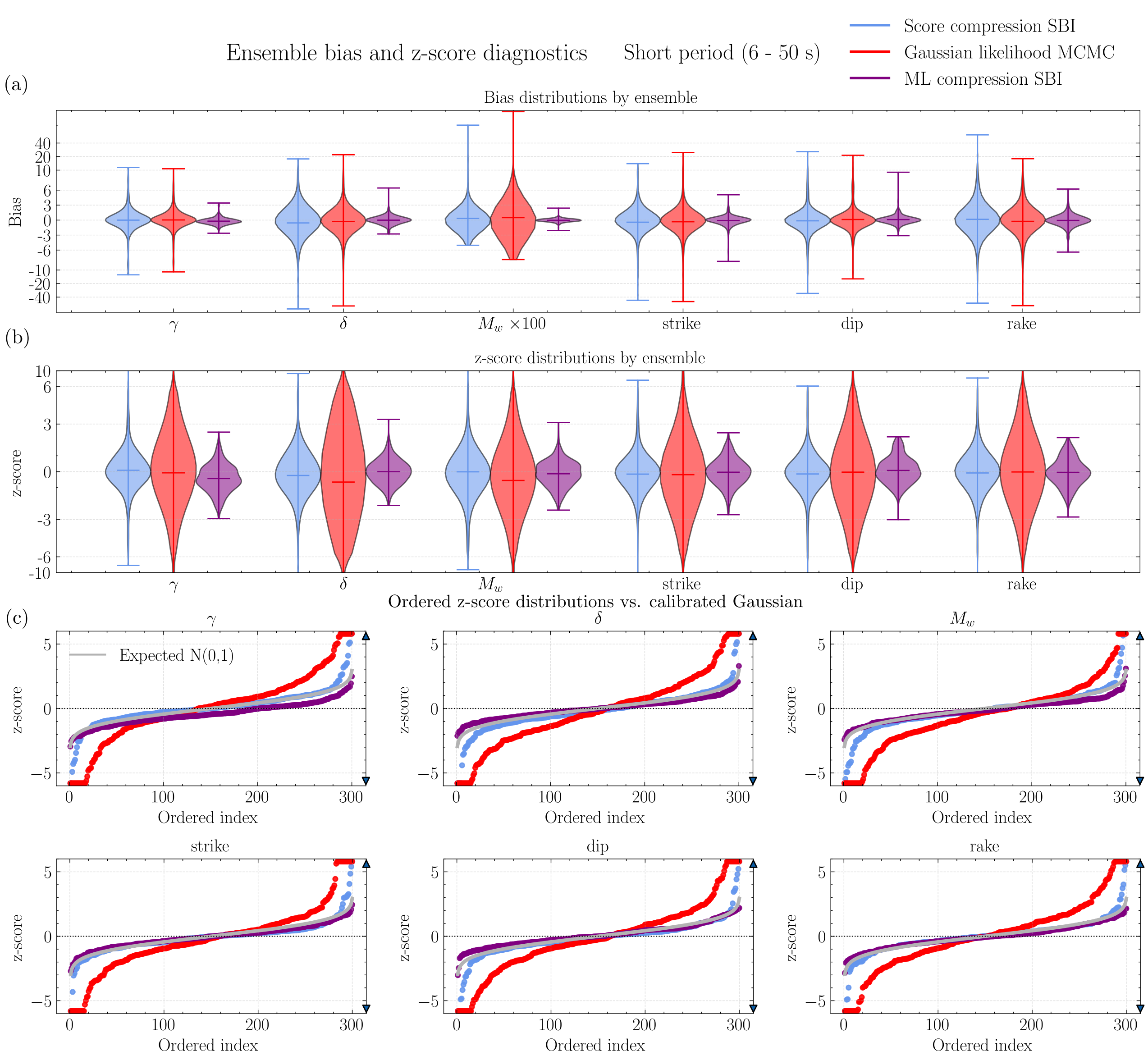}
   \caption{A detailed look at the posterior solutions for the $300$ artificial inversions used to perform coverage testing, for the short period example. Panel (a) shows the per-parameter distributions of the bias between the posterior mean and the true solution. Panel (b) shows the same for the $z$-score distribution. Panel (c) shows the ordered per-parameter $z$-scores for all inversions against the expected distribution for well-calibrated (Gaussian) posteriors. }
   \label{fig:per_parameter_biases_short}
\end{figure*}

We show the distributions of bias and $z$-score for each parameter independently for different experimental configurations in Figs. \ref{fig:per_parameter_biases_3}, \ref{fig:per_parameter_biases_5}, and \ref{fig:per_parameter_biases_short}. 

\Cref{fig:per_parameter_biases_3} shows that, for $\kappa=3\%$ Earth structure uncertainty, the degree of bias in the inferred solutions is similar across all approaches. The ML-based compression approach performs only slightly better in producing precise posterior solutions. However, the Gaussian likelihood approach produces poor uncertainty estimates that lead to over-confidence. The two SBI-based approaches produce better calibrated uncertainties, with closer alignment between the posterior solution $z$-scores and the expected statistical distribution in \Cref{fig:per_parameter_biases_3}(c). We see evidence of the mild bias of the score compression SBI approach, and the under-confidence (conservativeness) of the ML-based compression approach, which was discussed in the main text.

At $\kappa=5\%$, \Cref{fig:per_parameter_biases_5} shows that the degree of bias in both the Gaussian likelihood approach and the score compression approach increases significantly. While the score compression approach largely accounts for this with larger uncertainties, the Gaussian likelihood approach is very overconfident. On the other hand, the ML-based approach produces posterior solutions that are much more precise (i.e. they have a much narrower distribution of bias). These differences are apparent across all the focal mechanism parameters. 

We find similar results for the shorter period inversions, with filtering between $6-50$ s, in \Cref{fig:per_parameter_biases_short}. Again, the ML-based approach is much more precise than the alternatives in recovering the true source parameters. However, as highlighted in the main text, we also find some minor indications of bias and miscalibration: for instance, the CLVD component $\gamma$ is systematically under-estimated. 

We also report the mean per-parameter biases and posterior standard deviations of the $\kappa=3\%$ inversions in \Cref{table:uncertainties_and_bias_k3} and for the $6-50$ s filtering band inversions in \Cref{table:uncertainties_and_bias_short} ($\kappa=5\%$ is given in the main text). For $\kappa=3\%$, the Gaussian likelihood and score compression SBI biases are small but significant. However, for the shorter period inversions the Gaussian likelihood approach has major biases. These underscore the findings above and in the main text. In particular, $M_w$ is systematically underestimated by over one standard deviation of the posterior ($\sigma=0.37\times10^{-2}$, $\textrm{bias}=0.50\times10^{-2}$). The positive isotropic component is also systematically under-estimated by around $50\%$ of $\sigma$.

\newpage

\section{Real data inversions}

\subsection{Croatia 2-D composite Earth model}

\begin{figure*}
 \centering
    \includegraphics[width=\textwidth]{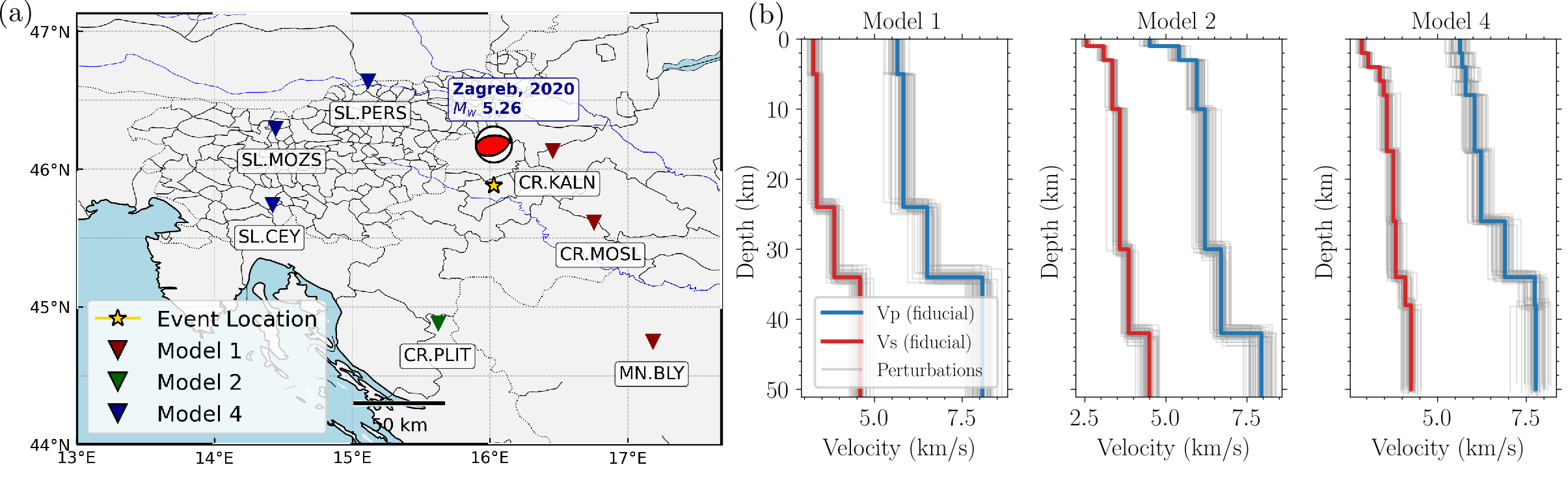}
   \caption{The 2-D composite Earth model adapted from \citet{hu2025bayesian} to include 1-D velocity model perturbations. Panel (a) shows the network configuration and which model was used in each region. Panel (b) shows the model velocities as a function of depth and the corresponding $\kappa=3\%$ perturbations used in the main text. We keep model numbering consistent with \citet{hu2025bayesian}.}
   \label{fig:croatia_composite_model}
\end{figure*}

\Cref{fig:croatia_composite_model} shows the 2-D composite Earth model used to analyse the Zagreb, Croatia event described in the main text. It has been adapted from \citet{hu2025bayesian} to include 1-D velocity model perturbations. The simplistic treatment of velocity model perturbations used in this work (with independent random perturbations applied to layer thickness, $V_p$ and $V_s$) can lead to unphysical effects such as occasional slower layers at greater depths. We therefore smoothed Model 4 and reduced the number of layers relative to the model used in \citet{hu2025bayesian}; \Cref{fig:croatia_composite_model} shows the smoothed Model 4 used in this work. 

These unrealistic perturbations could be addressed in future work, ideally using data-derived tomographic ensembles \citep{chiang2025bayesian}.

\begin{table*}
\centering
\begin{tabular}{l l l c c c c c c}
\hline
Method & Compression & Metric & $\gamma$ & $\delta$ & $ M_w (10^{-2})$ & strike & dip & rake \\
\hline
\multirow{4}{*}{SBI}
 & \multirow{2}{*}{Optimal score} & $\sigma$ & 0.42 & 0.77 & 0.28 & 0.64 & 0.49 & 0.78 \\
 &  & Bias & -0.06 & -0.07 & -0.02 & -0.09 & -0.03 & -0.02  \\
 & \multirow{2}{*}{Deep learning} & $\sigma$ & 0.53 & 0.65 & 0.26 & 0.69 & 0.56 & 0.87 \\
 &  & Bias & -0.01 & -0.04 & -0.01 & -0.02 & 0.05 & -0.01 \\

\hline
\multirow{2}{*}{Gaussian}
 & \multirow{2}{*}{ -} & $\sigma$ & 0.26 & 0.42 & 0.17 & 0.38 & 0.30 & 0.47 \\
 &  & Bias & -0.05 & 0.01 & -0.08 & -0.07 & -0.11 & -0.09  \\
\hline
\end{tabular}
\caption{Average posterior standard deviations ($\sigma$) and biases for each inference method over the 300 artificial inversions with Earth structure uncertainty of $\kappa=3\%$. Bias is computed per inversion by computing the difference between the mean of the modelled posterior and the true source parameters. $M_w$ values are reported in units of $10^{-2}$, all other values in units of degrees. }
\label{table:uncertainties_and_bias_k3}
\end{table*}

\begin{table*}
\centering
\begin{tabular}{l l l c c c c c c}
\hline
Method & Compression & Metric & $\gamma$ & $\delta$ & $ M_w (10^{-2})$ & strike & dip & rake \\
\hline
\multirow{4}{*}{SBI}
 & \multirow{2}{*}{Optimal score} & $\sigma$ & 1.49 & 1.77 & 1.12 & 3.02 & 1.95 & 3.04  \\
 &  & Bias & -0.01 & -0.57 & 0.34 & -0.43 & -0.14 & 0.16 \\
 & \multirow{2}{*}{Deep learning} & $\sigma$ & 0.75 & 1.01 & 0.40 & 1.12 & 0.99 & 1.50\\
 &  & Bias & -0.21 & -0.02 & -0.05 & -0.09 & 0.10 & -0.07 \\

\hline
\multirow{2}{*}{Gaussian}
 & \multirow{2}{*}{ -} & $\sigma$ & 0.52 & 0.68 & 0.37 & 0.92 & 0.58 & 0.90 \\
 &  & Bias & 0.03 & -0.29 & 0.50 & -0.33 & 0.11 & -0.26  \\
\hline
\end{tabular}
\caption{Average posterior standard deviations ($\sigma$) and biases for each inference method over the 300 artificial inversions with Earth structure uncertainty of $\kappa=5\%$ and shorter period ($6-50$ s) data. Units as before. }
\label{table:uncertainties_and_bias_short}
\end{table*}

\subsection{Quantitative evaluation of the posterior predictive checks}

For both the real event inversions in the main manuscript, we perform 10,000 posterior predictive simulations. We use these to quantify how closely each of the modelled posteriors can reproduce the observations. 

We emphasise that this analysis is not intended as a conventional posterior predictive check in the sense of assessing whether the observed data are typical under the posterior predictive distribution. Instead, our goal is to evaluate the capability of each inferred posterior to reproduce the salient features of the observations when marginalising over a large number of nuisance parameters. This is useful since our ML-based compression treats station-specific time-shifts as extra nuisance parameters during the inversion

Concretely, we evaluate a set of complementary distance or misfit metrics between each realisation and the observed data. For a given metric, we then retain the subset of posterior predictive samples that minimise that metric, and report the mean misfit over this ``best-matching'' subset. This procedure provides a lower-bound estimate on the discrepancy achievable by each posterior under a given notion of similarity, and allows us to compare different inference strategies in terms of their ability to reproduce the observations. We use the following metrics to serve as somewhat independent checks on waveform similarity:

\begin{enumerate}
    \item \textbf{Reduced chi-squared ($\chi^2_\mathrm{red}$).} 
    Covariance-weighted waveform misfit (see main text for details).

    \item \textbf{Correlation misfit.} 
    One minus the Pearson correlation between observed and synthetic waveforms to quantify phase and shape agreement.

    \item \textbf{Envelope misfit.} 
    Mean absolute relative difference between the signal envelopes.

    \item \textbf{Power misfit.} 
    Relative difference in total per-station-component waveform power.

    \item \textbf{Mean squared error (MSE).} 
    Unweighted elementwise $\ell_2$ misfit between observed and synthetic waveforms.
\end{enumerate}

\begin{table*}
\centering
\begin{tabular}{lccccc}
\hline
Method & $\chi^2_\textrm{red}$ & Correlation misfit & Envelope misfit & Power misfit & MSE \\
\hline
Score compression SBI & 0.563 & 0.052 & 0.421 & 0.001 & 3.72e-13 \\
ML compression SBI & 0.590 & 0.095 & 0.455 & 0.001 & 5.04e-13 \\
Gaussian likelihood & 0.606 & 0.110 & 0.419 & 0.001 & 7.09e-13 \\
\hline
\end{tabular}
\caption{LV2: Best-subset posterior predictive performance. Each column reports the mean value of a given metric computed on the top 50 posterior predictive samples selected according to that metric. Lower values indicate closer fits for all metrics.}
\label{table:best_subset_ppc}
\end{table*}

\begin{table*}
\centering
\begin{tabular}{lccccc}
\hline
Method & $\chi^2_\textrm{red}$ & Correlation misfit & Envelope misfit & Power misfit & MSE \\
\hline
Score compression SBI & 2.648 & 0.055 & 0.482 & 0.008 & 1.53e-11 \\
ML compression SBI & 2.808 & 0.083 & 0.458 & 0.005 & 1.90e-11 \\
Gaussian likelihood & 3.575 & 0.182 & 0.438 & 0.002 & 3.97e-11 \\
\hline
\end{tabular}
\caption{Croatia: Best-subset posterior predictive performance. Each column reports the mean value of a given metric computed on the top 50 posterior predictive samples selected according to that metric.  Lower values indicate closer fits for all metrics.}
\label{table:best_subset_ppc_croatia}
\end{table*}

Across both Tables S3 and S4, we find the SBI approaches produce moment tensor solutions that are competitive with or improve upon the Gaussian likelihood approach in recovering the observation. In Table S3, the Gaussian likelihood approach produces worse fitting ``closest'' solutions than the SBI approaches across all metrics but the envelope misfit. Interestingly, the score compression SBI approach produces better fitting synthetics than the ML-based approach across all metrics, despite its lower effective model dimensionality (since the ML-based compression approach has greater expressivity through its sampling over station-specific time-shifts). This provides some evidence that for the LV2 event, the score compression approach gives marginally more plausible solutions than the ML-based compression approach. 

In Table S4, the results are more mixed, with each approach performing better or worse depending on the metric used. This indicates that all approaches produce roughly comparable solutions in terms of reproducing the observation.

We emphasise that these are not traditional posterior predictive checks, which we avoid since we used a different (broader) prior for the ML-based compression approach. In particular, models with broad posterior predictive variance may achieve low best-subset misfits despite assigning low probability to observation-like realisations. For this reason, the results in Tables S3-S4 are best interpreted as a diagnostic of model expressivity and solution ability to reproduce the observations. They are sufficient to show that the ML-based compression SBI solutions yields comparable fits to the observations.  

\newpage
\bibliographystyle{gji}
\bibliography{bibliography}